\documentclass[useAMS,usenatbib]{mn2e}

\def\aap{AA}

\def\apjl{ApJL}
\def\apss{APSS}
\def\mnras{MNRAS}
\def\apj{ApJ}
\def\apjs{ApJS}
\def\aj{AJ}

\def\pasj{PASJ}
\def\nat{Nat}

\def\dgemail{gilmanda@ucla.edu}

\usepackage[titletoc]{appendix}
\usepackage{graphicx}
\usepackage{float}
\usepackage{amssymb}
\usepackage{amsfonts}
\usepackage{amsmath} 
\usepackage{color}
\usepackage{wrapfig}
\usepackage{hyperref}

\title{Strong lensing signatures of luminous structure and substructure in early-type galaxies } 
\author[Gilman et al.]{\parbox{\textwidth}{
  Daniel Gilman$^{1}$\thanks{\dgemail},
  Adriano Agnello$^{1,2}$,
  Tommaso Treu$^{1,\dag}$,
  Charles R. Keeton$^{3}$,
  Anna M. Nierenberg$^{4}$}
  \\
  \\
\parbox{\textwidth}{
$^{1}$Department of Physics and Astronomy, University of California,
Los Angeles, CA 90095, USA \\
$^{2}$ESO-European Southern Observatory, D-85748 Garching bei
M\"unchen, Germany \\
  $^\dag$ Packard Fellow.\\
$^{3}$ Department of Physics and Astronomy, Rutgers, the State University of New Jersey, 136 Frelinghuysen Road, Piscataway, NJ 08854 \\
$^{4}$ Center for Cosmology and AstroParticle Physics, The Ohio State University, Columbus OH 43204, USA \\
}
}

\begin{document}

\voffset-.6in

\date{Accepted . Received }

\pagerange{\pageref{firstpage}--\pageref{lastpage}} 

\maketitle

\label{firstpage}

\begin{abstract}
The arrival times, positions, and fluxes of multiple images in strong lens systems can be used to infer the presence of dark subhalos in the deflector, and thus test predictions of cold dark matter models. However, gravitational lensing does not distinguish between perturbations to a smooth gravitational potential arising from baryonic and non-baryonic mass. In this work, we quantify the extent to which the stellar mass distribution of a deflector can reproduce flux ratio and astrometric anomalies typically associated with the presence of a dark matter subhalo. Using Hubble Space Telescope images of nearby galaxies, we simulate strong lens systems with real distributions of stellar mass as they would be observed at redshift $z_d=0.5$. We add a dark matter halo and external shear to account for the smooth dark matter field, omitting dark substructure, and use a Monte Carlo procedure to characterize the distributions of image positions, time delays, and flux ratios for a compact background source of diameter 5 pc. By convolving high-resolution images of real galaxies with a Gaussian PSF, we simulate the most detailed smooth potential one could construct given high quality data, and find scatter in flux ratios of $\approx 10\%$, which we interpret as a typical deviation from a smooth potential caused by large and small scale structure in the lensing galaxy. We demonstrate that the flux ratio anomalies arising from galaxy-scale baryonic structure can be minimized by selecting the most massive and round deflectors, and by simultaneously modeling flux ratio and astrometric data.  
\end{abstract}

\section{Introduction}

One of the most robust predictions of cold dark matter models is that
galaxy and cluster scale halos should host a large number of subhalos,
described by a steep mass function of the form $dn/dM\propto M^{-1.9}$ \citep{Kly++99,Moo++99,Kra09}. Observational evidence against this prediction would force a revision of the standard model in favor of more exotic kinds of dark matter. For example, dark matter models with non-negligible free streaming lengths, such as keV scale sterile neutrinos are expected to manifest themselves as a cutoff in the subhalo mass function \citep{Colombi++96,Vogelsberger++16,Bose++16,Lovell++16,Nie++13,Menci++16}.

The standard test of this prediction consists of measuring the abundance of luminous satellites around galaxies such as the Milky Way. Significant efforts over the past decades have shown that indeed the abundance of luminous satellites is lower than what is predicted for subhalos. However, the interpretation of this tension is ambiguous. Low mass subhalos might not exist in sufficient numbers, or could simply not be capable of forming stars, and thus be invisible \citep[]{Nie++13,Nie++16,Guo++2011,Starkenburg++13,Wetzel++16,Sawala++16,DespVeg16}.

For almost two decades it has been recognized that strong gravitational lensing offers an alternative and potentially very clear observational test of this fundamental cosmological prediction, whereby the properties of dark matter subhalos are probed directly by their impact on the arrival times, positions, and flux ratios of lensed images. A variety of techniques have been developed over the years to carry out these measurements, and applied to a variety of datasets. Broadly speaking, the measurements obtained so far are consistent with cold dark matter predictions, although their sensitivity has been limited by sample sizes and quality of the data. Fortunately, sample size and data quality are rapidly improving, and it is therefore important to explore all sources of potential systematic errors in the applications of this technique.

The goal of this paper is to study the impact of baryonic substructure on the application of the so-called lensing anomalies (in time delays, positions, and fluxes) to the study of dark matter substructure.  The term anomalies arises from the standard approach in strong lensing communities where the mass distribution of a galaxy is described as the superposition of a `smooth' mass distribution representing most of the luminous and dark matter, plus a clumpy distribution of dark substructures typically in the range $10^6 - 10^9 M_{\odot}$. This approach is motivated by the fact that a simple smooth component is generally sufficient to capture the main features of the lensing observables, while substructure below a certain threshold effectively behaves as smooth for given bakcground source size.

Typically, the positions and arrival time delays between lensed images are reproduced by a smooth lens model, while the ratios of the magnifications (also known as flux ratios) may or may not be recovered. If the observed flux ratios cannot be recovered with `smooth' lens models, the flux ratios are deemed anomalous, and the discrepancy is attributed to the presence of a compact, massive perturbing mass near an image, such as a dark subhalo. Similarly, the inability of smooth models to reproduce image arrival times and astrometry (both for compact and extended sources) gives rise to the so-called time delay and astrometric anomalies \citep{Che++07,K+M09,Inoue++12}. Both astrometric and flux ratio anomalies have been used to characterize the distribution, abundance, mass function, and density profile of subhalos \cite[e.g.,][]{M+M01,D+K02,Chi02,Vegetti:2010p30241,Veg++12,FadelyKeeton12,Veg++14,V+V14,Nie++14,Hezaveh++16}.

However, the presence of dark subhalos is not the only possible explanation for the observed anomalies.  Stellar microlensing \citep{Schechter:2003p701} and matter along the line of sight \citep{Met05,McCullyEtal2016,Xu++09} can give rise to anomalies in the positions and flux ratios of compact sources. The astrophysical noise from these features can be mitigated by observing sources that are sufficiently extended to smooth away microlensing, by observing at wavelengths unaffected by dust, and by carrying out multiplane lensing analysis.

In this study we focus on astrophysical noise arising from inhomogeneities in the stellar mass distribution of the lensing galaxy that may not be resolved at typical lens redshifts, and could potentially cause anomalies that could be conflated with the presence of dark subhalo. A clear and recent example is given by \citet{HsuehEtal16}, who show that the apparent flux ratio anomaly in the system B1555 can be readily explained by the presence of an elongated disk in the deflector, which is detected in high resolution imaging of the system.

This potential noise term was recognized early on. For example, \citet{M+S98} and \citet{Chi02} calculated the impact of globular clusters based on simple analytic models. \citet{MHB03} highlighted the importance of disk components in the statistics of flux ratios, considering their occurrence in early-type galaxies within nearby galaxy clusters. With improvements in sample size and data quality it is important to revisit theses issues and perform quantitative, systematic, and realistic calculations of the overall distribution of the anomalies induced by the stellar component on arrival times, positions, and fluxes of the multiple images. In this context, using numerical simulations, \citet{XuEtal2016} have shown that the density profiles in the vicinity of the Einstein radius of simulated galaxies are not as simple as those traditionally used to model galaxy-scale lenses, which could amplify the impact of the baryonic mass component of a lens.

In this work, we address this problem by using real Hubble Space Telescope (HST) observations of nearby galaxies to build mock lenses with realistic baryonic mass distributions, and varying degrees of morphological complexity.  We complement this baryonic mass component with an NFW dark matter halo, omitting dark substructure in order to isolate the effect of luminous matter. From the degree to which flux ratios from our mock lenses can be recovered with smooth lens models, we quantify the anomalies that can be attributed to the baryonic mass of a deflector (we identify stars with baryons but neglect the contribution of gas, which is assumed to be smooth on the relevant scales). 

This paper is structured as follows. In Section~\ref{sect:setup}, we detail our procedure for building mock lenses from HST images of nearby galaxies, the type of lens models considered in this work, and our fitting methodology. In Section~\ref{sect:results}, we present the results of our comparison between smooth models and realistic simulated lenses.  In Section~\ref{sect:conclusions}, we summarize the results of our analysis, and discuss the lessons learned in the context of ongoing and future strong lensing studies of dark matter.  
When needed to compute distances, we adopt a standard concordance cosmology with $\Omega_m=0.3$,  $\Omega_{\Lambda}=0.7$ and $h=0.7$, even though our results are independent of this choice. All of the lens simulations, ray-tracing and computation of lensing observables (positions, time-delays, magnifications) are performed using the {\tt{lensmodel}} software \citep{KeetonGLens2003}.

\section{Building and fitting mock lens systems}
\label{sect:setup}

In this Section we describe in detail our procedure to build mock lens
systems and then fit them with lens models. We begin by describing our
source of high resolution images about the surface brightness of
early-type galaxies in Section~\ref{ssec:sb}. In
Section~\ref{sect:lenses} we summarize how we obtain the global
structural parameters for the lens galaxies, either from the
literature or our own fits to the light. In Section~\ref{ssec:tomass} we describe
how we convert surface brightness into lensing potential, accounting
for the dark matter halo and external shear. In
Section~\ref{ssec:models} we describe the ingredients of our five
different mass models used to produce mock lenses and fit them. In
Section~\ref{ssec:mocks} we describe the process of generating data sets with our mock lenses for two of our models that are derived from the real HST images, and in Section~\ref{sect:fitting} we describe the process of fitting two smooth lens models to data obtained from the mock lenses.

\subsection{The stellar surface brightness of early-type galaxies at high resolution}
\label{ssec:sb}

The starting point for our mocks is archival Hubble Space Telescope observations early-type galaxies from the nearby Virgo and Coma clusters \citep{Fer++06b,COMAsurvey08}. In order to obtain a sample that is representative of lensing galaxies we select all the elliptical and lenticular galaxies with available HST images, central velocity dispersions between 165 and 320 km s$^{-1}$, and ellipticities in the range 0.05-0.43. We limit our selection to galaxies imaged with the Advanced Camera for Survey with filters F814W or F850LP, in order to minimize the effects of dust, and map the stellar light as closely as possible, while taking advantage of the wider field of view of view and finer pixel scale than the infrared channel of Wide Field Camera 3. The sample displays a variety of interesting features, including globular clusters, disks, tidal tails and shells, which we take as representative of the kind of baryonic structure and substructure that we are interested in studying. In Table~\ref{table:gal_list}, we list the galaxies used in our data set, along with their relevant physical parameters.

We avoid galaxies with prominent dust lanes, and sources of visual contamination obvious to the naked eye, as these features would be problematic in our procedure for assigning mass to light, which we discuss in the next section. There are often bright galaxies or stars in the line of sight, which we replace with a smooth interpolation of the main lens profile. We do not expect this to significantly affect our results, however, as we avoid generating lenses where an image would be located near one of these defects. When the computation of the lensing potential, according to our normalization procedure, requires information from pixels outside the ACS field of view, we extrapolate a smooth model fit to the light into these regions. After solving the lens equation, we ensure that no lensed images land in an interpolated region.

We note that real lens samples tend to be dominated by high velocity dispersion galaxies above 240 kms$^{-1}$\citep{Aug++10,PaperIV}, due to their favorable lensing cross section. Surveys of high velocity dispersion galaxies \cite[e.g.][]{MassiveV} show that the most massive ellipticals tend to be slow rotators, while low velocity dispersion galaxies, which are more likely to be fast rotators and host disks, are over-represented in our mock sample. As such, our sample is not representative of that of typical lens galaxies, and will likely result in an over-estimate of the contribution to time delay, astrometric, and flux ratio anomalies by the baryonic mass component of a deflector. In light of this, we interpret the fraction of anomalous systems in our analysis as an upper limit to the frequency with which one expects to encounter baryon-induced anomalies in a survey of real lensed quasars.
\begin{figure*}
{\includegraphics[trim=0cm 0.6cm 0cm
0cm,clip,width=.48\textwidth]{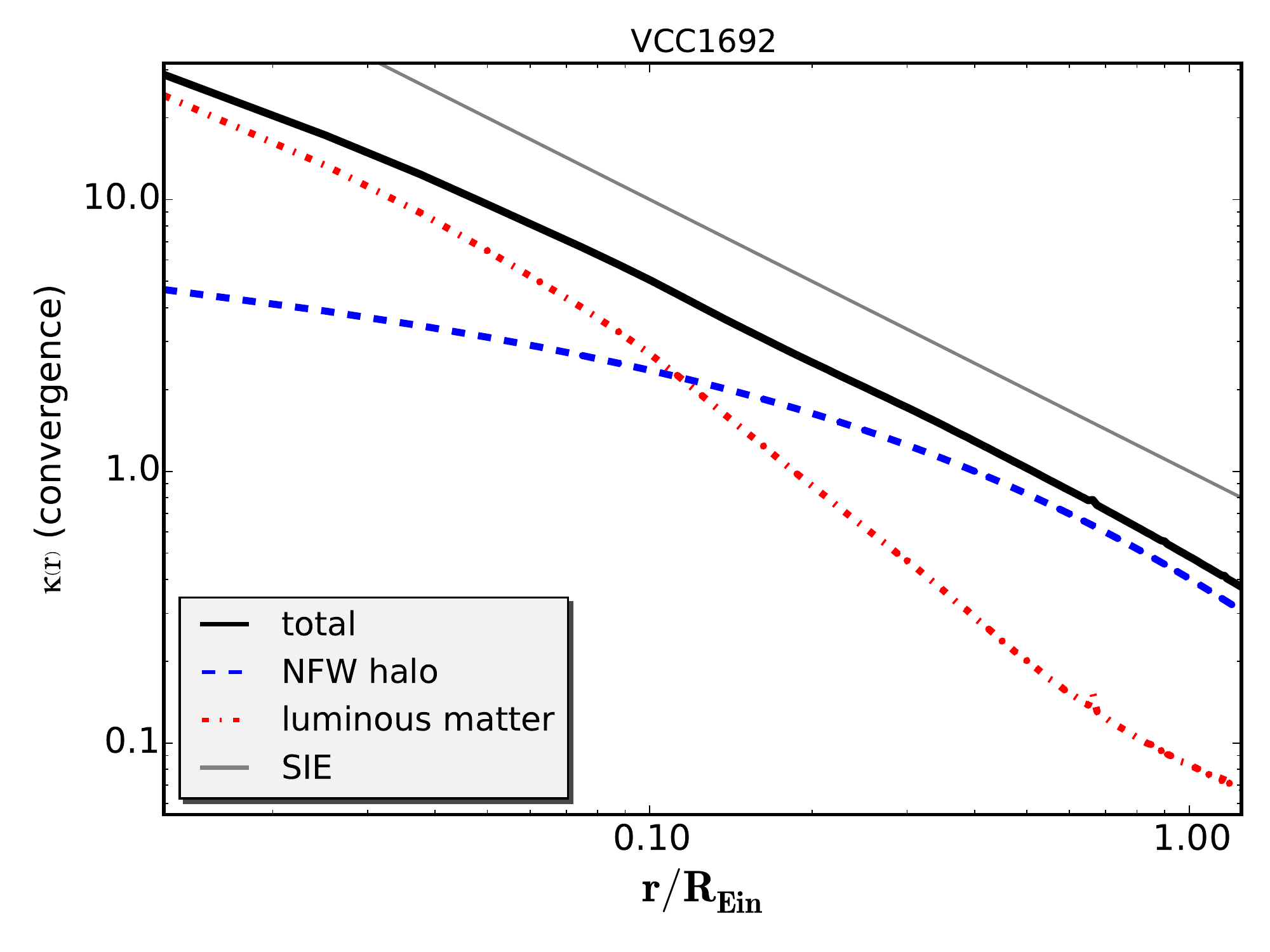}}
{\includegraphics[trim=0cm .6cm 0cm
0cm,clip,width=.48\textwidth]{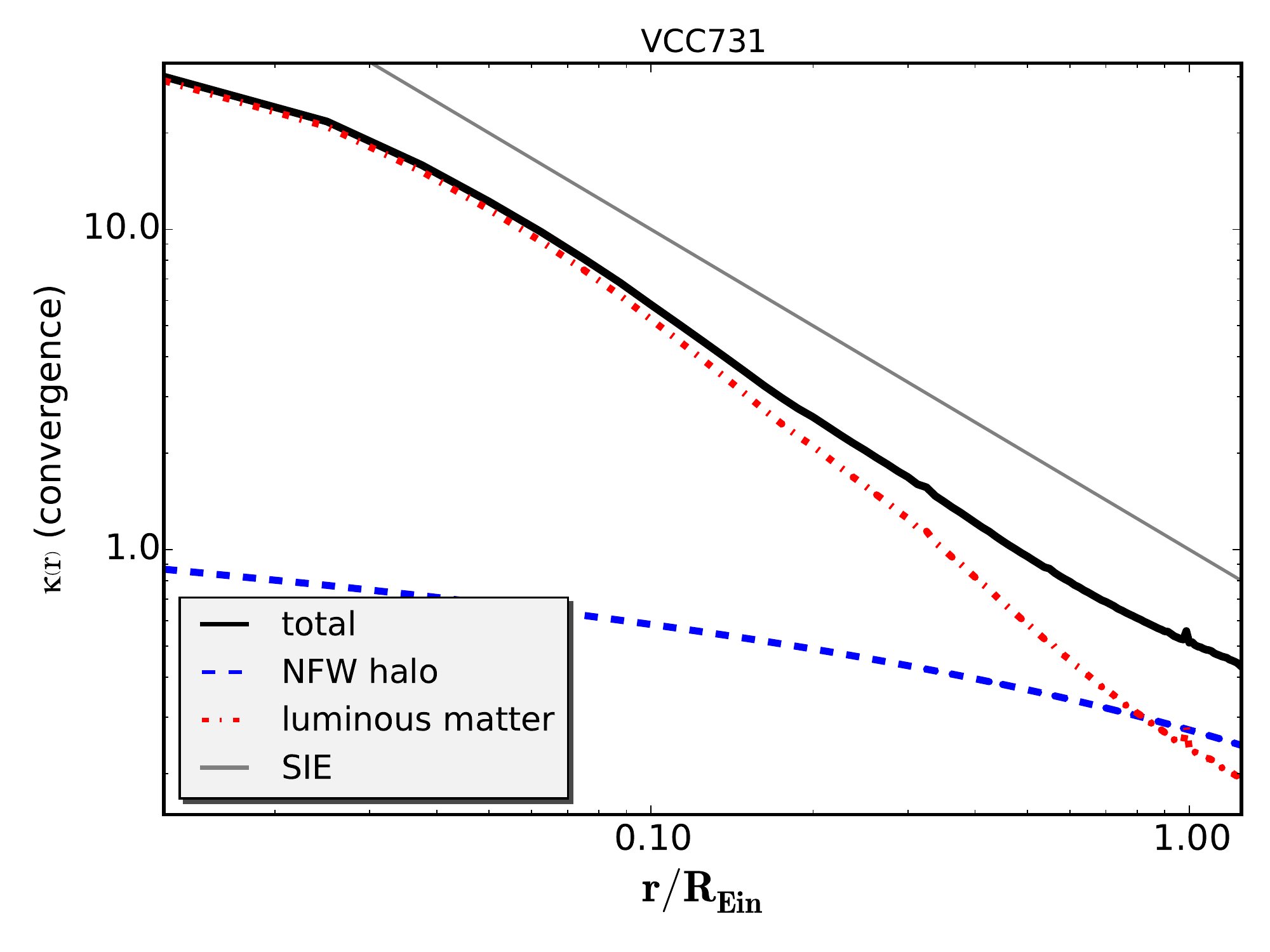}}
\caption{\label{fig:r_vs_kap}Convergence as a function of radius for a normalized map of surface mass density for the deflectors VCC1692 and VCC731. For reference, the slope of an SIE mass density profile (with arbitrary normalization) is shown in grey.}
\end{figure*} 
\subsection{Structural Parameters of the Sample Galaxies}
\label{sect:lenses}

In order to simulate the lensing properties of the galaxies in our sample, we require a measurement of central stellar velocity dispersion $\sigma_*$, half-light radius $R_{1/2}$, ellipticity $\epsilon$, position angle $\theta_{\epsilon}$, and a S{\'e}rsic index $n$ for each host galaxy. We draw measurements of the central velocity dispersion from the HyperLeda online catalog \citep{Makarov++2014} and from \cite{Ma++2014}, while measurements of the half light radii, ellipticity and position we obtain for Virgo objects from \citet{Fer++06b} and from HyperLeda.

When the parameters describing the host light distribution are not available in the literature, we derive them by fitting the light profiles with a single S{\'e}rsic component using {\tt {galfit} }  \citep{Peng:2002p217} and derive the parameters ourselves, mimicking the efforts of an observer attempting to model the luminous matter of a strong lens. The parameters that we adopt for each galaxy are summarized in 
Table~\ref{table:gal_list}.

\subsection{From surface brightness to surface mass density}
\label{ssec:tomass}

We transform the surface brightness maps of the galaxies into maps of surface mass density (convergence) in order to determine the gravitational lensing properties. In translating between surface brightness and surface mass density, we assume that light traces luminous matter in the field of view, with a constant stellar mass-to-light ratio. This is a conservative approach as it will assign higher masses to young star populations which tend to populate disky areas, relative to the old star populations which tend to populate the smooth elliptical component. Thus, by adopting a uniform stellar mass to light ratio we tend to increase the lensing signal of disky strucutures, consistent with our interpretation of our results as upper limits on the perturbative effect of baryonic structure on lensing data. 

For simplicity, we simulate all our systems as they would be observed for typical deflector and source redshifts $z_{d}=0.5$ and $z_{s}=1.5$. The smooth dark matter component of each deflector is described by a circular NFW halo, whose scale radius $R_s$ is taken to be 5$R_{1/2}$, where $R_{1/2}$ is the half-light radius of the target galaxy, in projection. We do not expect this choice for the dark matter normalization to effect our main results, as our choice for $R_s$ simply reflects the different spatial scales over which the smooth dark matter and baryonic mass component vary. While real NFW halos are unlikely to be circular, the NFW halo in our analysis serves only to boost the convergence within the Einstein radius to that of a typical deflector. Further, ellipticity in the NFW halo is, to some extent, degenerate with external shear, which we add as a separate component.

We compute the Einstein radius of each mock lens by exploiting the observational fact \citep{Tre++06,Koo++09} that in lens galaxies the stellar velocity dispersion $\sigma_*$ approximates, within a few percent, the velocity dispersion $\sigma_{\rm{SIE}}$ of the best fitting singular isothermal ellipsoid (SIE), for which the Einstein radius is given by
\begin{equation}
R_{\rm E} = 4\pi\left(\frac{\sigma_{\rm{SIE}}}{c}\right)^2\frac{D_{ds}}{D_s},
\end{equation} 
where $D_s$, and $D_{ds}$ are the angular diameter distances to the source, and from the deflector to the source, respectively. This equation is one of the consequences of the so-called bulge-halo conspiracy \citep{T+K02a,T+K04,Koo++06,Koo++09,D+T14}: the projected total mass density profile of early type galaxies is well described by a single power law with logarithmic slope $-1$. As a consistency check, we verify that the total convergence (after adding stellar mass to the light and a dark matter component) of our mock galaxies is well approximated by an isothermal profile, as shown in Figure~\ref{fig:r_vs_kap}. Also, we check that the stellar masses derived from our convergence maps are consistent with those reported by \citet{Gal++08b}. Details of the normalization procedure, based on empirical measurements of the relative abundances of stellar mass and dark matter, are given in Appendix~\ref{app:A}. 
In order to mimic the tidal field of the large scale structure expected at intermediate redshifts, we add, at random position angles, external shears of magnitude 0.05 or 0.08, which are typical shear magnitudes in strong lens systems \citep{H+S03}. 
\subsection{Description of the lens models}
\label{ssec:models}

In order to carry out our quantitative analysis of the lensing effects of unresolved stellar structures, we compare the lens configurations obtained from the high resolution mass maps (the ``truth''), with two models based on lower resolution data, and two simply parametrized smooth models commonly used in the literature. The two models based on a low resolution version of the ``truth'' are intended to simulate the best data that one could hope to extract from a distant lens using HST. The two simply parametrized lens models are meant to represent the models typically used as a reference to detect anomalies due to dark substructure.

Thus, in total, we consider five lens models, with the following
characteristics:

\begin{itemize}
\item Model 1 (real data) - \textit{``Truth''}. This model directly uses the image of the galaxy obtained by HST, converted to a convergence map as described in the previous section and Appendix \ref{app:A}. 
\end{itemize}
\begin{figure*}
\includegraphics[clip,trim=4.1cm 1.5cm 4.1cm 2cm,width=.325\textwidth]{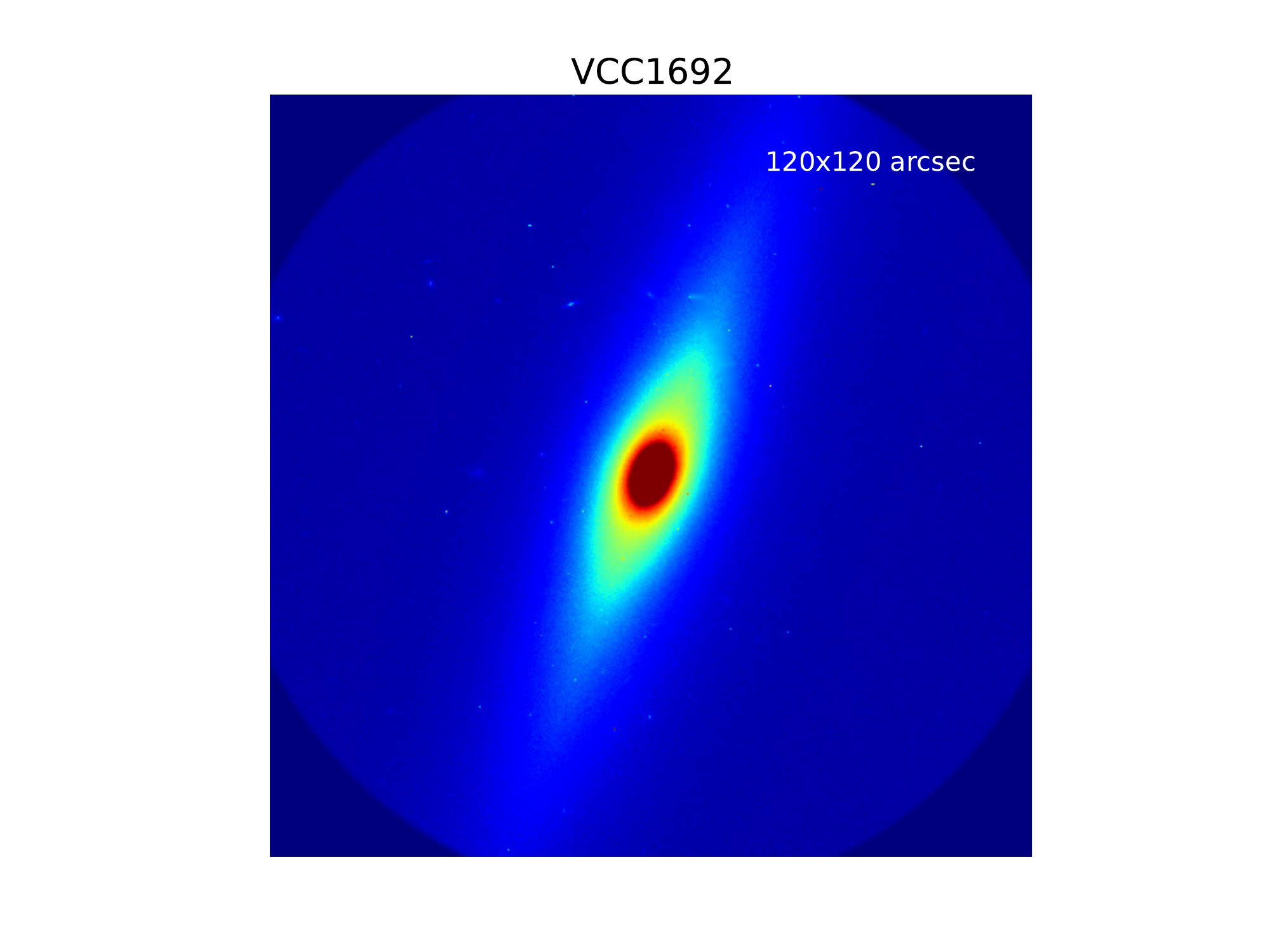}
\includegraphics[clip,trim=4.1cm 1.5cm 4.1cm 2cm,width=.325\textwidth]{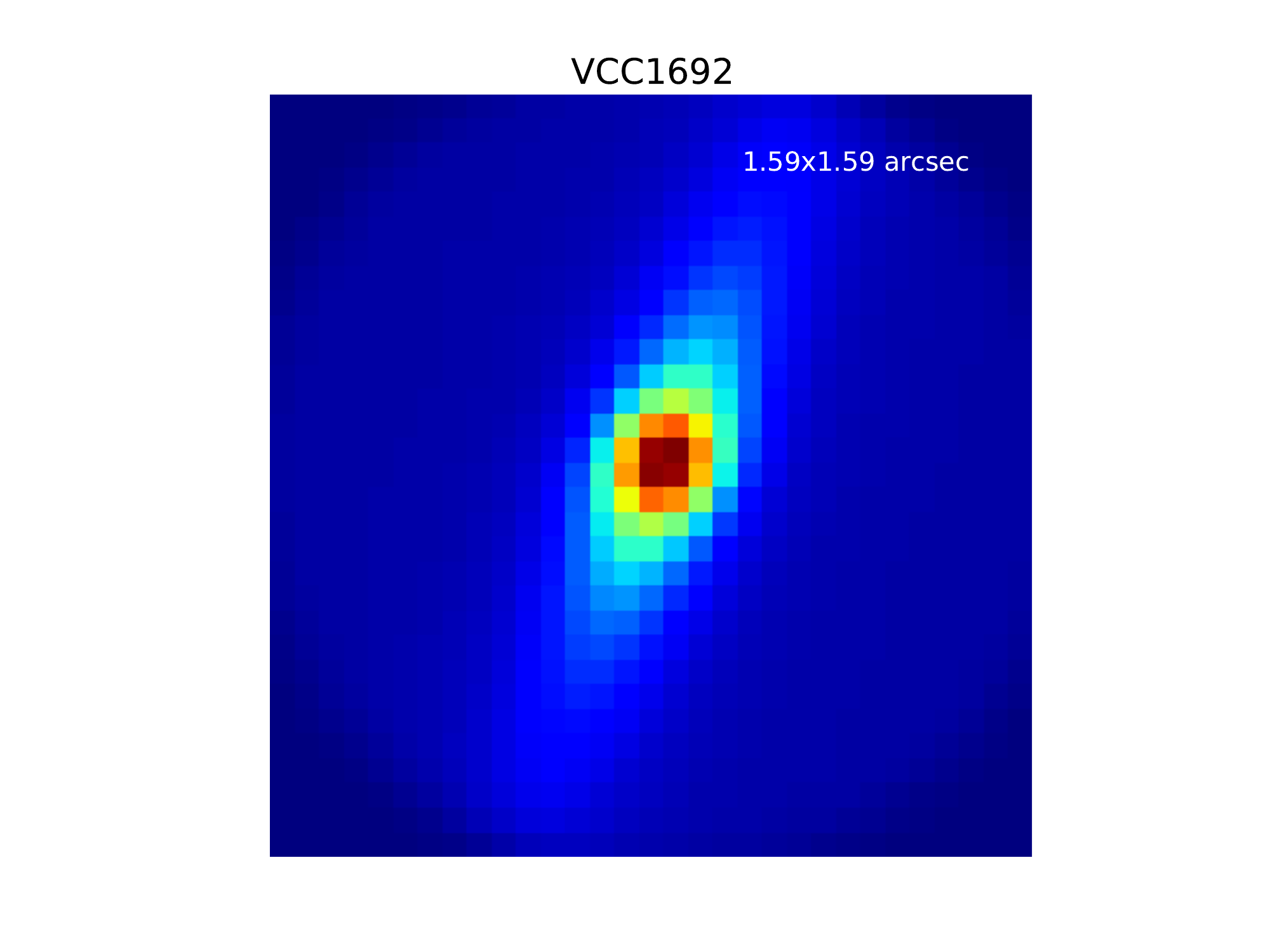}
\includegraphics[clip,trim=4.1cm 1.5cm 4.1cm 2cm,width=.325\textwidth]{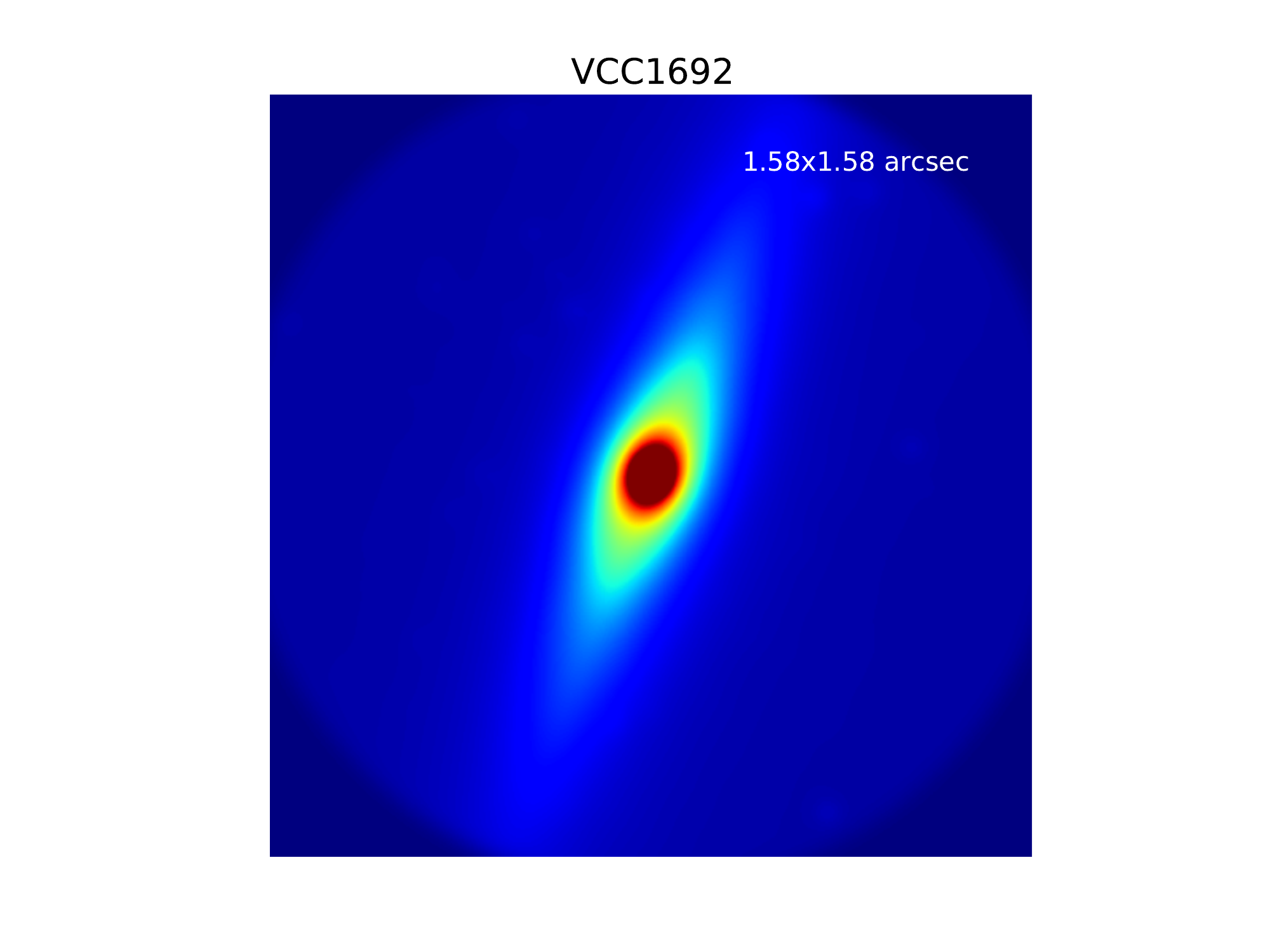}
\caption{\label{fig:VCC1692real_vs_mods}{\bf{\emph{Left:}}} Surface mass density of VCC1692 as it appears at a distance of 16 Mpc. \newline {\bf{\emph{Center:}}} The galaxy as it appears at redshift 0.5 (1280 Mpc) after rebinning pixels 80x80 to account for a loss of spatial resolution. \newline {\bf{\emph{Right:}}}  The galaxy after convolving with a PSF with FWHM of 80 pixels to simulate an observation of the galaxy where sub-pixel information has been recovered via dithering, effectively the best smooth model one could construct given HST data.}
\end{figure*}
\noindent We evaluate the following four models by their ability to reproduce the `real' data of Model 1:

\begin{itemize}
\item Model 2 (relies on real data) - \textit{``real HST"}.  This is a simulated single exposure of an HST image, including the effects of a Gaussian PSF, and pixelization. First, we rebin pixels of the \textit{Truth} model by a factor corresponding to the loss of spatial resolution going to $z_d=0.5$ from the native redshift of the galaxy. For example, translating the angular diameter distance of the Virgo cluster ($z=0.0038$) to $z=0.5$ changes image resolution by a factor of 80, so the image used in the \textit{Truth} model is rebinned 80x80. We then convolve the rebinned map with a Gaussian Point-Spread-Function (PSF) of FWHM = 2 pixels. We checked that the order of operations of rebinning and convolving does not affect the results. This model is meant to represent an attempt to fit the stellar mass of the lens by scaling the pixel values observed by HST. An example is shown next to the \textit{Truth} stellar mass distribution in Figure~\ref{fig:VCC1692real_vs_mods}.

\item Model 3 (relies on real data) - \textit{``HST Interpolated"}. This model simulates an HST image where the sub-pixel information has been recovered via dithering, thereby representing the best possible data set obtainable for these systems at a redshift of 0.5, approximating the Hubble PSF with a Gaussian PSF. In effect, this data has been smoothed over at a scale comparable to the Hubble PSF at redshift 0.5, thereby erasing structure on scales smaller than rebinning factor at redshift 0.5, thereby erasing structures on scales $<0.32$ kpc for Virgo galaxies, and $<0.51$ kpc for Coma galaxies. In practice, this model represents the best description of the stellar component that one could build from HST observations, using a smooth interpolation or a fit to the pixel data. As such, the degree to which this model reproduces the flux ratios from the \textit{Truth} model represents a noise floor for flux ratio data.  An example of the stellar mass distribution corresponding to this model is shown in Figure~\ref{fig:VCC1692real_vs_mods}.
\end{itemize}
The following two models are different from the previous three, as they are analytic functions fit to the data obtained from the \textit{Truth} model.
\begin{itemize}
\item Model 4 (fit to \textit{Truth} positions, time delays) - Singular isothermal ellipsoid with external shear (SIE). This model is physically motivated by the fact that the combined mass profile of baryons and a NFW halo is well approximated by an isothermal power law, as shown in Figure \ref{fig:r_vs_kap}. We do not include information about image magnification when performing the fit, and use positional and time delay uncertainties of 0.003" and 2 days to simulate the best data currently available.
\item Model 5 (fit to \textit{Truth} positions, time delays) - S{\'e}rsic + NFW halo (SNFW). We fit an elliptical S{\'e}rsic \citep{Sersic63} mass distribution and a NFW with external shear to image positions and time delays, with the same observational uncertainties as Model 4. The SNFW model has nearly double the number of free parameters as the SIE, which at face value suggests it would be a more adaptable functional form than the former, and better suited to representing a possibly complex distribution of baryonic and dark matter. However, models with too many free parameters are prone to degeneracies given the limited constraints available. We will consider this point again in Section 2.6. This model is meant to represent a practical approach which might be as close as possible to the best one can do, especially in the presence of bright lensed quasar images. 
\end{itemize}

We stress that because we do not explicitly add dark substructure to our mock lenses, the only source of small scale structures or non-smooth features, akin to the clumpy nature of dark matter substructure, is that of the baryons in the lensing galaxy, luminous satellites of the deflector, and background galaxies. Therefore, any discrepancy in flux ratios between the ``Truth'' model and models 4-5 is due entirely to a baryonic mass component that cannot be absorbed by the SIE or SNFW functions.

Similarly, with data of extraordinary quality, one could imagine using more flexible and complicated smooth lens models to describe the stellar mass component. This is captured in by the \textit{HST Interpolated} model, which provides a reasonable upper limit on the capability of a smooth lens potential to fully account for the baryonic structure of a lensing galaxy.

\subsection{Generating mock data sets}
\label{ssec:mocks}

For each of the three lens models based on real images (\textit{Truth}, \textit{Real HST}, \textit{HST Interpolated}), we manually place the source position within the astroid caustic so as to produce a cusp and a fold lens configuration. While the light traces mass hypothesis allows us to efficiently normalize and assemble realistic mock lenses, it introduces a significant complication. Shot noise in the HST images and discontinuities due to pixelization cause small scale variation in surface mass density that introduce a small scale pattern in the local magnification map. For a point source this would introduce a microlensing-like signal, which could introduce spurious scatter in the fluxes predicted by the \textit{Truth} model. We avoid this by modeling the background quasar as an extended source 5 parsecs in diameter, a procedure we describe in detail in Appendix~\ref{app:B}. For reference, this source size is roughly the size of a radio jet source (1-10 pc), but smaller than the narrow-line region (10-100 pc) \citep{M+M03}, and is large enough to avoid micro-lensing effects while preserving sensitivity to small scale structure in the image plane, and corresponds to 0.265 mas$^2$ in the source plane.
\newline \indent For the three mock deflectors (Models 1-3), we apply a Monte Carlo procedure: for each image configuration (cusp and fold), we randomly sample 250 source positions from a circular area in the source plane, centered on a reference source position guaranteed to produce a cusp or a fold lens. For each of the 250 new source positions, for each of the \textit{Truth}, \textit{Real HST}, and \textit{HST Interpolated} convergence maps we directly solve the lens equation to obtain 250 new sets of positions, time delays, and flux ratios. We do not add measurement noise in this process, as we are only interested in the effects of baryonic mass on these data. 
\newline \indent For the simply parametrized lens models (Models 4 and 5), we use the software package {\tt{lensmodel}} to fit an SIE and SNFW model to each of the 250 data sets, corresponding to each of the 250 sampled source positions, constraining the models by only astrometric and time delay data and demanding that the S{\'e}rsic halo and NFW halo are centered at the same location. We introduce a $\chi^2$ penalty to discourage {\tt{lensmodel}} from adopting unphysical characteristics, such as an NFW halo with a scale radius smaller than the stellar half-light radius.
\newline \indent We plot the resulting data for each of our models as histograms that characterize the distributions for each lensing observable, taking into account small variations in the the unknown source position. The scatter in the distributions of the \textit{Real HST} and \textit{HST Interpolated} data we obtain can be attributed to variation in the source position, since the process of rebinning pixels and convolving with a PSF wipes out small scale features in the lensing potential, which could lead to flux ratio perturbations. On the other hand, the variance of the \textit{Truth} data is affected by variations in the source position \textit{and} perturbations from small scale features in the lensing potential, resulting in a systematically larger scatter. To account for this, we interpret significant offsets in the means of these distributions as evidence for flux ratio perturbations by luminous matter.
\newline \indent In Figures \ref{fig:fluxratios} and \ref{fig:fluxratios2}, we show distributions of flux ratios obtained for the 6 lens systems in our mock sample with the largest $R_{\rm{cusp}}$ or $R_{\rm{fold}}$ values. The frequency and magnitude of flux ratio and astrometric anomalies across our full sample of mock lenses, and the physical characteristics that give rise to these phenomena, characterize what properties of lensing galaxies are likely to perturb flux ratios and other lensing data. We will return to interpret the results of these figures in more depth in Section \ref{sect:results}.
\subsection{Fitting simply parametrized lens models to mock data}
\label{sect:fitting}
\subsubsection{Adopted uncertainties}
We assume astrometric uncertainties of 0.003 arcseconds, time delay uncertainties of 2 days, i.e. comparable to the best data currently available. For the magnification ratios we adopt uncertainties of a factor of 100 which ensures that we fit the smooth potentials only to image positions and time delays. This approach is motivated by the current standard procedure, where the flux ratios are normally not used as constraints for smooth models in order to bypass the effects of substructure and astrophysical noise arising from dust, microlensing, and variability. 

\subsubsection{Fitting procedure}
When fitting the SIE, we vary the Einstein radius, position, ellipticity, shear, and the two corresponding position angles. We optimize these parameters simultaneously, first optimizing numerous random realizations of an SIE profile in the source plane, and then keeping and re-optimizing the best model in the image plane \citep[see][]{Keeton2010}.

\begin{figure*}

\includegraphics[width=.48\textwidth]{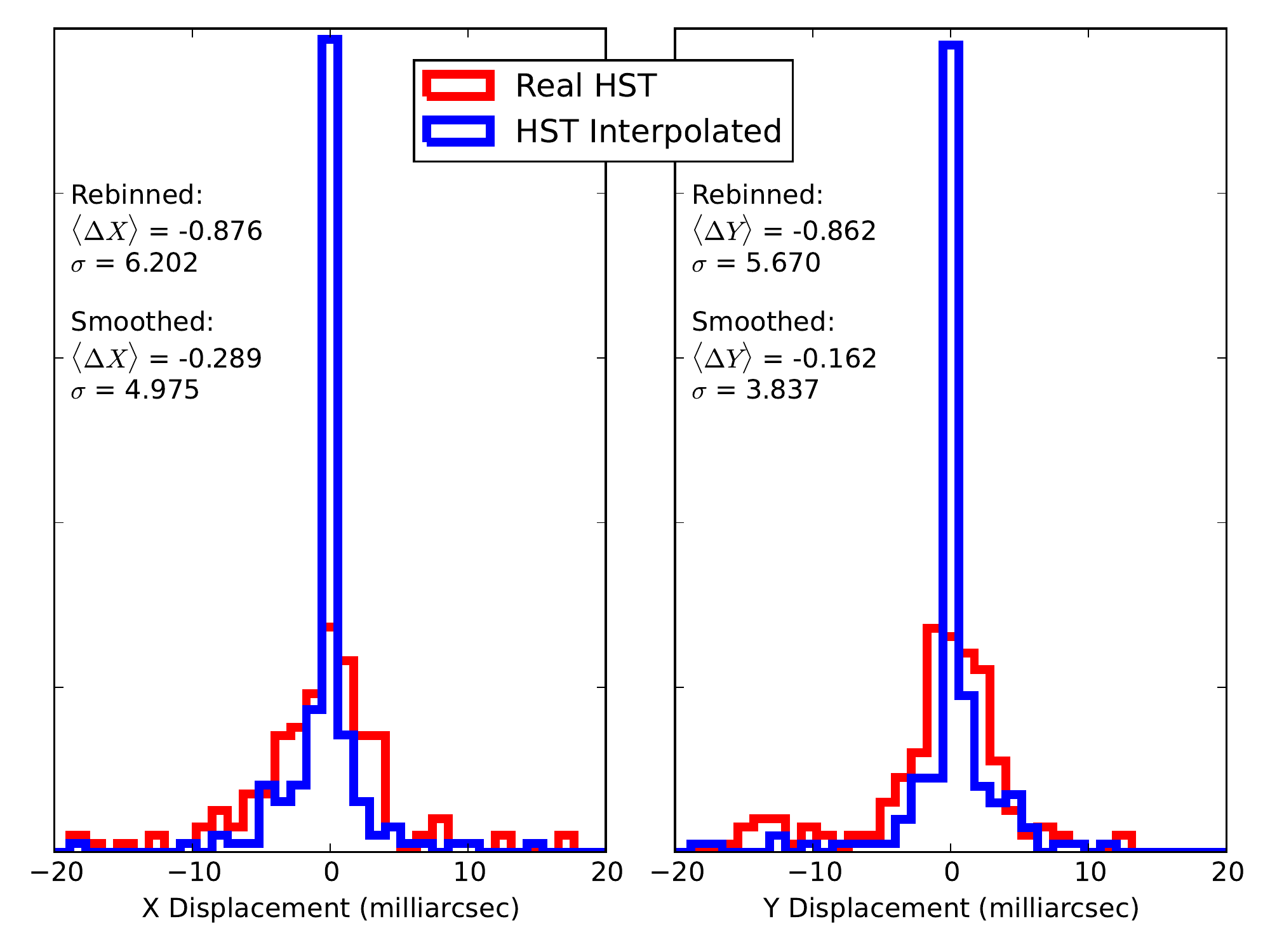}
\includegraphics[width=.48\textwidth]{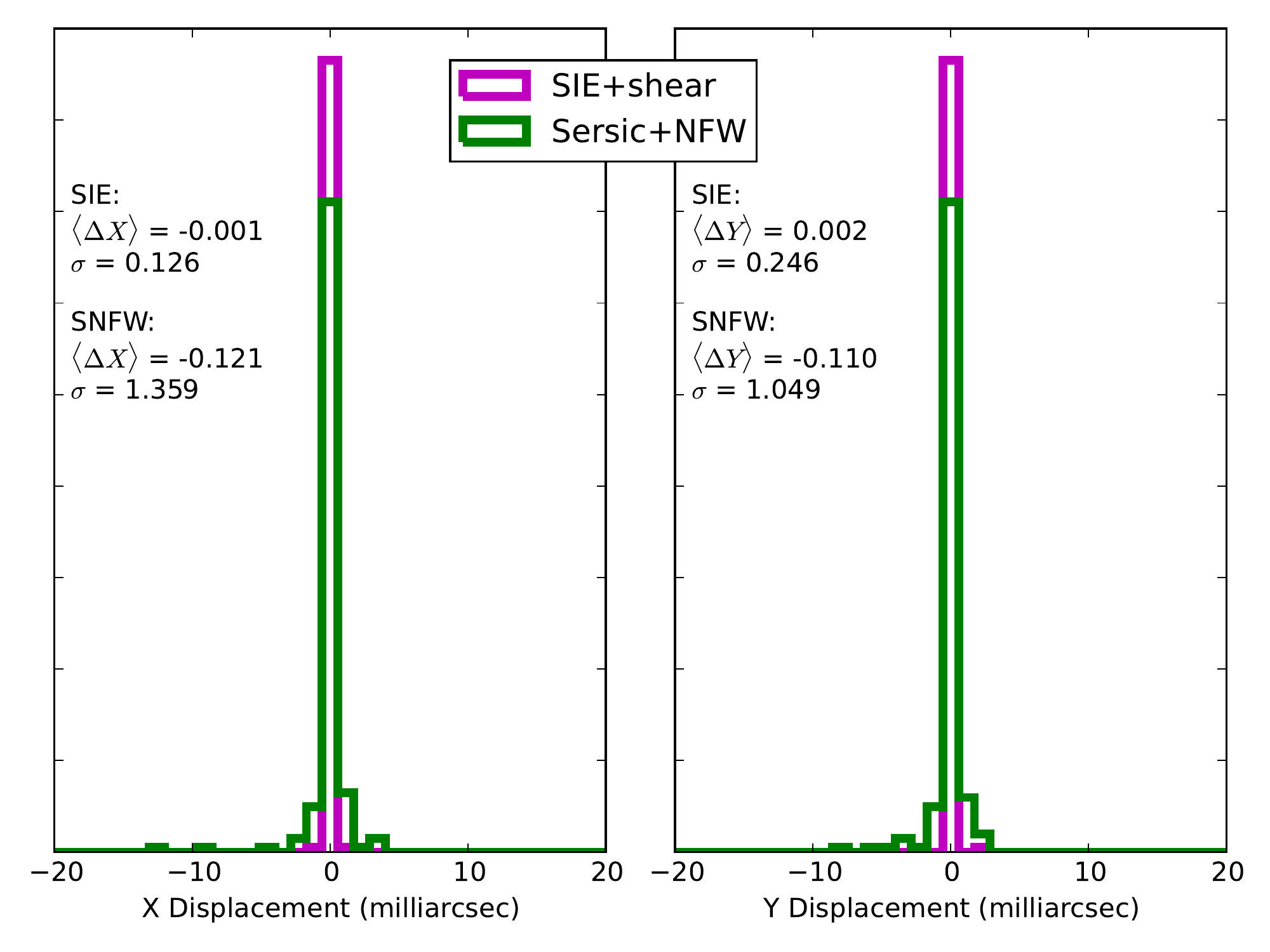}
\includegraphics[clip,trim=1cm 0cm 1.5cm 1cm,width=.49\textwidth]{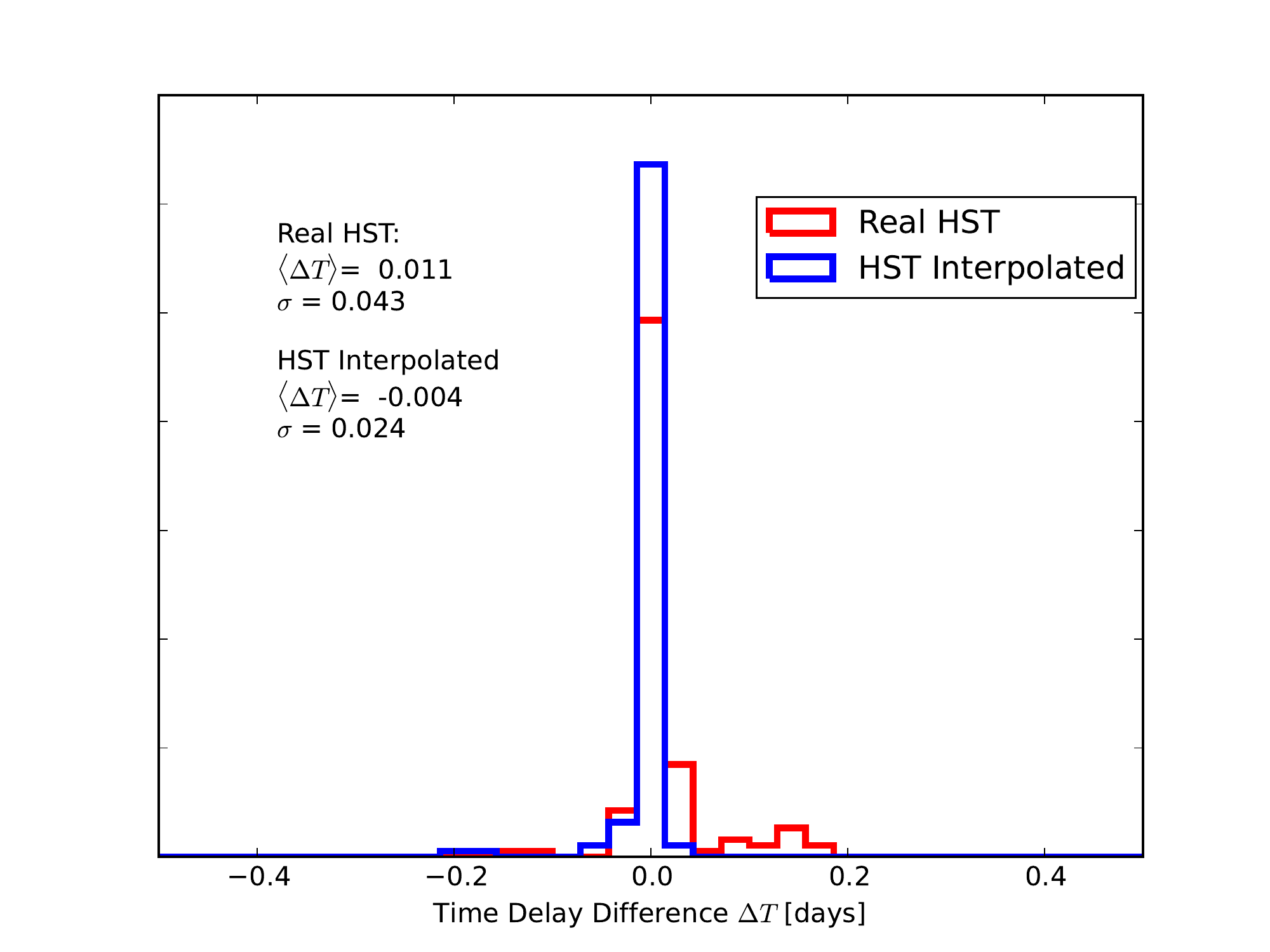}
\includegraphics[clip,trim=1.5cm 0cm 1cm 1cm,width=.49\textwidth]{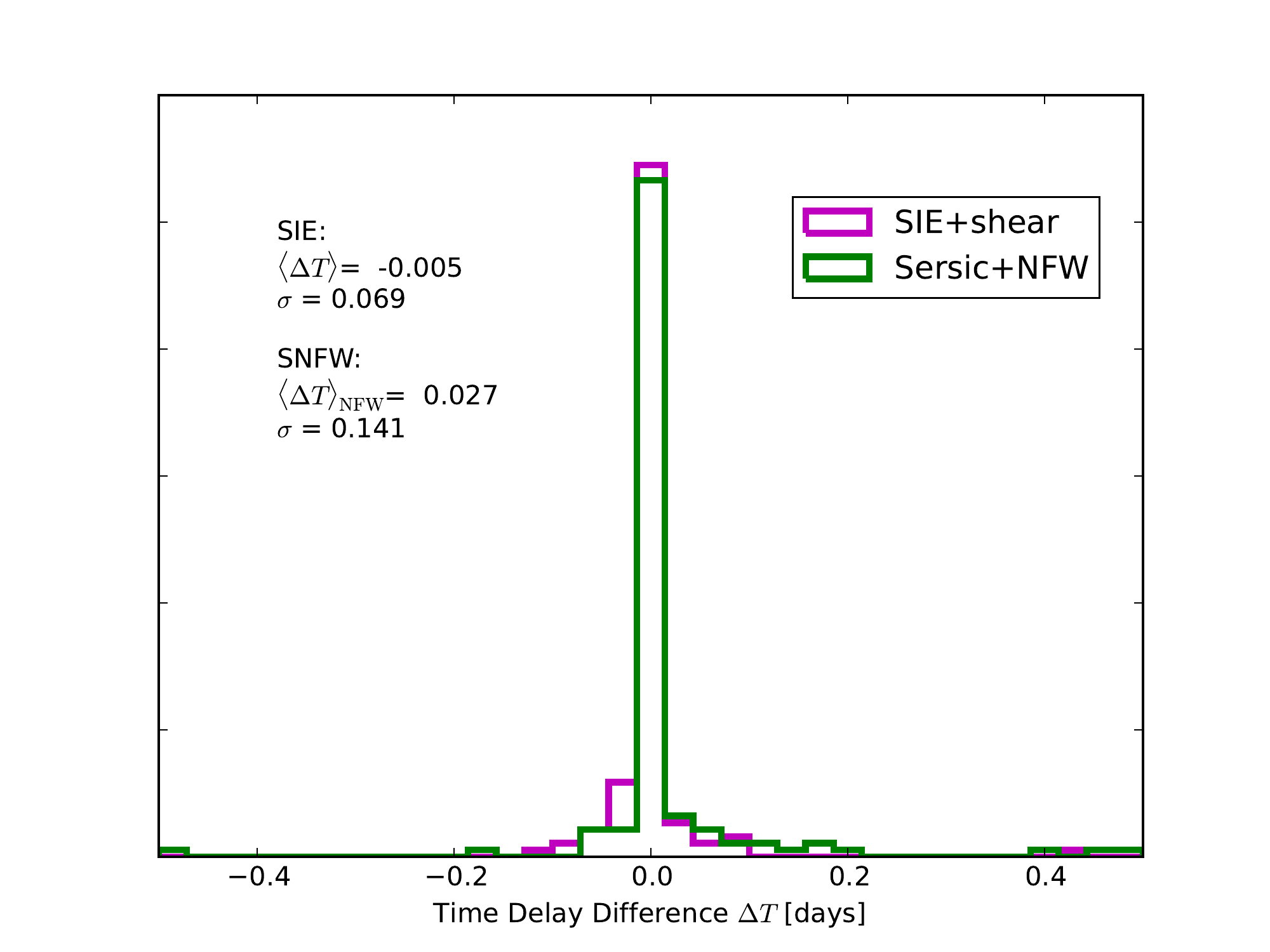}
\caption{\label{fig:pos_tdel}Distributions of the difference in positions (top) and times delays (bottom) from the mean of the \textit{Truth} distributions. Standard deviation, denoted by $\sigma$ is displayed for each data set. The absence of measurement noise in our mock data results in the narrow distributions, whose width is determined by specific lensing properties of each model.}
\end{figure*}
\begin{figure*}
%\begin{tabular}{cc}
\includegraphics[clip,trim=.85cm 0cm .1cm
0cm,width=0.48\linewidth,keepaspectratio]{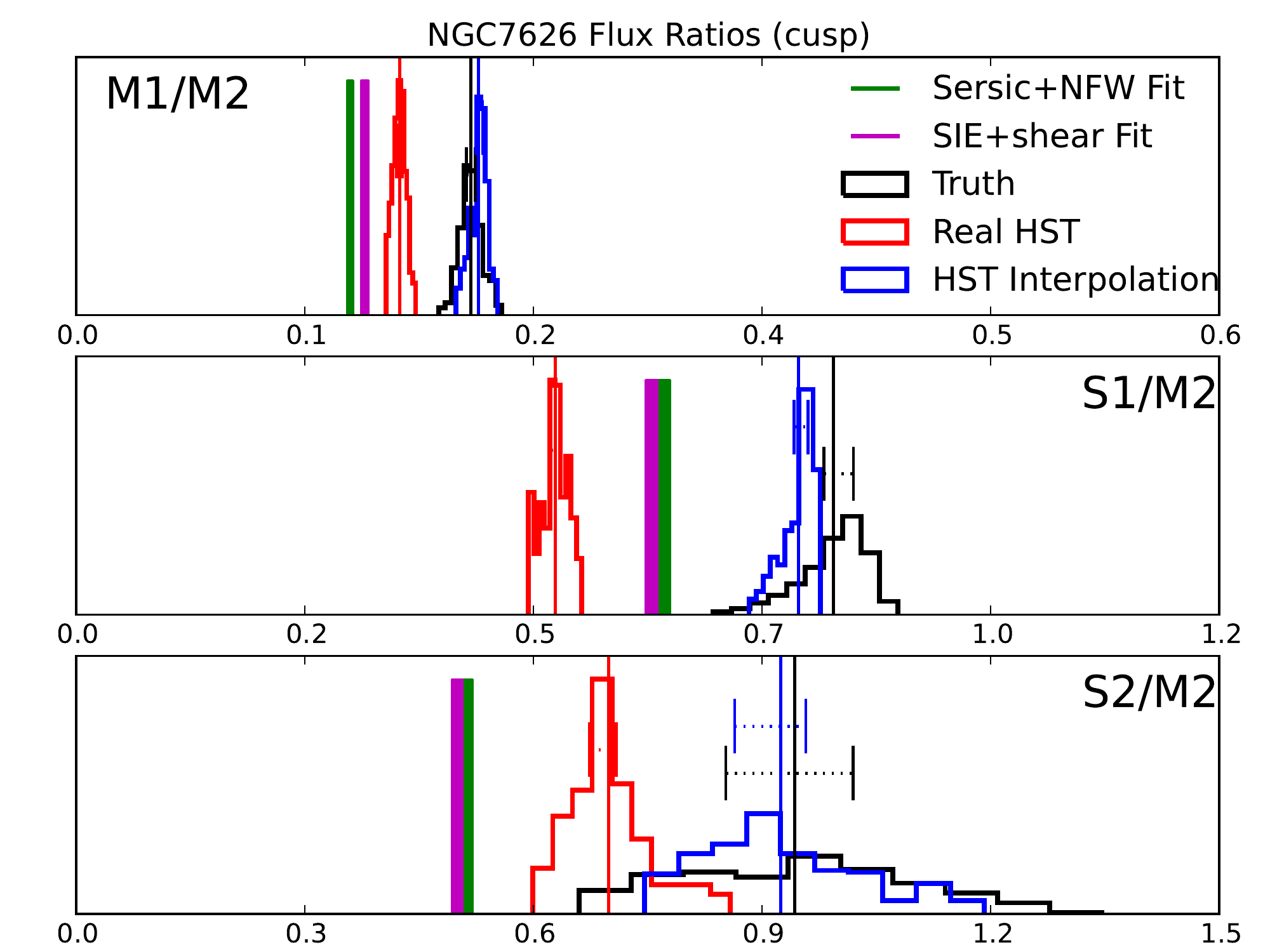}
\includegraphics[clip,trim=.9cm 0cm .2cm
0cm,width=0.48\linewidth,keepaspectratio]{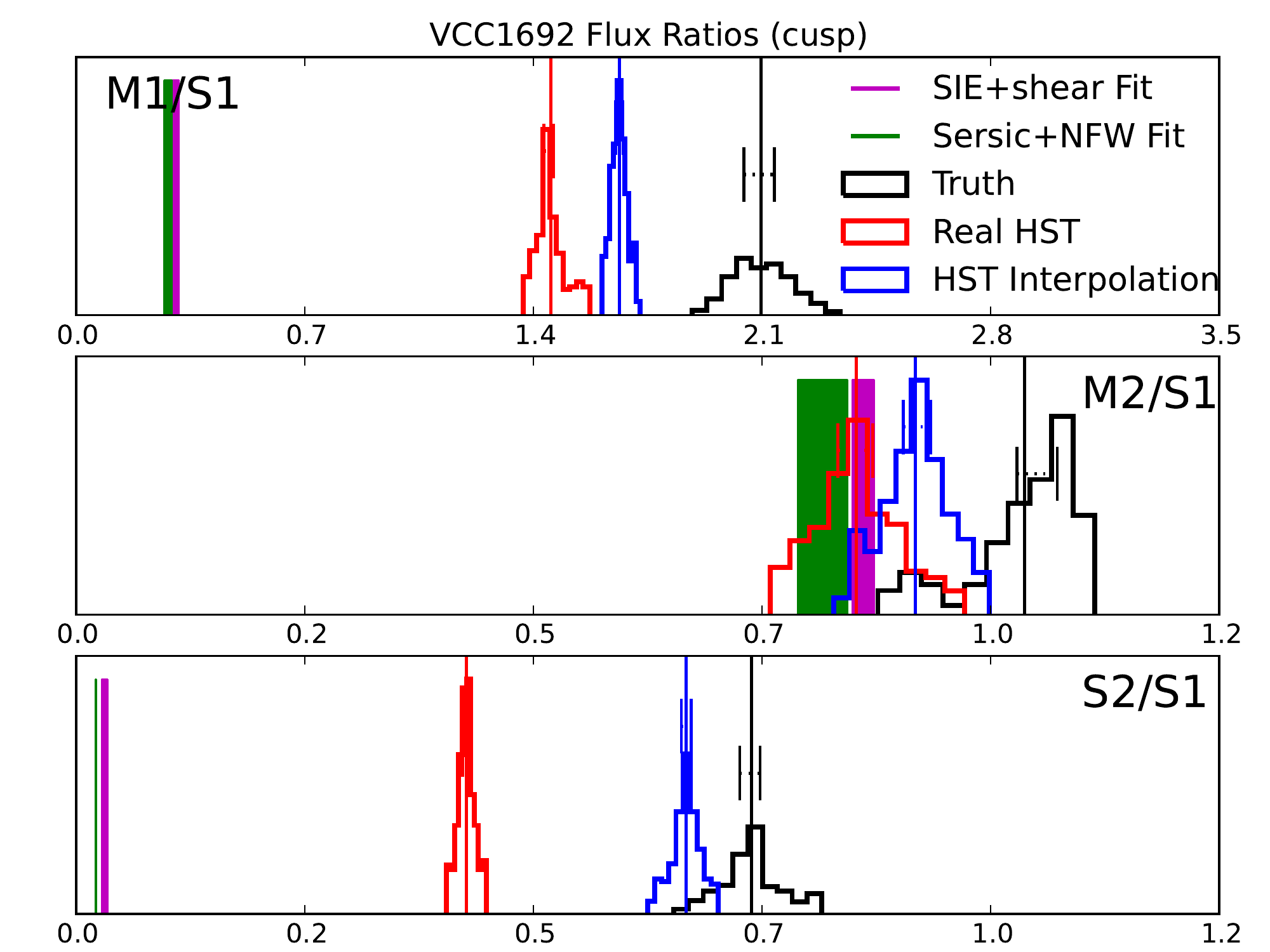}
\includegraphics[trim=3cm 0cm 3cm 0cm,clip,width=.24\textwidth]{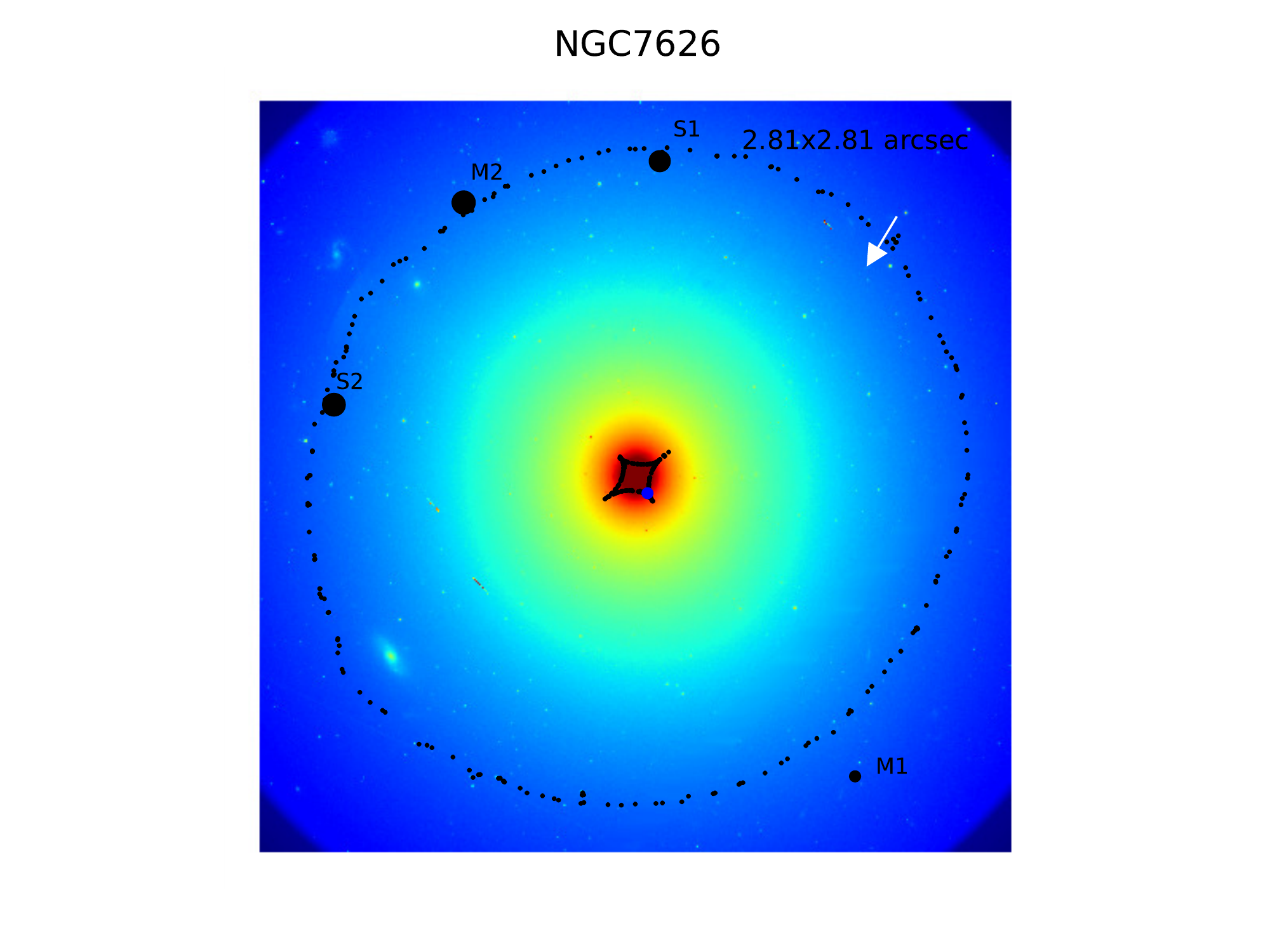}
\includegraphics[trim=3cm 0cm 3cm 0cm,clip,width=.24\textwidth]{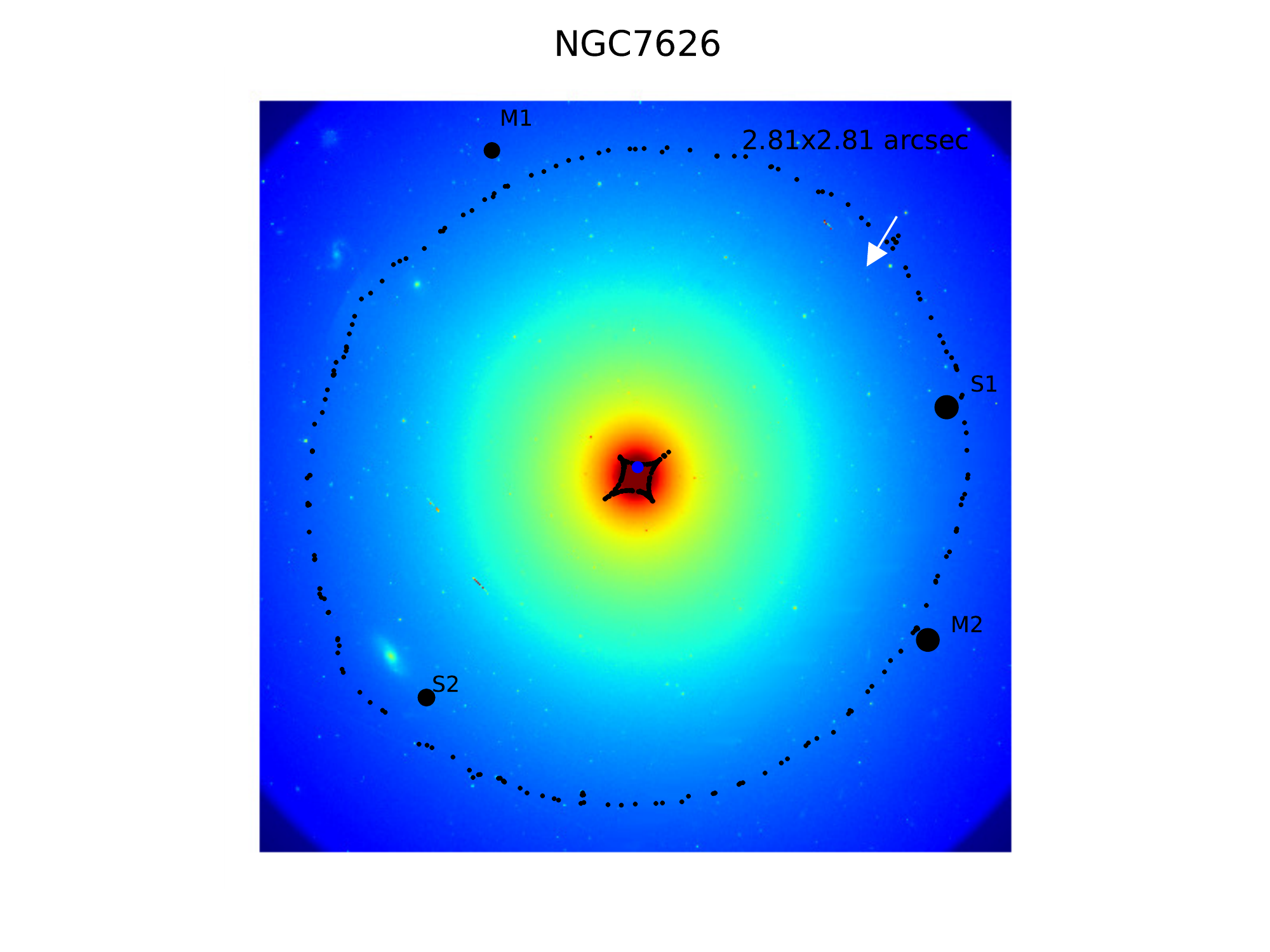}
\includegraphics[trim=3cm 0cm 3cm 0cm,clip,width=.24\textwidth]{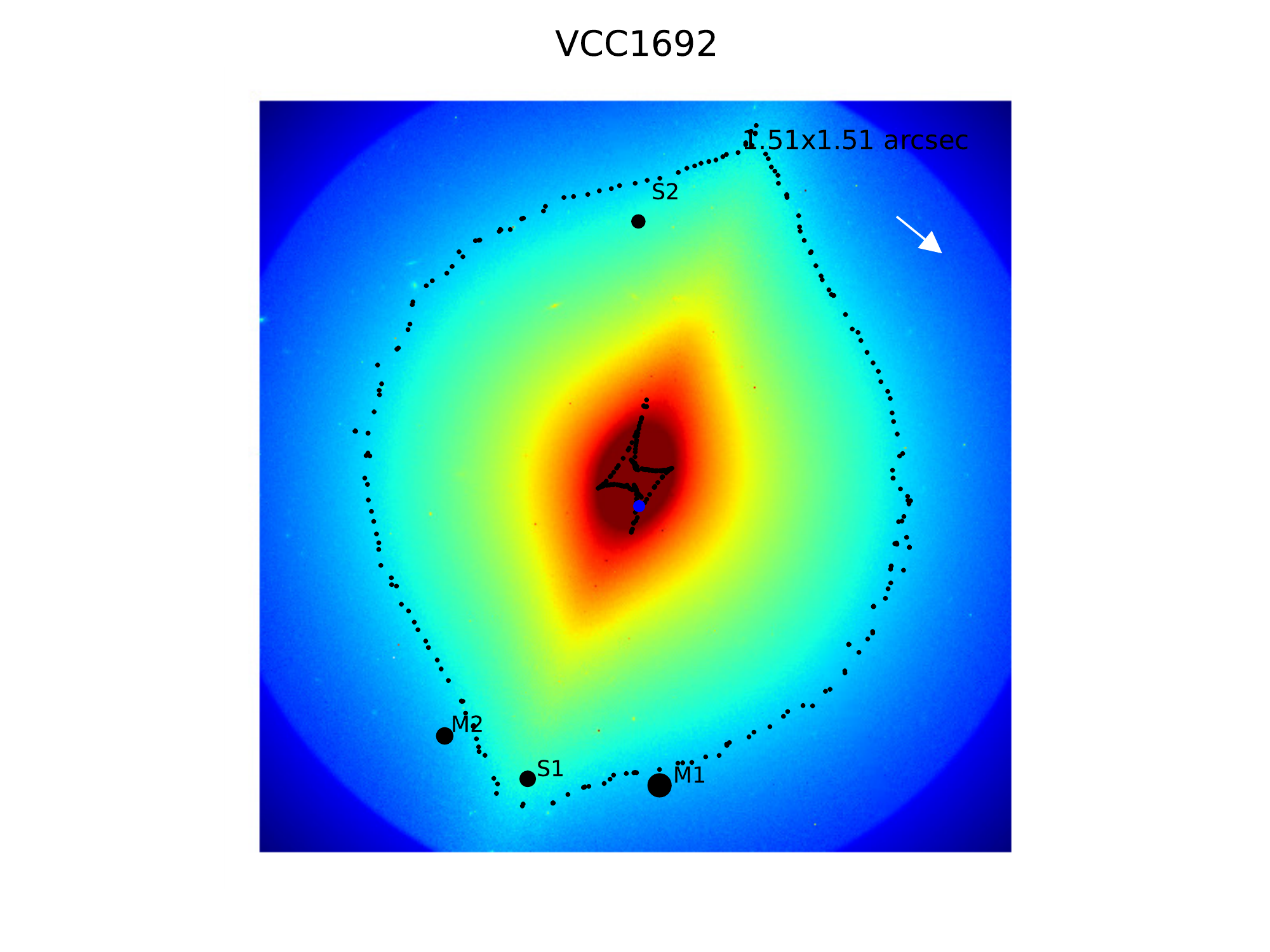}
\includegraphics[trim=3cm 0cm 3cm 0cm,clip,width=.24\textwidth]{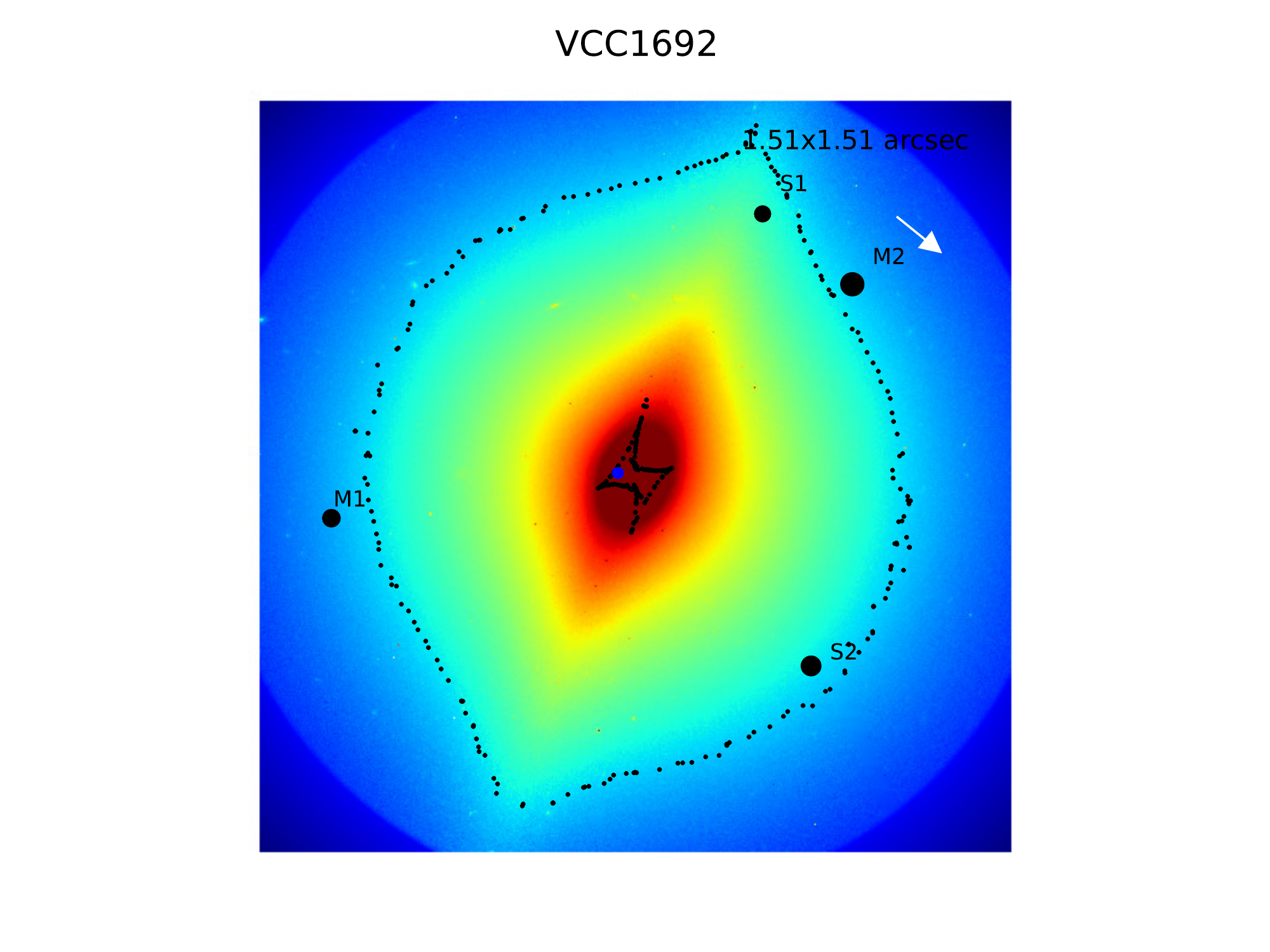}
\includegraphics[clip,trim=.85cm 0cm .1cm
0cm,width=0.48\linewidth,keepaspectratio]{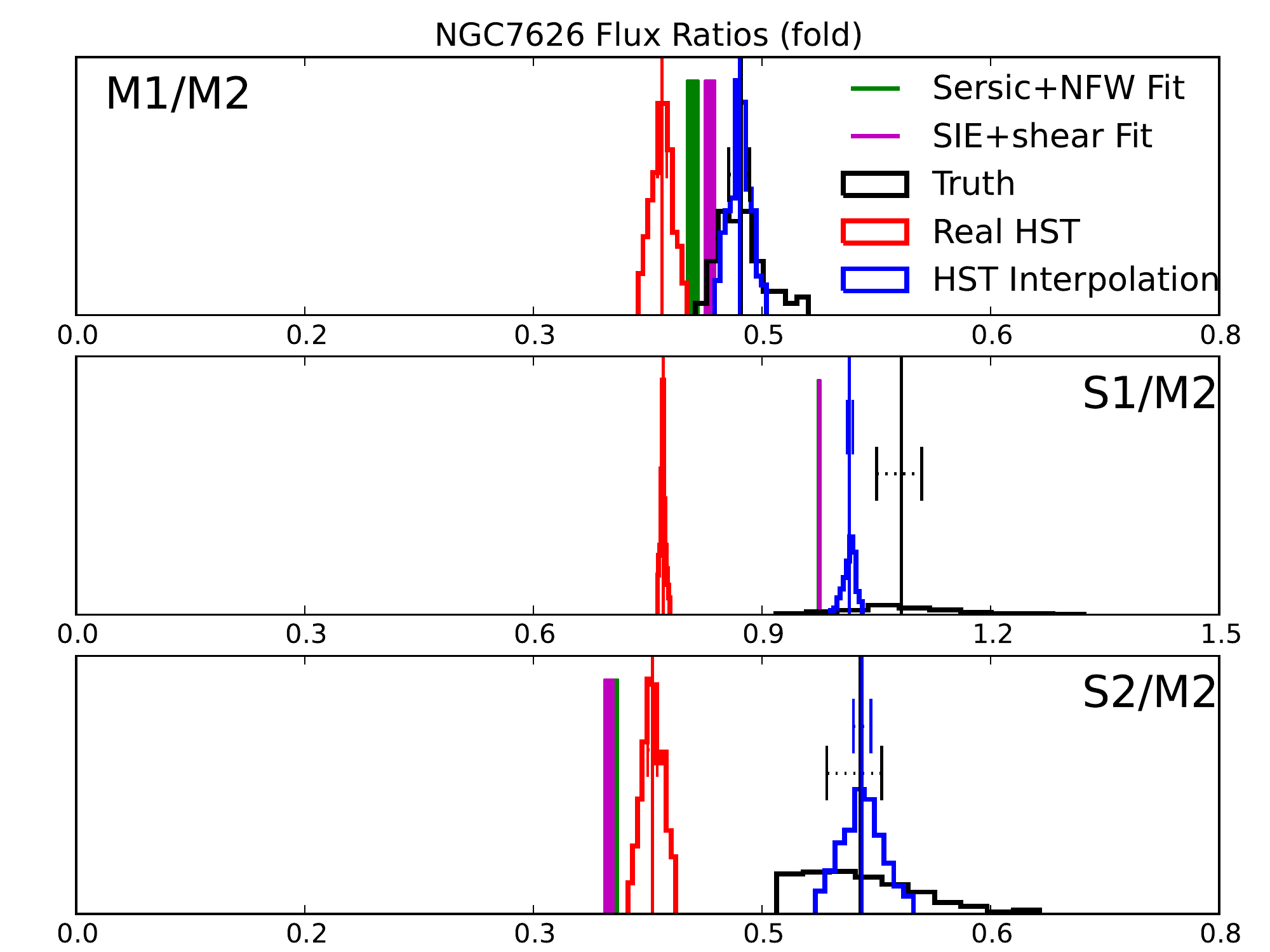}
\includegraphics[clip,trim=.9cm 0cm .2cm
0cm,width=0.48\linewidth,keepaspectratio]{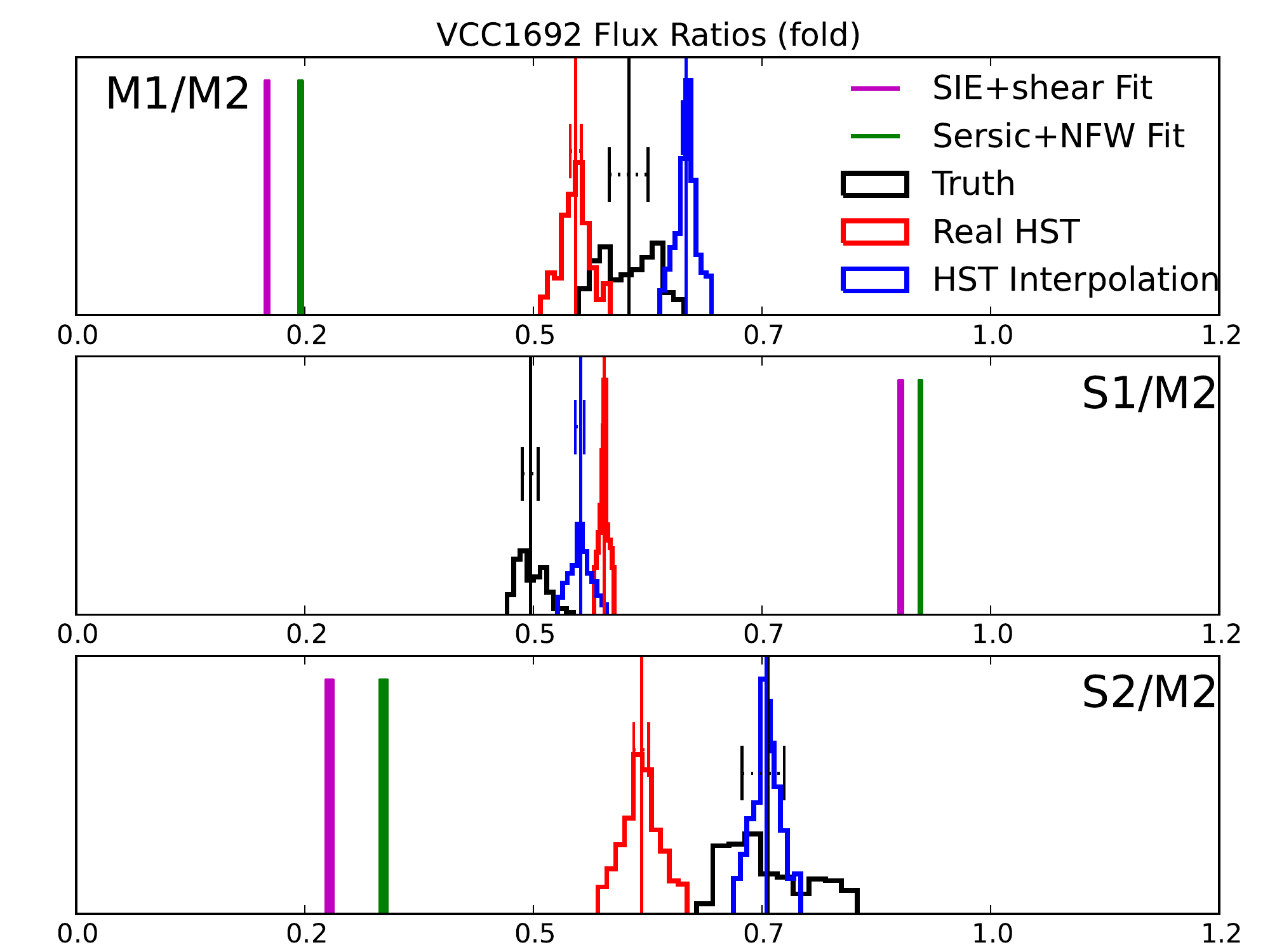}
\caption{\label{fig:fluxratios} Flux ratio distributions for two anomalous systems, NGC7626 and VCC1692. Images in the middle panels are classified as minima (M) or saddle points (S) of the time delay surface. The outer critical curve and astroid caustic are marked by black points, while the source position is marked as a blue point. The line in the upper right corner indicates the direction of the applied external shear. \newline {\bf{\emph{Left}}}: NGC7626 displays an anomalous $R_{\rm{cusp}} = 0.26$. The system NGC7626 has no obvious large scale structure, such as a stellar disk or boxy isophotes, that would contribute to the observed $R_{\rm{cusp}}$ anomaly. However, there are several compact structures near the critical curves between S2, M2, and S1, whose effects we will discuss in more detail in Section 3.3.1. The agreement between the \textit{HST Interpolated} model and the \textit{Truth} suggests the anomaly is not entirely caused by a compact structure, as the convolution process would smooth over features smaller than the convolution kernel (which in this case is 22 native pixels or about 0.5 kpc). While the fold configuration does not display an anomaly between the two merging images M2/S1, a satellite galaxy near S2 induces an anomaly with respect to the SIE and SNFW models of $\approx 8 \%$. In the S1/M2 flux ratio, the vertical bar marking the mean of the SNFW fit is hidden by the bar marking the SIE centroid, as both models produce the same flux ratio. \newline {\bf{\emph{Right}}}: VCC1692 is the most anomalous mock lens in our sample with $R_{\rm{cusp}} = 0.51$ and a relative flux ratio anomaly in M1/S1 of $\approx 80 \%$. The $R_{\rm{fold}} = 0.35$ of VCC1692 is also abnormally high, with an S1/M2 flux ratio anomaly of $\approx 30 \%$. In VCC1692, a stellar disk at a nearly orthogonal position angle results in the warped astroid caustic and abnormal image configuration, as well as likely inducing the observed anomaly.}. 
\end{figure*}
\begin{figure*}
%\begin{tabular}{cc}
\includegraphics[clip,trim=.85cm 0cm .1cm
0cm,width=0.48\linewidth,keepaspectratio]{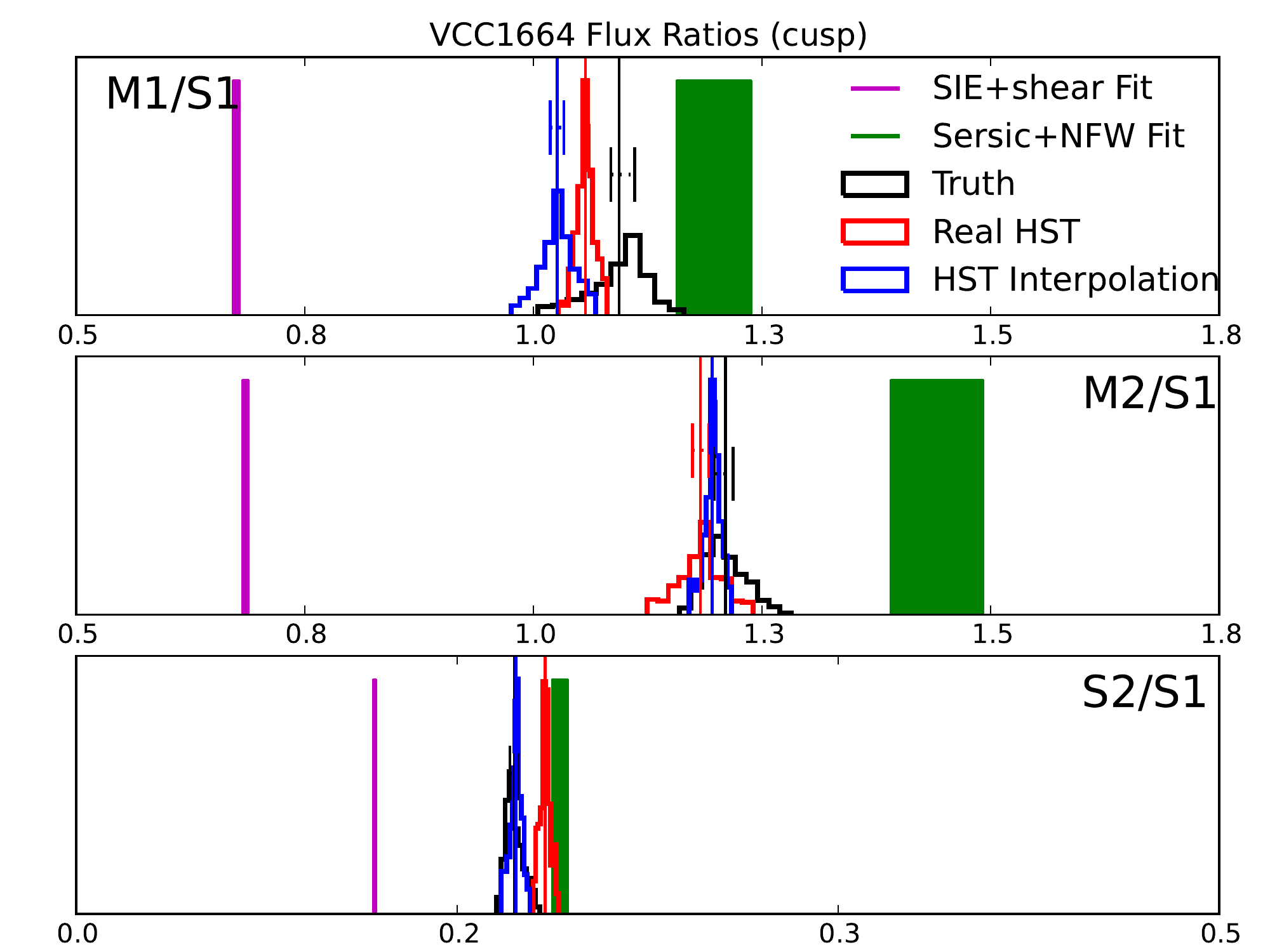}
\includegraphics[clip,trim=.9cm 0cm .2cm
0cm,width=0.48\linewidth,keepaspectratio]{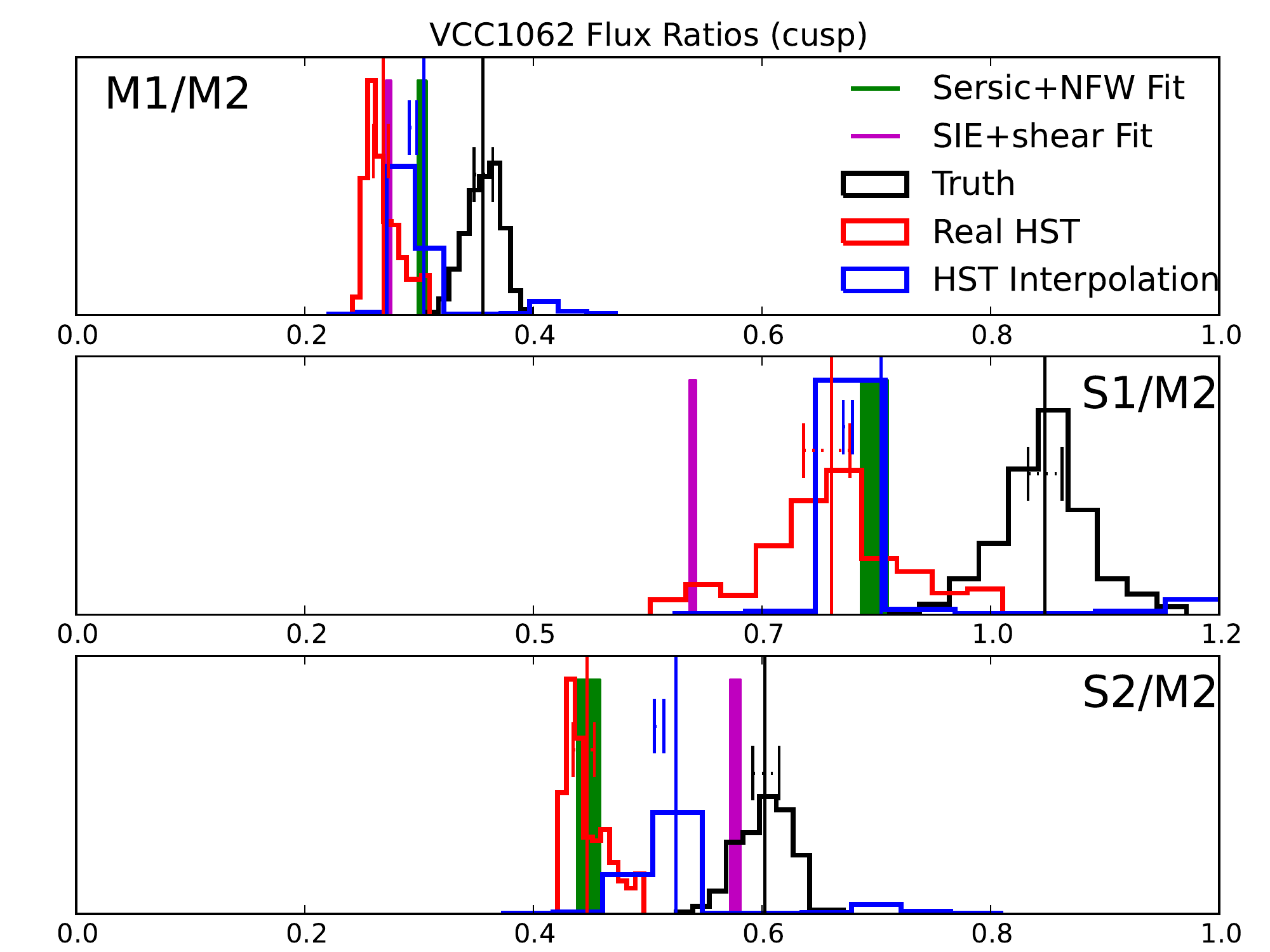}
\includegraphics[trim=3cm 0cm 3cm 0cm,clip,width=.24\textwidth]{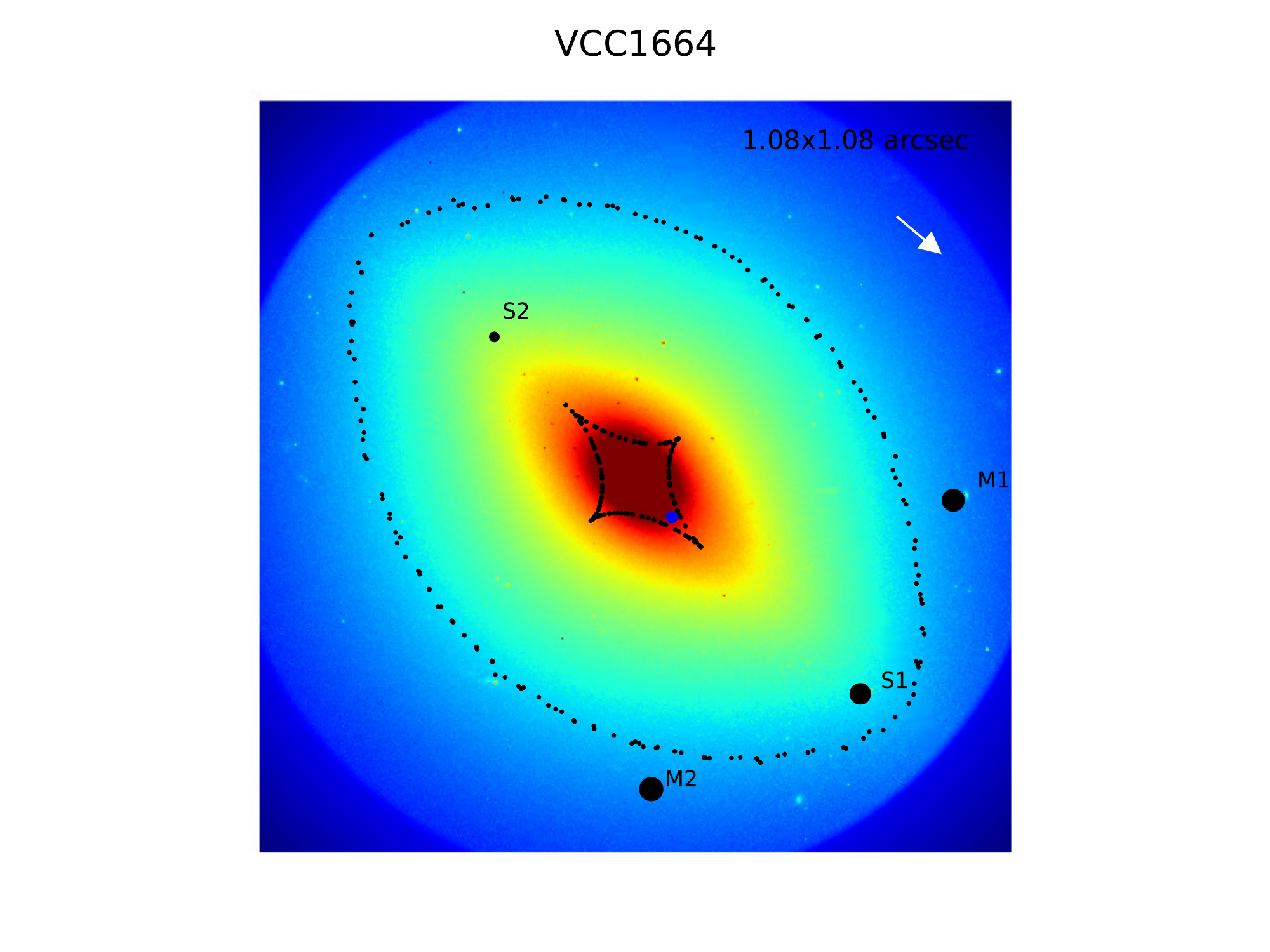}
\includegraphics[trim=3cm 0cm 3cm 0cm,clip,width=.24\textwidth]{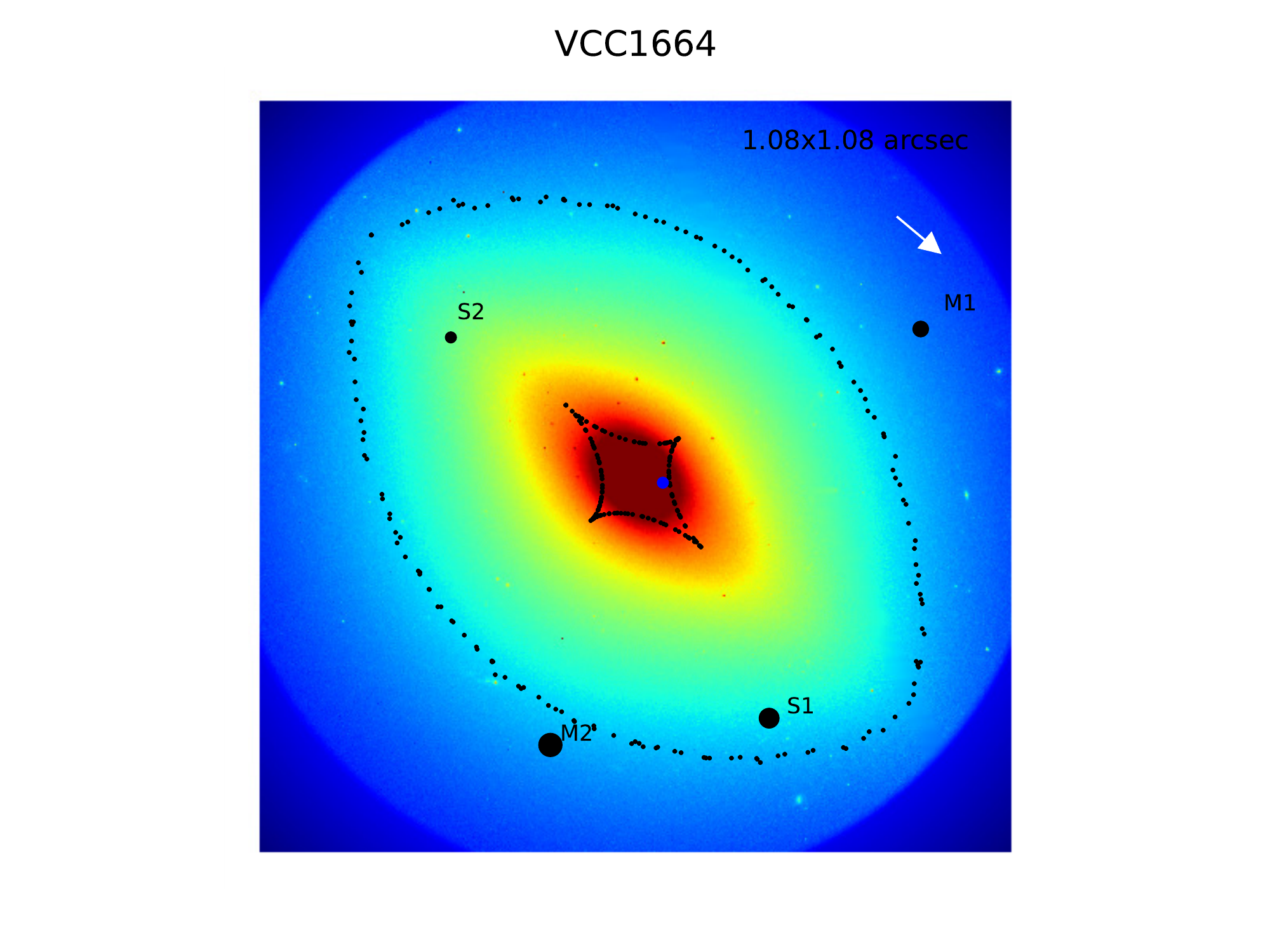}
\includegraphics[trim=3cm 0cm 3cm 0cm,clip,width=.24\textwidth]{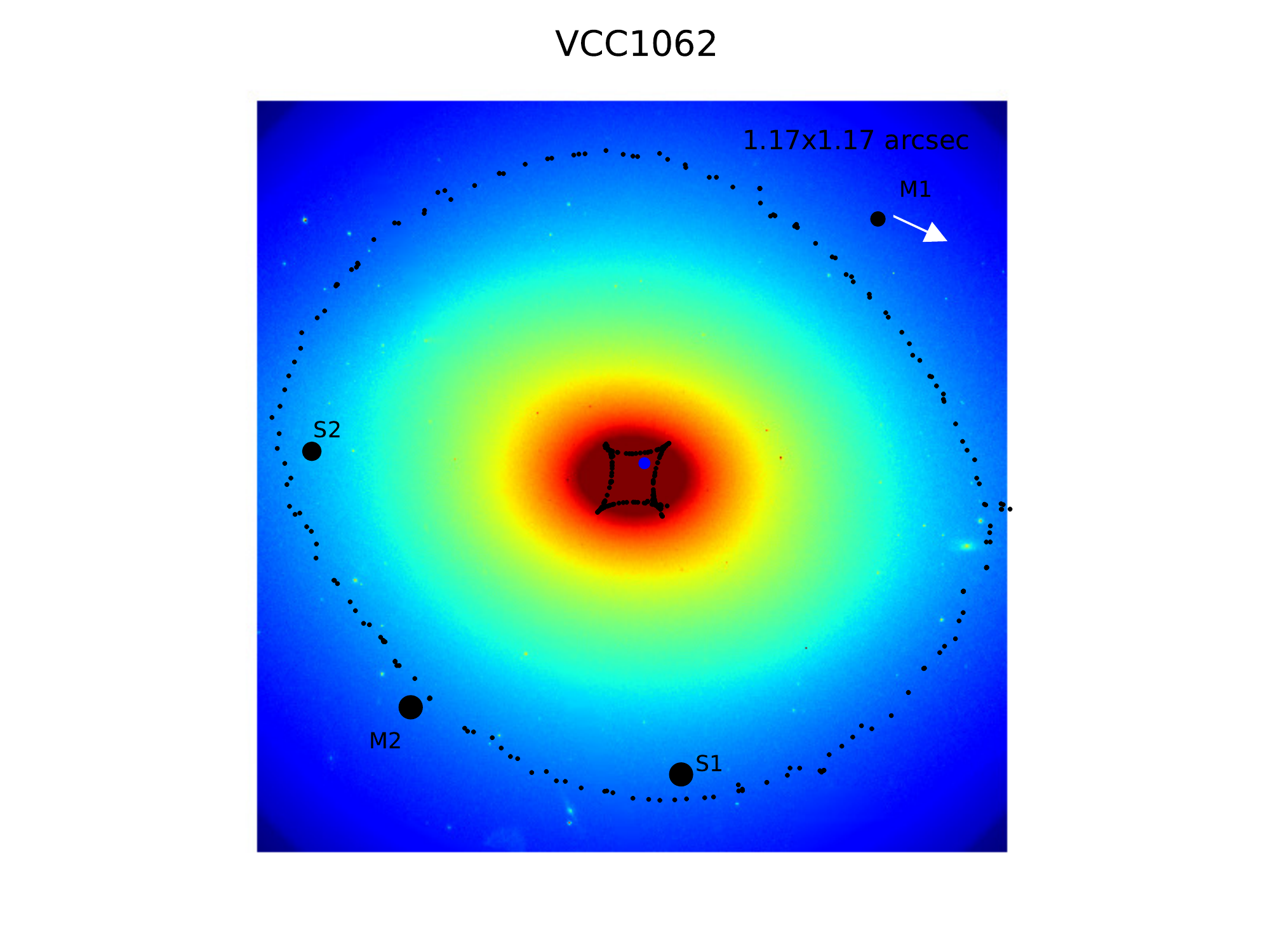}
\includegraphics[trim=3cm 0cm 3cm 0cm,clip,width=.24\textwidth]{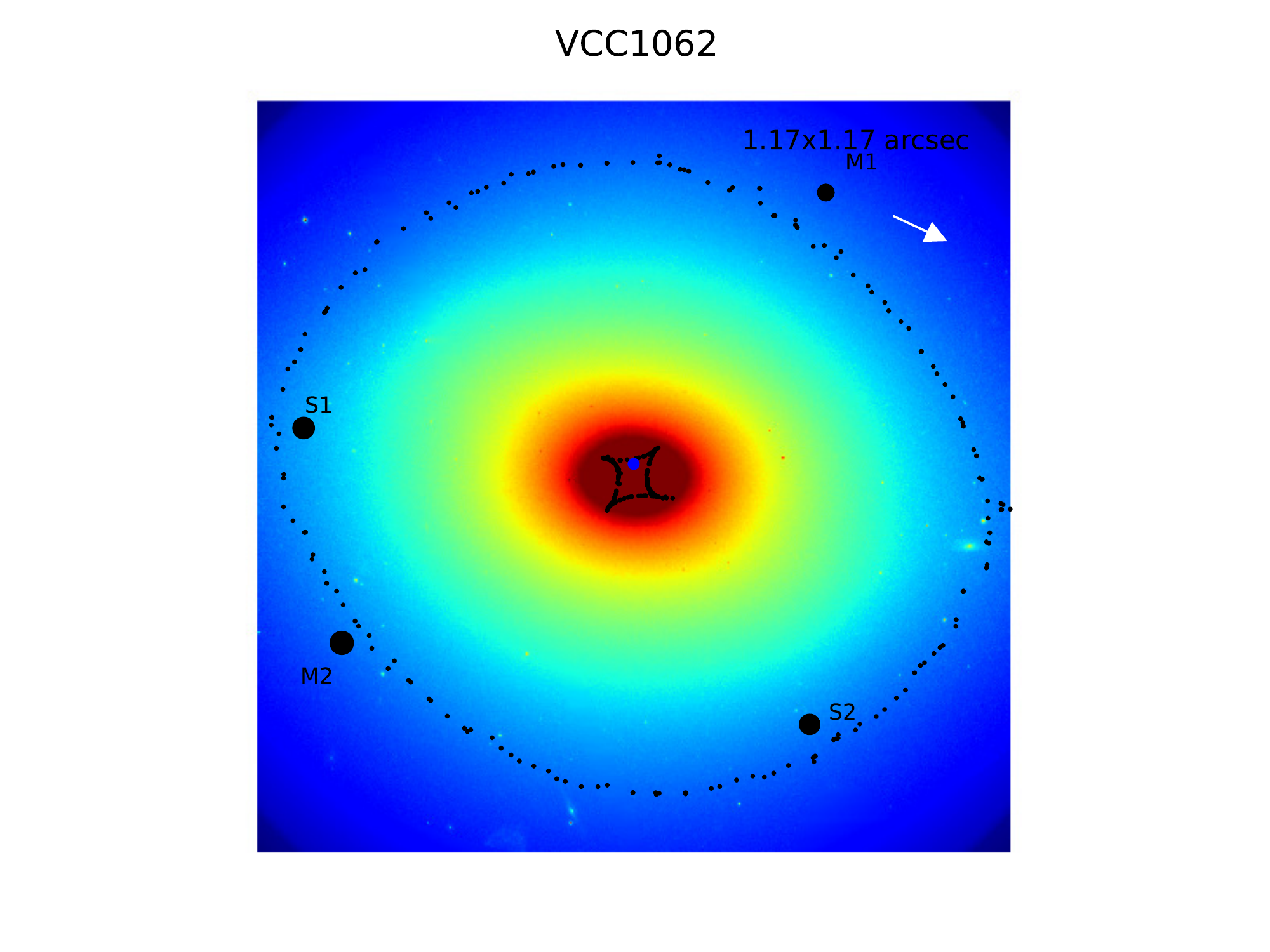}
\includegraphics[clip,trim=.85cm 0cm .1cm
0cm,width=0.48\linewidth,keepaspectratio]{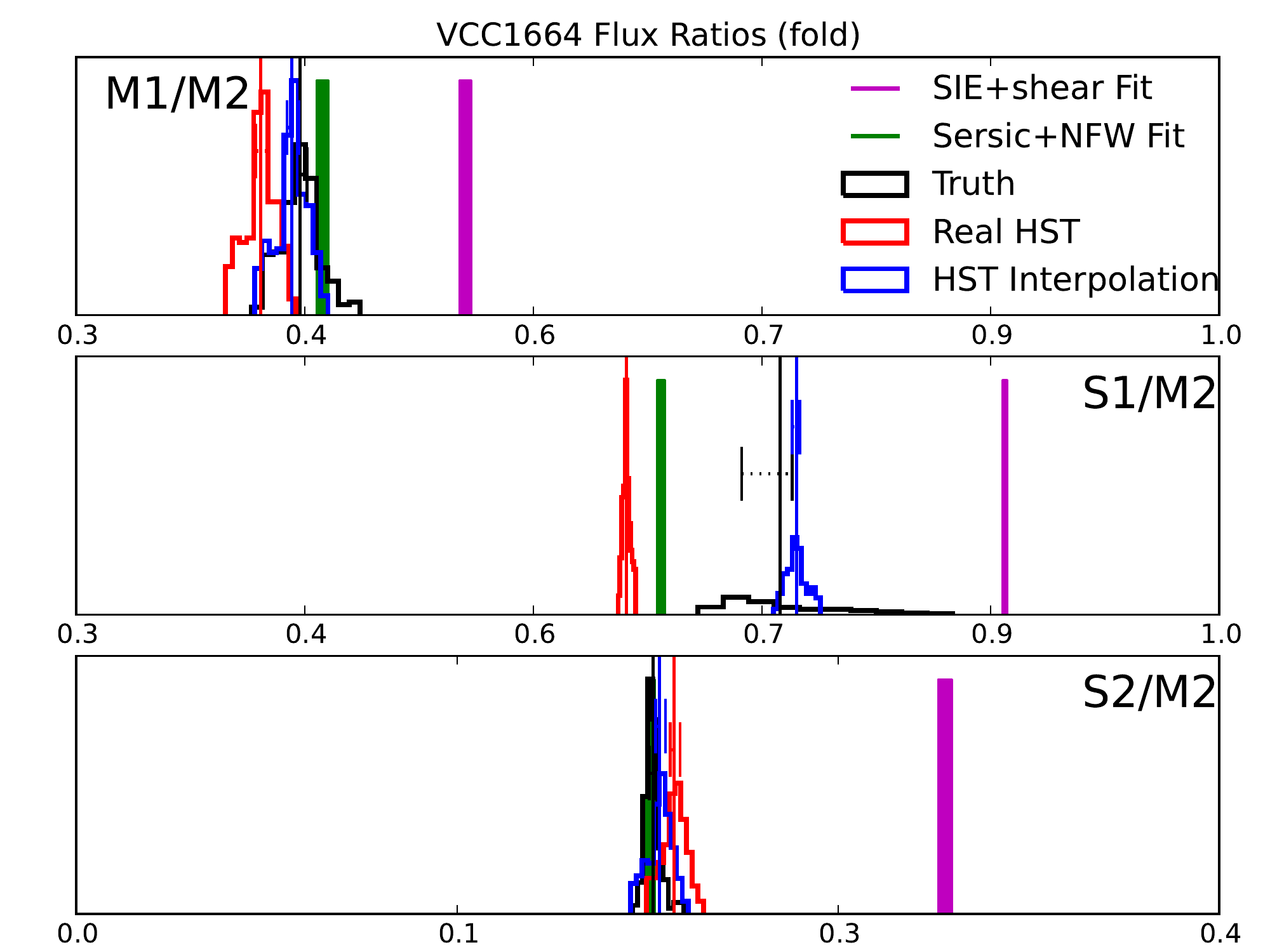}
\includegraphics[clip,trim=.9cm 0cm .2cm
0cm,width=0.48\linewidth,keepaspectratio]{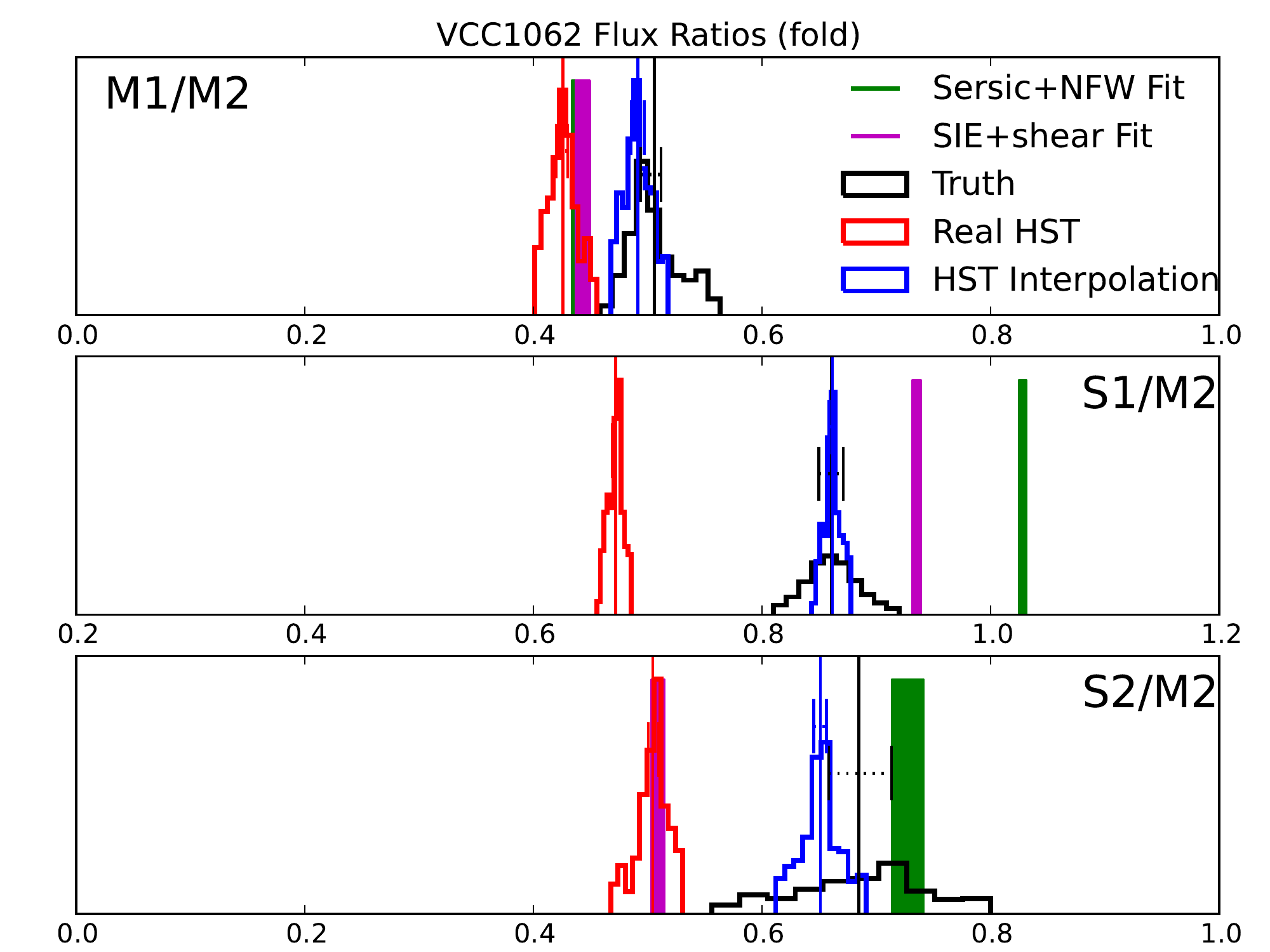}
\caption{\label{fig:fluxratios2} The distributions of flux ratios for VCC1664 and VCC1062, two anomalous systems. Images in the middle panels are classified as minima (M) or saddle points (S) of the time delay surface. The outer critical curve and astroid caustic are marked by black points, while the source position is marked as a blue point. The arrow in the upper right corner indicates the direction of the applied external shear. \newline {\bf{\emph{Left:}}} VCC1664 displays an anomalous $R_{\rm{cusp}} = 0.40$, with large anomalies between the best fit SIE and \textit{Truth} models in both the M1/S1 and M2/S1 image pairs. Like VCC1692, the disky nature of VCC1664 is likely responsible for the anomalous flux ratios. \newline {\bf{\emph{Right:}}} VCC1062 has an anomalous $R_{\rm{cusp}} = 0.25$ and an anomaly between the S1/M2 image pair. Different external shear magnitude (0.05 and 0.075 for cusp and fold, respectively), result in differently shaped astroid caustics. The VCC1062 is one of the few cases where the \textit{HST Interpolated} model does not agree with the \textit{Truth} model, suggesting that irregularities in stellar structure or other small scale features perturb the flux ratios in the system, although close visual inspection of the convergence map near each image does not reveal a obvious culprit.} 
\end{figure*}
In contrast, when fitting with the SNFW, we attempt to fit the lens by holding the parameters of the S{\'e}rsic profile describing each galaxy fixed while varying the properties of the NFW halo. We obtain the S{\'e}rsic parameters either from literature (see the references in Table 1), or by measuring them ourselves using {\tt {galfit}}. Specifically, in the first iteration we vary only the normalization of the S{\'e}rsic profile, the normalization of the NFW halo, the scale radius of the NFW, and the external shear and position angle. 

In most cases, this approach fails to fit the positions and time delays with a reduced $\chi^2 < 2.5$, which we take to be the threshold acceptable $\chi^2$ fit. We choose this $\chi^2$ to permit individual astrometric and time delay $\chi^2$ values greater than 1, resulting in a conservative measure of the degree to which our smooth potentials can recover image positions and time delays. As we do not add measurement noise to our data, the SIE and SNFW frequently fit the data almost exactly, resulting in reduced $\chi^2$ much less than unity. After the first iteration of fitting, for the systems with unacceptable model fits, we allow the S{\'e}rsic index, ellipticity and position angle of the S{\'e}rsic to vary, and attempt to fit the lens again. If this approach fails, we vary the effective radius, ellipticity and position angle. 

The complications we encounter trying to fit lenses with a single S{\'e}rsic model suggests that, for the purpose of lens modeling, a more complicated lens model is required to fit the luminous matter of a lens, e.g. a bulge+disk of different ellipticity or two components at different position angles. On the other hand, the success of a simple SIE model suggests that this simple model is sufficient in most cases to caputure the lensing effects of baryonic matter. Interestingly, fold configurations required more flexible models (with varying S{\'e}rsic index, ellipticity, and position angle) than cusp configurations. Overall, we find that an SIE model absorbs the combined properties of stellar mass and dark matter as well as, or better than, the SNFW model. In Table~\ref{table:fitting_list} we summarize the parameters we allow to vary when fitting with the SIE and SNFW models, and the results of the fit for each galaxy in our sample.
\section{Results}
\label{sect:results}
In this section, we compare the data obtained for the \textit{Truth}, \textit{Real HST}, and \textit{HST Interpolated} mock lenses, and the two analytic models, the SIE and SNFW. First, in 3.1 we investigate the extent to which positions and time delays vary between the \textit{Truth} data set and the four comparison models. Section 3.2 reviews the $R_{\rm{cusp}}$ and $R_{\rm{fold}}$ statistics, and present the values for these statistics we obtain for the \textit{Truth} model, side by side with observed $R_{\rm{cusp}}$ and $R_{\rm{fold}}$ statistics from real strong lenses, and characterize baryonic flux ratio anomalies by their coupling to astrometric anomalies. In 3.3, we examine in detail each anomalous system to understand the source of flux ratio anomaly, making use of magnification maps derived from our convergence maps. Finally, in 3.4 we discuss the the degree to which the \textit{Real HST}, \textit{HST Interpolated}, SIE and SNFW models recover the flux ratios, $R_{\rm{cusp}}$ and $R_{\rm{fold}}$ values of the data of the \textit{Truth} model. 

\subsection{Image Positions and Time Delays}

Since time delays and image positions depend on the total
gravitational potential, and the gradient of the potential,
respectively, one can expect these data to be relatively insensitive
to small perturbations to the gravitational potential, whether by dark
subhalos or by baryonic features. Therefore, it is expected that the
rebinned and smoothed models will yield similar image positions and
time delays as the \textit{Truth} model, and that both the SIE
and SNFW models we fit to the \textit{Truth} will also
accurately recover these data.

Our results are consistent with these expectations. In Figure~\ref{fig:pos_tdel} we plot the distributions of offsets between
means of the \textit{Truth} model, the three models based on
the unfiltered data (Models 1,2,3), and the SIE and SNFW fits to the
\textit{Truth} model. The standard deviations in the
distributions of image positions are comparable to the 0.003 arcsecond
uncertainty we assume in our lens models, while the standard
deviations in the distributions of time delays is an order of
magnitude smaller than our assumed uncertainty. For the
\textit{Real HST} and \textit{HST Interpolated} models, this indicates that
the information lost in the process of rebinning pixels, or convolving
with the smoothing kernal, does not, in most cases, significantly impact the predicted image positions and arrival times. For the SIE and SNFW fits, these results confirm that the image positions and time delays of real lenses are consistent with those produced by a smooth lens model, as expected.
\newline \indent While these results were expected a priori, this should not undermine their significance. The agreement of astrometric data between the different models we consider implies that luminous matter is highly unlikely to result in astrometric anomalies. Conversely, many real lenses with flux ratio anomalies similar to those we observe in our mocks exhibit flux ratio anomalies accompanied by astrometric anomalies, suggesting an avenue by which a baryonic lensing signal may be distinguished from other sources of anomaly. We will revisit this point in detail later in this section.
\subsection{Flux ratios and the $\bf{R_{cusp}}$ / $\bf{R_{fold}}$ relations, \textit{Truth} model vs. real lens systems}
In order to quantify the flux ratio anomaly across the full ensemble
of mock lenses, we consider the differences in the mean of each
distribution (explicitly, each $F_{i}$ denotes the mean of a
distribution for an image generated from a lens described by model
$i$). The mean is marked as vertical bar in
Figures~\ref{fig:fluxratios} and \ref{fig:fluxratios2}. In this work we consider the relative flux ratio anomalies $\delta F_i$, with respect to the \textit{Truth} data:
\begin{figure*}
\includegraphics[trim=0cm 0cm 0cm 0cm,clip,width=.6\textwidth]{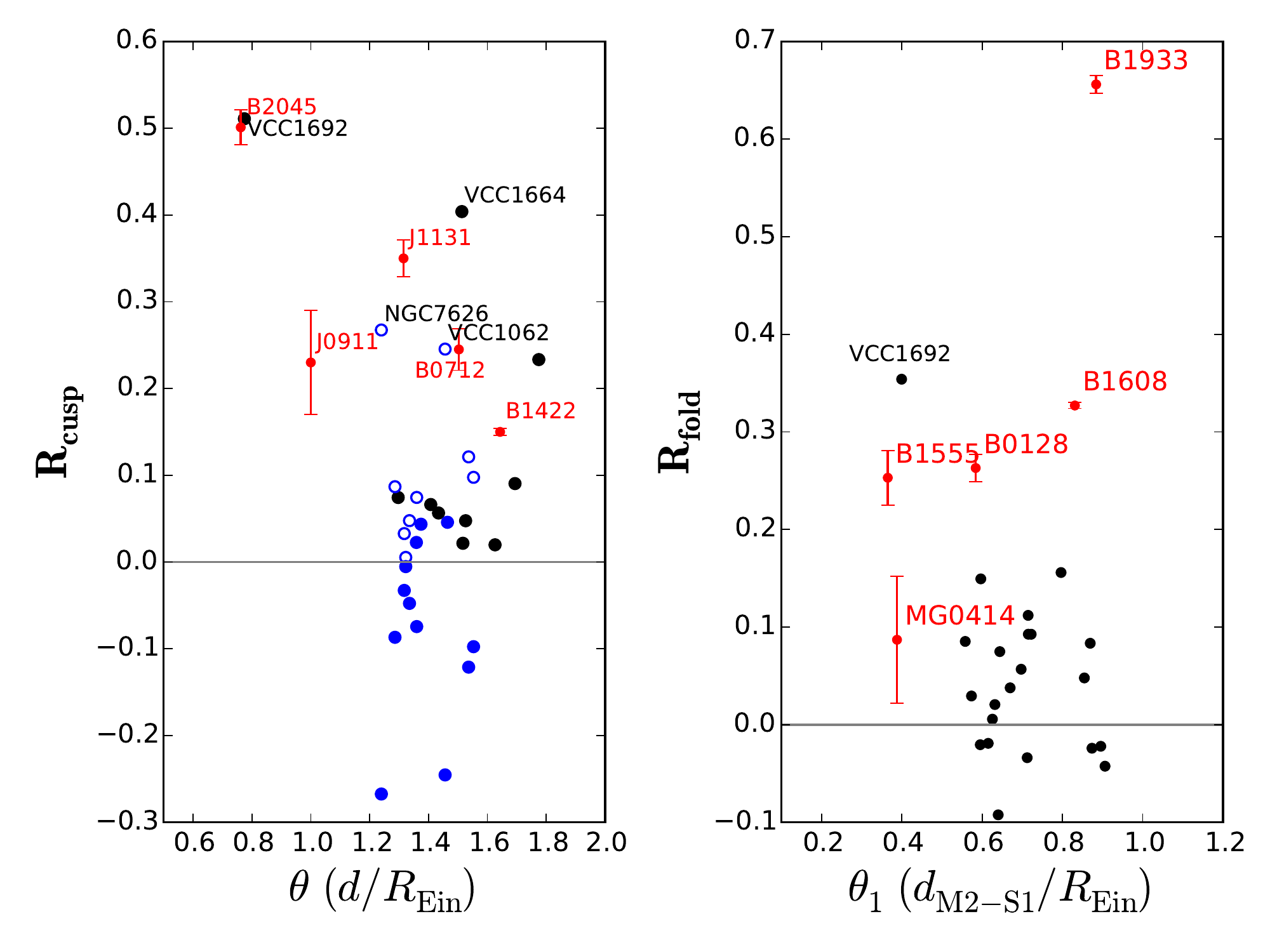}
\includegraphics[trim=5cm 1cm 6.5cm 2cm,clip,width=.3\textwidth]
{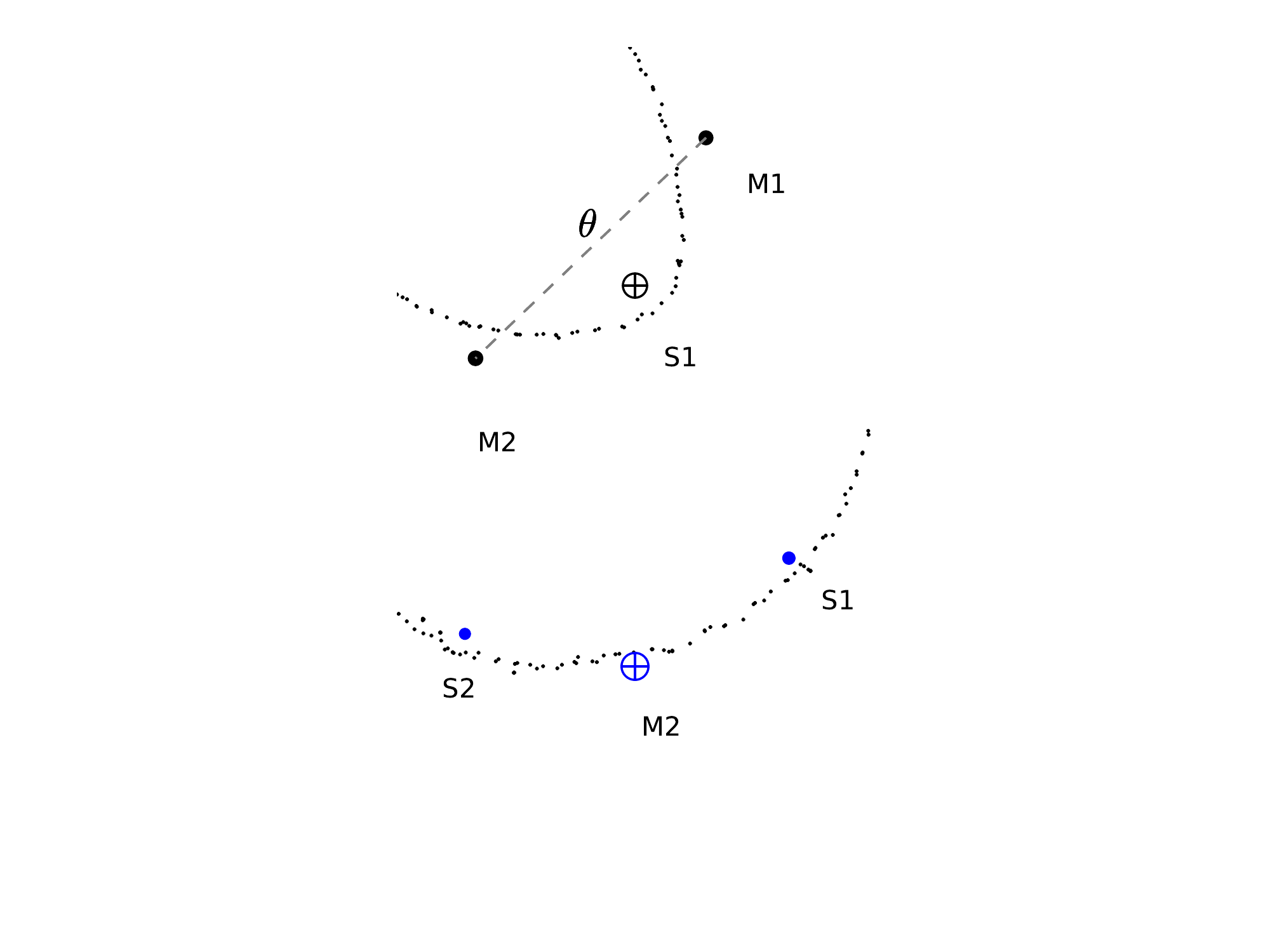}
\caption{\label{fig:Rcuspfold_vs_real}$R_{\rm{cusp}}$ and $R_{\rm{fold}}$ statistics of the \textit{Truth} data, compared with a set of $R_{\rm{cusp}}$ and $R_{\rm{fold}}$ statistics of real lenses. An empty circle indicates the absolute value of an $R_{\rm{cusp}}$ that turns out to be negative. Two classes of cusp configurations, with different numbers of minima and saddle points are shown in black and blue, with the corresponding image configurations shown on the right. Minor axis cusps (blue points) are defined by a minimum as the central cusp image.}
\end{figure*}
\begin{equation}
\nonumber \delta F_i = \frac{| F_{Truth} - F_{model} |}{F_{Truth}}
\end{equation}

\noindent We calculate the commonly used $R_{\rm{cusp}}$ and $R_{\rm{fold}}$ statistics given by
\begin{eqnarray}
\nonumber R_{cusp} &=& \dfrac{M_{1} + M_{2} - S_{1}}{M_{1} + M_{2} + S_{1}} \quad \rm{(major \ axis)} \\ \nonumber &=& \dfrac{M_2 - S_{1} - S_{2}}{M_{2} + S_{1} + S_{2}} \quad \rm{(minor \ axis)} \\ \nonumber R_{fold} &=& \frac{M_{2} - S_{1}}{M_{2} + S_{1}}
\end{eqnarray}
for each model. Major axis and minor axis cusps are defined by whether the central cusp image is the first saddle S1 (major) or the second minimum M2 (minor). The different configurations are shown on the right hand side of Figure~\ref{fig:Rcuspfold_vs_real}. It has been shown \citep[see][]{Schechter:2002p699,Kee03a} that small, compact deflectors in the lensing galaxy, or local perturbations, tend to suppress the brightness of images appearing on saddle points of the time delay surface, while preferentially magnifying minima. On the other hand, perturbations to the gravitational potential on scales larger than the image separation, or global perturbations, do not discriminate between minima and saddle points. The different responses of these cusp configurations to lens structures suggest they could potentially be used to differentiate between different sources of flux ratio anomalies.
 
For models 2-5, we compute the offset between the unsigned $R_{\rm{cusp}}$ and $R_{\rm{fold}}$ statistics
\begin{equation}
\nonumber \Delta R_{model} = ||R_{Truth}|- |R_{model}|| 
\end{equation}
Smooth lens models will yield values close to zero in the limit of vanishing distance between neighboring images, with small variations depending on the image separation and properties of the main lens model \citep[see][]{KGP03,Keeton:2005p591}. Large values of $R_{\rm{cusp}}$ and $R_{\rm{fold}}$ are typically associated with perturbations to the lens potential on scales smaller than the image separation, and as such, these statistics are often used as a indicators of small scale structure near a lensed image.
\subsubsection{Distribution of $R_{\rm{cusp}}$ and $R_{\rm{fold}}$ for target galaxies}
\begin{figure}
\includegraphics[trim=0cm 0cm 0cm 0cm,clip,width=.48\textwidth]{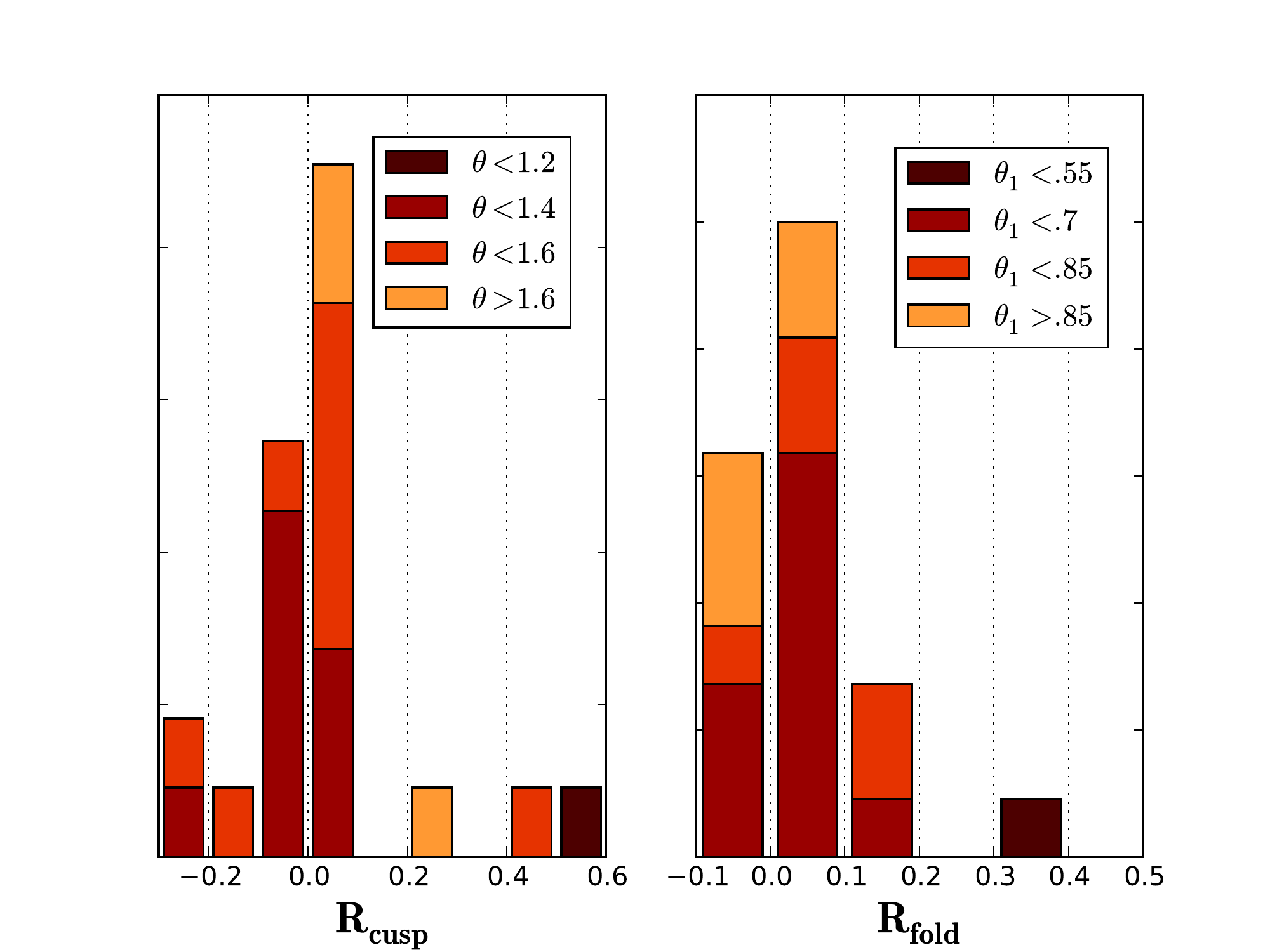}
\caption{\label{fig:Rcuspfold_dis}Distributions of $R_{cusp}$ and $R_{\rm{fold}}$ statistics of the \textit{Truth} data, color coded by the largest separation between the three merging images $\theta$ (for cusp configurations) or the separation between the merging image pair $\theta_1$ (for fold configurations), normalized by the Einstein radius.} 
\end{figure}

In Figure \ref{fig:Rcuspfold_dis}, we plot the distributions of the $R_{\rm{cusp}}$ and $R_{\rm{fold}}$ statistics of the \textit{Truth} data for each of our target galaxies as a function of the largest distance (normalized by $R_{\rm{Ein}}$) between the three merging images $\theta$ (cusp lenses) or the merging pair $\theta_1$(fold lenses). Our sample of mock lenses contains $R_{\rm{cusp}}$ and $R_{\rm{fold}}$ statistics as high as 0.5 for cusp lenses, and as high as 0.35 for fold lenses.
In Figure \ref{fig:Rcuspfold_vs_real}, we plot the $R_{\rm{cusp}}$ values of ten real lenses, with data as reported in \citet{KGP03} and \citet{XuEtal2016}. Several of these lenses, notably B2045 and B1933 have $R_{\rm{cusp}}$ and $R_{\rm{fold}}$ that are inconsistent with lensing by a smooth potential \citep{KGP03,McK++07}, while \citet{XuEtal2016} showed some of these anomalies could be accounted for by introducing a population of dark subhalos and multipole potential terms in smooth lens models, while noting that the observed anomalies were unlikely to caused entirely by dark subhalos. 
\begin{figure}
\includegraphics[trim=0cm 0cm 0cm .4cm,clip,width=.48\textwidth]{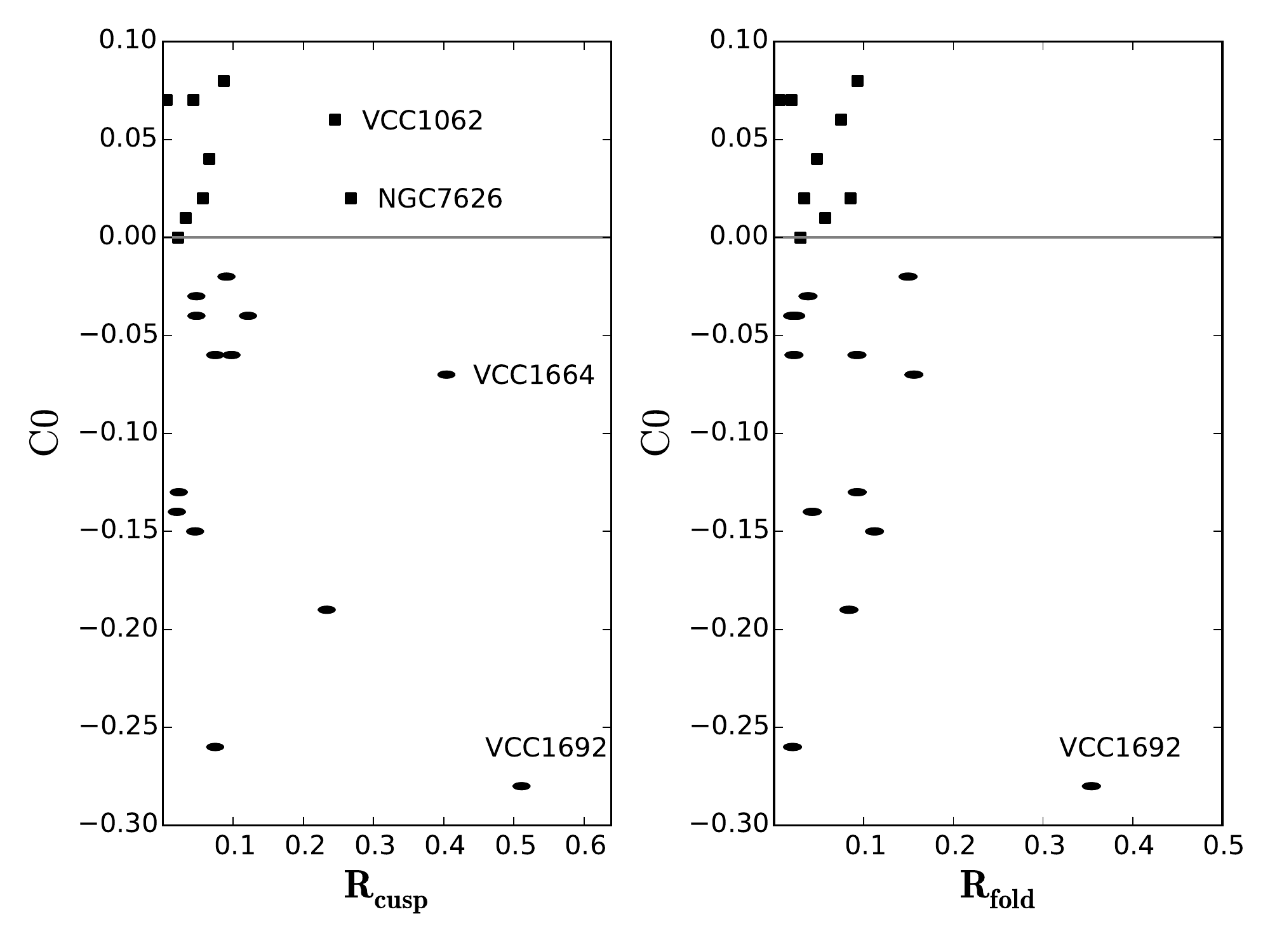}
\caption{\label{fig:C0_vs_Rcuspfold} Diskyness and boxyness for each mock deflector, described by the {\tt {galfit} }parameter $C0$. A positive $C0$ corresponds to boxy isophotes, while a negative value indicates diskyness. The anomalous systems shown in Figures \ref{fig:fluxratios}, \ref{fig:fluxratios2}, and \ref{fig:Rcuspfold_vs_real} are labeled.}
\end{figure}
In the following paragraphs, we will focus attention on the 4 systems with large $R_{\rm{cusp}}$ and $R_{\rm{fold}}$ anomalies in order to understand their origin, and to gain insight into how a baryon-induced induced flux ratio anomaly might reveal itself in an observational scenario.While we juxtapose real lens systems with our mock lenses in Figure~\ref{fig:Rcuspfold_vs_real}, we do not argue that baryons are responsible for the anomalies seen in these systems. Rather, we emphasize that features of the luminous matter in a lensing galaxy can give rise to the large values of $R_{\rm{cusp}}$ and $R_{\rm{fold}}$ typically associated with non-baryonic substructure - albeit rarely - especially in systems with a stellar disk or other irregularities. The stellar mass components and flux ratio distributions for these lenses are shown in Figures~\ref{fig:fluxratios} and \ref{fig:fluxratios2}, while maps of the magnification surfaces are shown in Figures~\ref{fig:magmaps_VCC1692c}, \ref{fig:magmaps_VCC1664}, \ref{fig:magmaps_NGC7626}, \ref{fig:magmaps_VCC1062}, and \ref{fig:magmaps_VCC1692f}.
Unsurprisingly, three of the most anomalous mock lenses show evidence for disky or boxy isophotes (see Figure \ref{fig:C0_vs_Rcuspfold}). The remaining system, NGC7626, has an $R_{\rm{cusp}}$ anomaly that can can be partially accounted for by the presence of a background galaxy and globular clusters near the images. Three out of of four anomalous mock lenses have velocity dispersion below 200 km s$^{-1}$, unlike the high central velocity dispersion deflectors most likely to act as strong lenses. In light of the inflated percentage of low stellar velocity dispersion targets in our lens sample, which are more likely to contain disks and irregular morphological features than high velocity dispersion deflectors, the non-detection of flux ratio anomalies through most of our lens sample illustrates the sub-dominant nature of flux ratio anomalies caused by luminous matter. 
\subsection{Analysis of Anomalous Systems}
Analysing the few mock deflectors where the luminous matter of the lensing galaxy influences the flux ratios serves to illustrate how stellar mass can affect lensing observables. We will first inspect the cusp configurations, followed by the fold configurations.
\subsubsection{Cusps}
\begin{itemize}
\begin{figure*}
\includegraphics[clip,trim=2.5cm .5cm 1cm
.25cm,width=.48\textwidth]{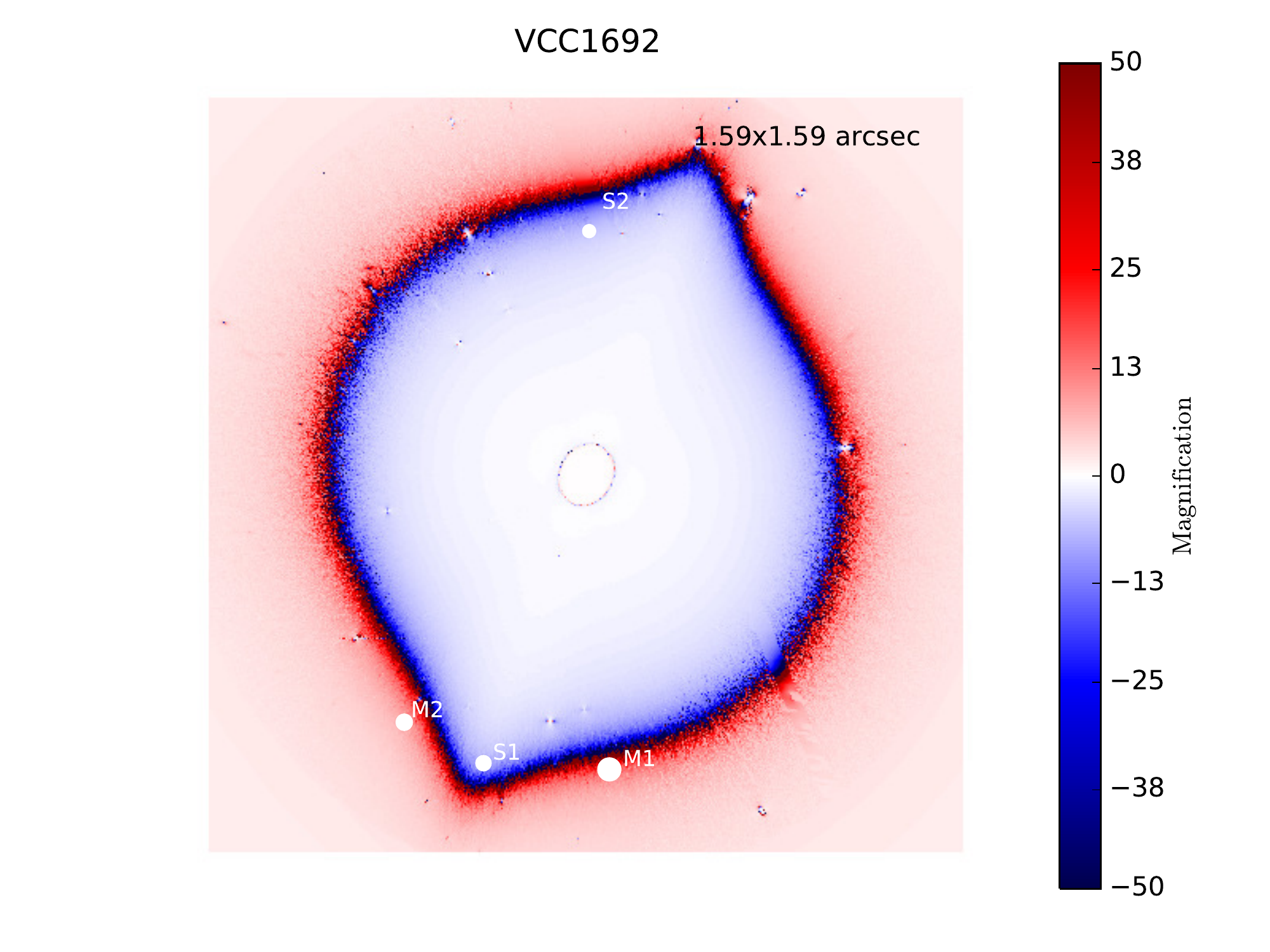}
\includegraphics[clip,trim=2.5cm .5cm 1cm
.25cm,width=.48\textwidth]{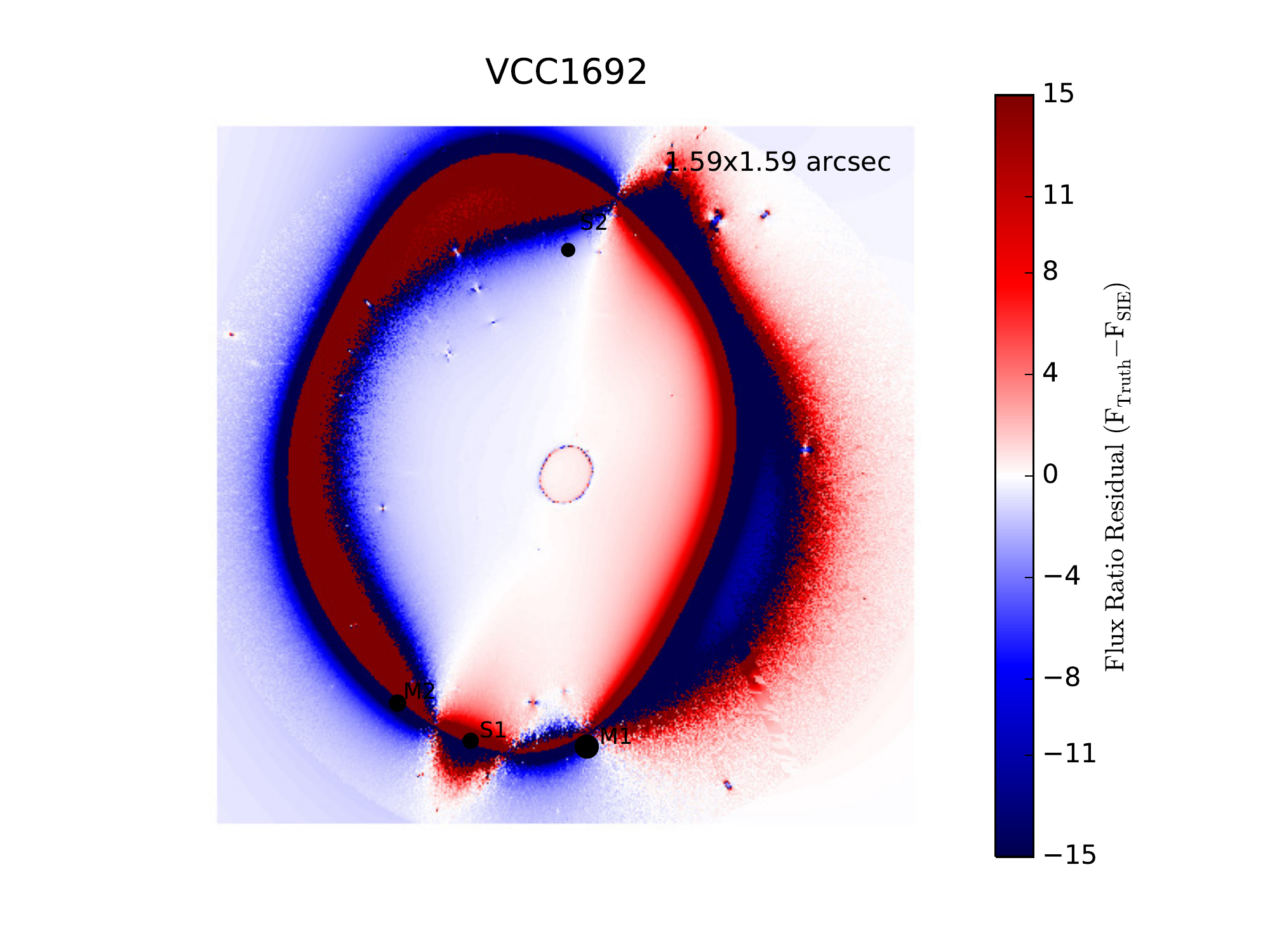}
\caption{\label{fig:magmaps_VCC1692c} $R_{\rm{cusp}}$ anomaly. Left: Magnification surface derived from the convergence map. Right: Residuals after subtracting the magnification mag of the best fit SIE model. The interplay of the disk and external shear, which is nearly orthogonal to the disk position angle in this system, creates large residuals in the magnification surface that create a strong flux ratio anomaly.}
\end{figure*}
\begin{figure*} 
\includegraphics[clip,trim=2.5cm .5cm 1cm
.25cm,width=0.48\linewidth,keepaspectratio]{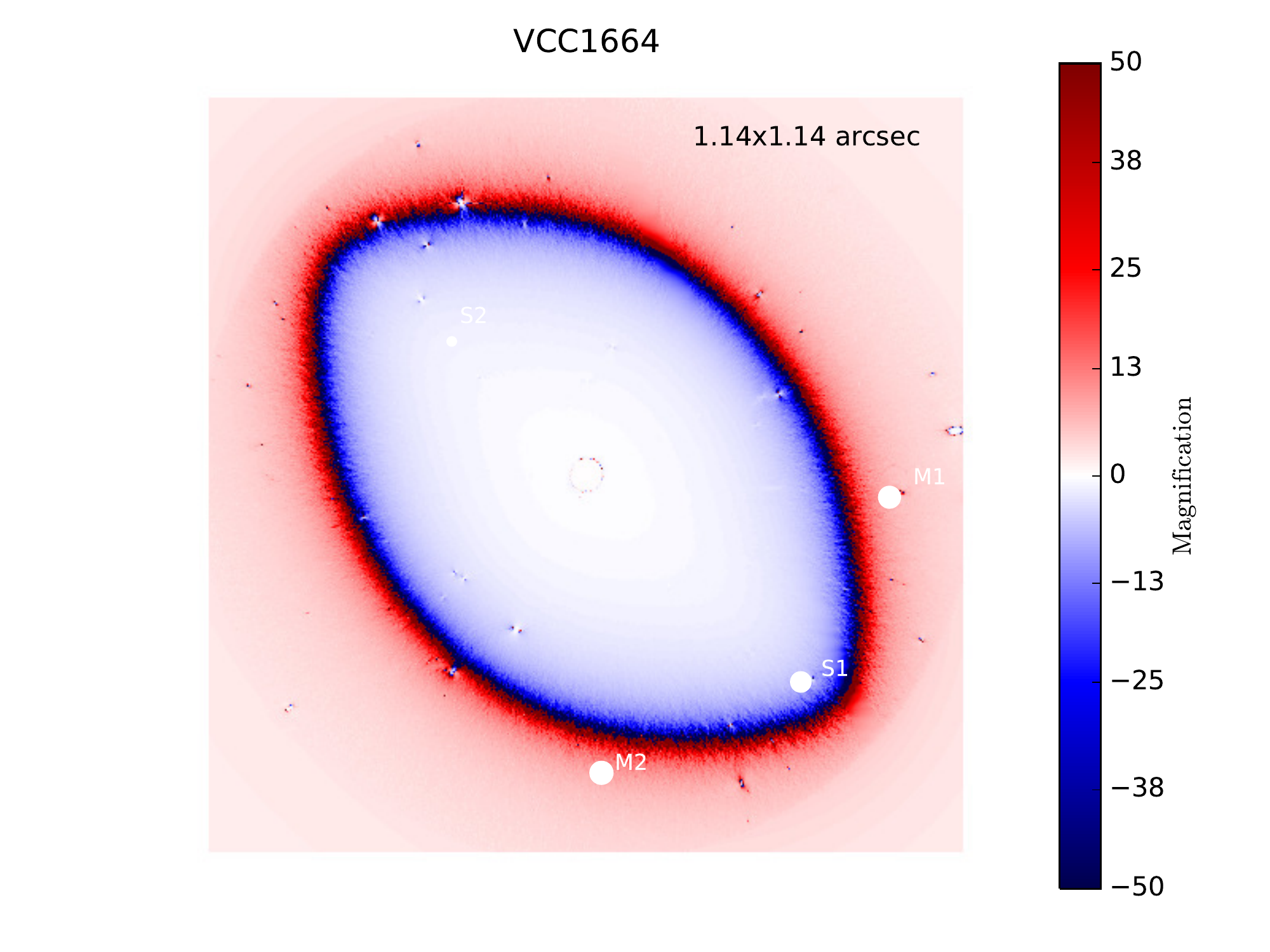}
\includegraphics[clip,trim=2.5cm .5cm 1cm
.25cm,width=0.48\linewidth,keepaspectratio]{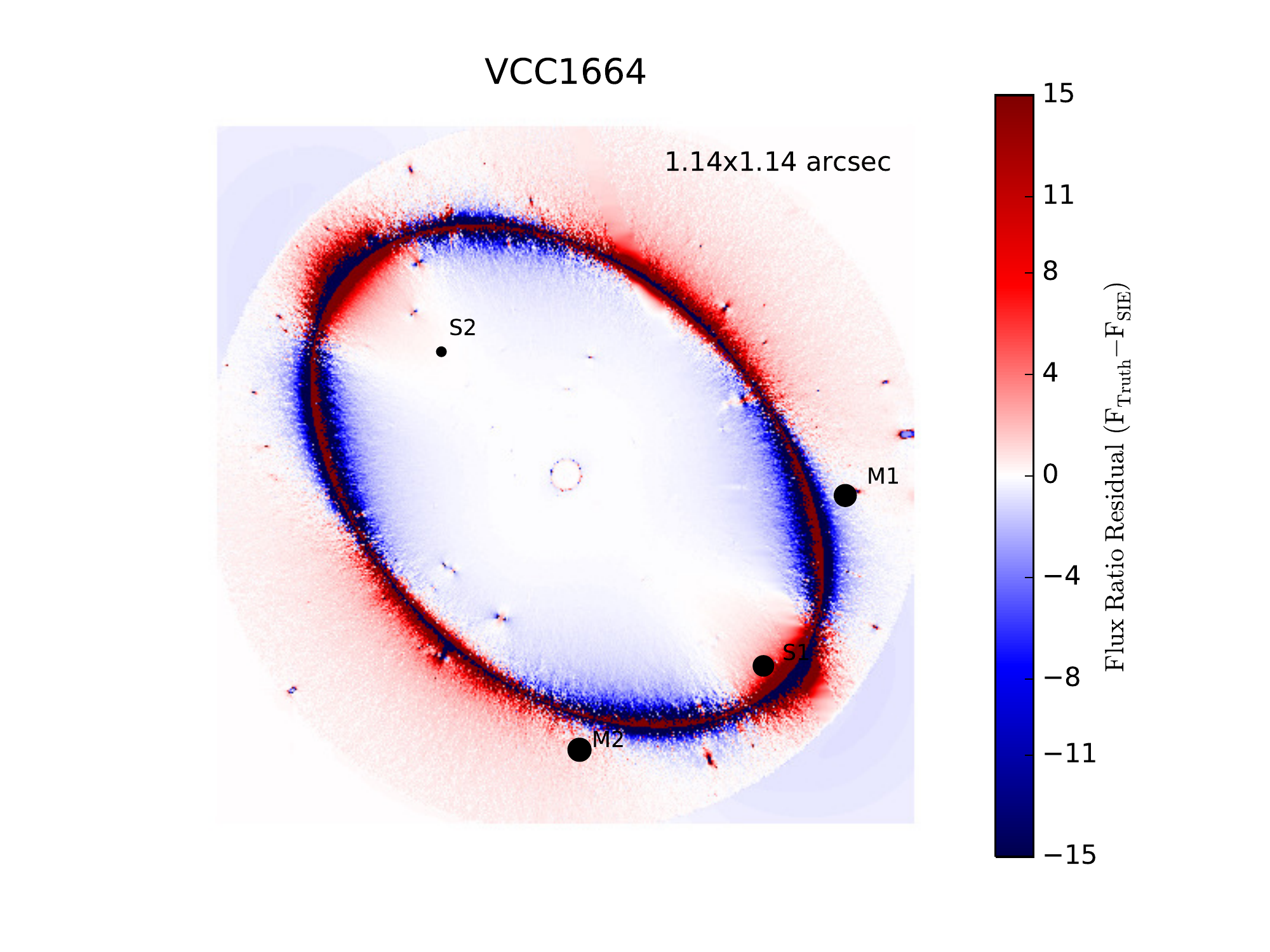}
\caption{\label{fig:magmaps_VCC1664} $R_{\rm{cusp}}$ anomaly. Left: Magnification surface derived from the convergence map. Right: Residuals after subtracting the magnification mag of the best fit SIE model. The effect of the stellar manifests itself as a multipole pattern in the residual map.}
\end{figure*}
\item VCC1692: An elongated galaxy with a prominent disk, as seen in Figure \ref{fig:fluxratios}, with stellar velocity dispersion $\sigma_* = 187 \ \rm{km/sec}$. We embed this galaxy in an external shear with a 60 degree offset between the disk position angle and the position angle of the external shear, resulting in a warped astroid caustic. The residuals between the magnification surface of the mock lens and the best fit SIE, shown in Figure~\ref{fig:magmaps_VCC1692c}, dramatically displays this effect, with large residuals on the ends of the disk where images are located. The lens has an unusual image configuration, with the far image (S2) located off the symmetry axis of the cusp, while the three cusp images are very close to each other. Both the SNFW and SIE fit the positions reasonably well, although both models display M1/S1 flux ratio anomalies of 80\%. The proximity of images M2 and S1 to the stellar disk likely significantly the perturbs the flux ratios between these images. We can compare this with the closest real analog in Figure~\ref{fig:Rcuspfold_vs_real} B2045+265. Deep imaging of the system \citep{McK++07} shows that the deflector galaxy is almost perfectly round ($b/a=0.94\pm0.01$) when imaged with an F160W filter, while it displays irregular morphological features in F814W. McKean et al. also investigate the possibility that a luminous satellite located between the main deflector and the three cusp images could be responsible for the observed flux ratio anomaly. The Einstein radius of $1\farcs06$ corresponds to a velocity dispersion of 278 km s$^{-1}$, once the correct source redshift of $z_s=2.35$ is taken into account (Nierenberg et al. 2016, in preparation). Thus, the deflector galaxy in B2045 is very different than the one in the mock lens discussed here, consistent with a different origin of the anomaly, even though the amplitude is the same. In our sample of mock lenses, VCC1692 is the only lens with significant astrometric anomalies. In this sense it is an outlier, as our analysis shows that lensing by luminous matter typically does not result in image positions that cannot be fit by an SIE or SNFW, while they still may result in anamalous flux ratios. The interplay between the disk and external shear is likely to blame for this unique system, resulting in the asymmetric image configuration and peculiar shape of the astroid caustic. 
\begin{figure*} 
\includegraphics[clip,trim=2.5cm .5cm 1cm
.25cm,width=0.48\linewidth,keepaspectratio]{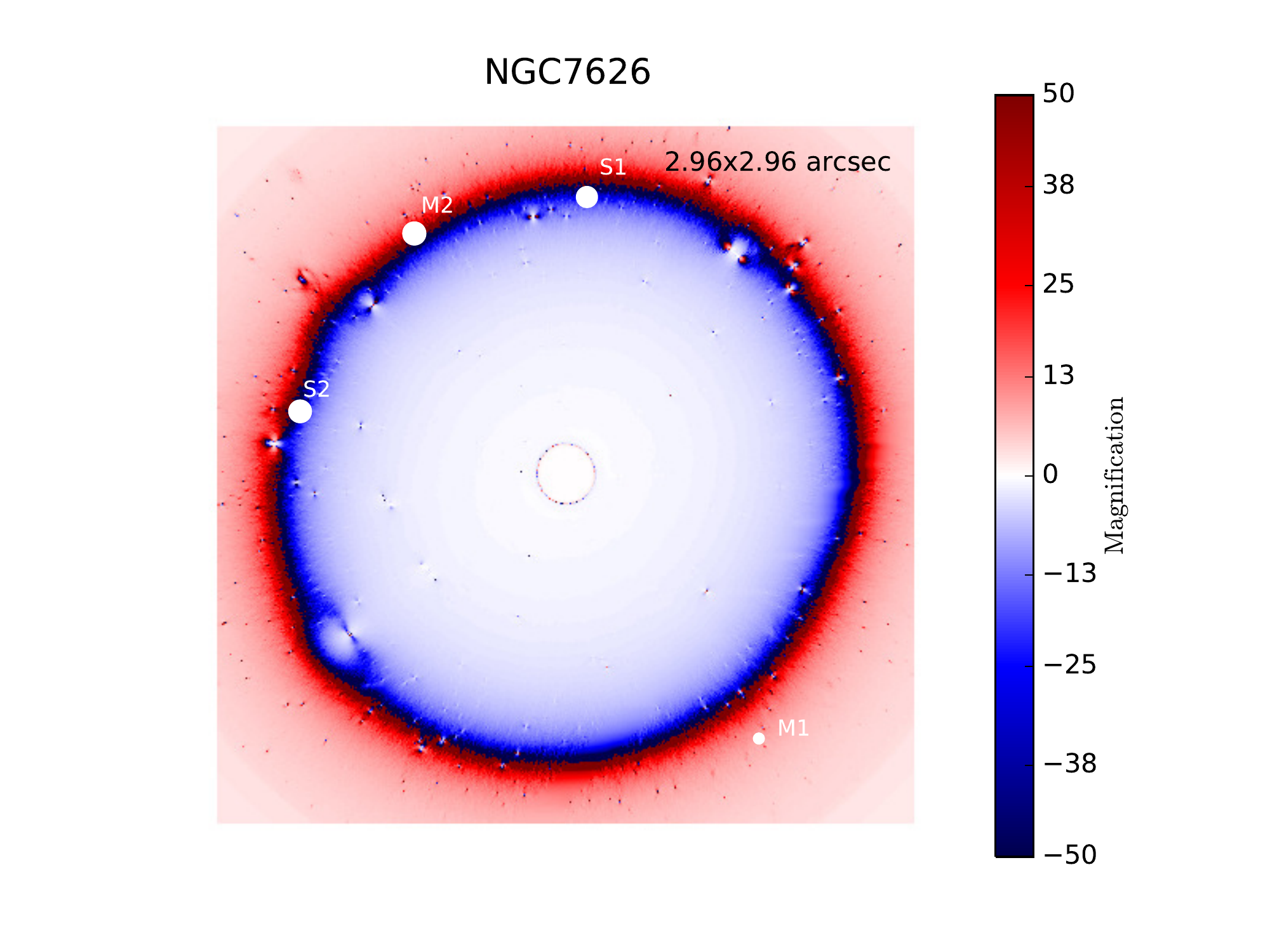}
\includegraphics[clip,trim=2.5cm .5cm 1cm
.25cm,width=0.48\linewidth,keepaspectratio]{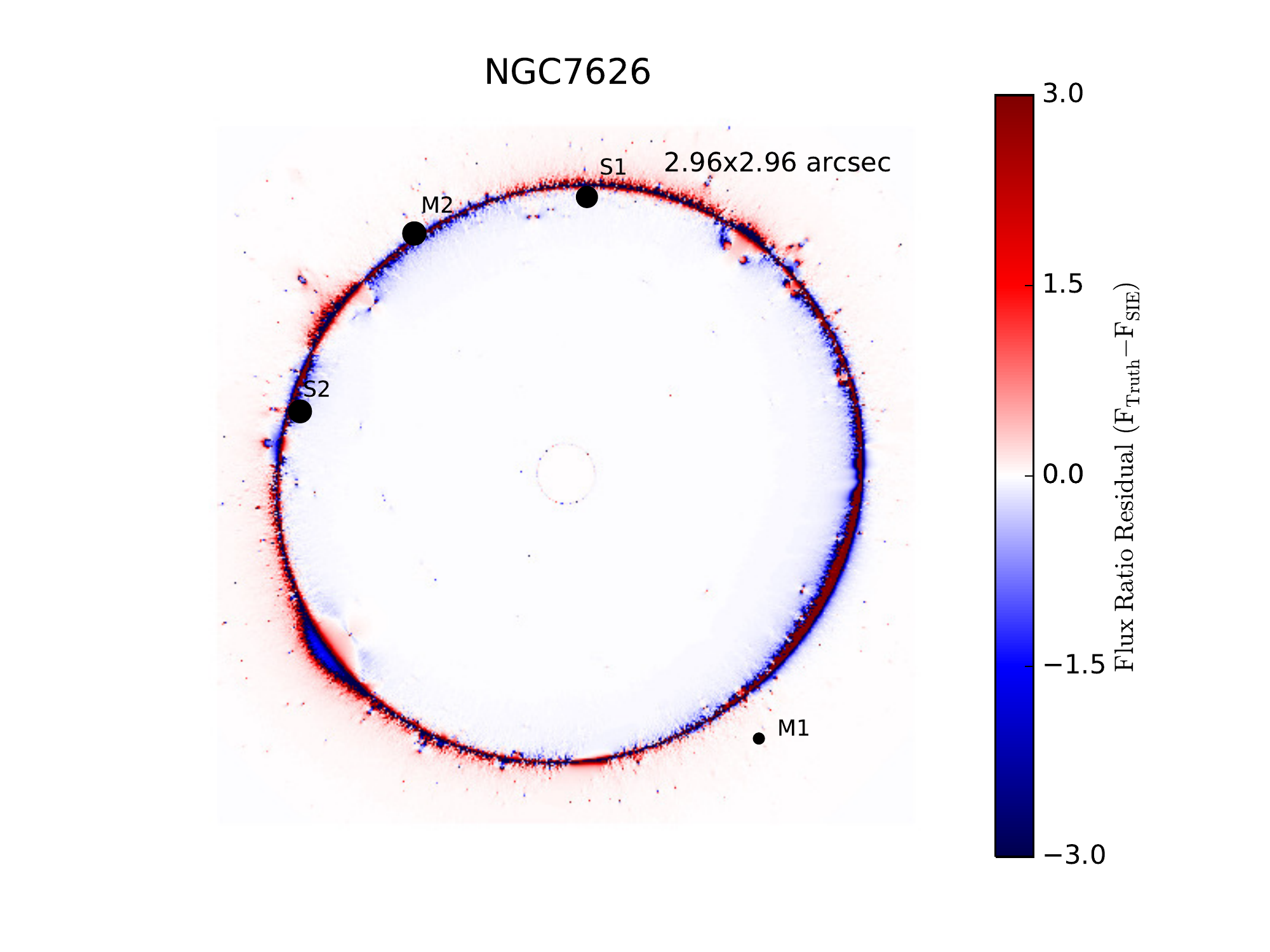}
\caption{\label{fig:magmaps_NGC7626} $R_{\rm{cusp}}$ anomaly. Left: Magnification surface derived from the convergence map. Right: Residuals after subtracting the magnification map of the best fit SIE model. This system does not exhibit any obvious large scale morphological irregularities, although the effect of a background galaxy in the lower left is clearly visible. The galaxy appears to contain many luminous substructures, some of which may be associated with a dark subhalo. However, it is possible that some of these features, such as the galaxy in the lower left, are in the background or foreground. In Section 3.3.1 we discuss the impact of luminous substructure on flux ratios and the $R_{\rm{cusp}}$ parameter.}
\end{figure*} 
\item VCC1664:  This is a small galaxy with velocity dispersion 155 km s$^{-1}$. As such, it is not representative of a typical deflector of lensed quasars. It is similar to VCC1692 in that it has a prominent disk that results in large M1/S1 and M2/S1 flux ratio anomalies that both the SIE and SNFW models fail to reproduce, as seen in the distributions of Figure \ref{fig:fluxratios2}. The magnification residuals between the \textit{Truth} model and the best fit SIE result in a multipole pattern around the critical curve, seen in Figure \ref{fig:magmaps_VCC1664}. Since there is a significant amount of small scale structure scattered around the deflector, some of which lay close to an image, we experimented with removing these potential sources of flux ratio anomaly, but found that this did not affect the flux ratios. A lens with a similar flux ratio anomaly, RXJ1131+1231 \citep{SluseEtal2003}, differs in several important ways. First, \citet{Suy++13} measured a stellar velocity dispersion in J1131 of $323 \pm 23$ km s$^{-1}$, which would likely result in images far enough from the majority of the stellar mass of the lens to be affected by morphological features of the luminous matter, especially as imaging of J1131 shows no evidence for the presence of a stellar disk or significant elongation. Second, the flux ratio anomaly in J1131 is accompanied by an astrometric anomaly; attempts to fit the lens with an single SIE with shear or a two-lens model both fail to recover the correct astrometry \citep{COSMOX}, whereas smooth potentials recover the image positions of VCC1664 almost perfectly. It should also be noted that because VCC1664 has a larger cusp image separation $\theta$ than J1131, the $R_{\rm{cusp}}$ statistic can naturally be larger without substructure. 
\begin{figure} 
\includegraphics[clip,trim=2.5cm 6.5cm 10cm
1.6cm,width=0.48\linewidth,keepaspectratio]{./figs/magmap_NGC7626_cusp_withshear-eps-converted-to.pdf}
\includegraphics[clip,trim=2.5cm 6.5cm 10cm
1.5cm,width=0.48\linewidth,keepaspectratio]{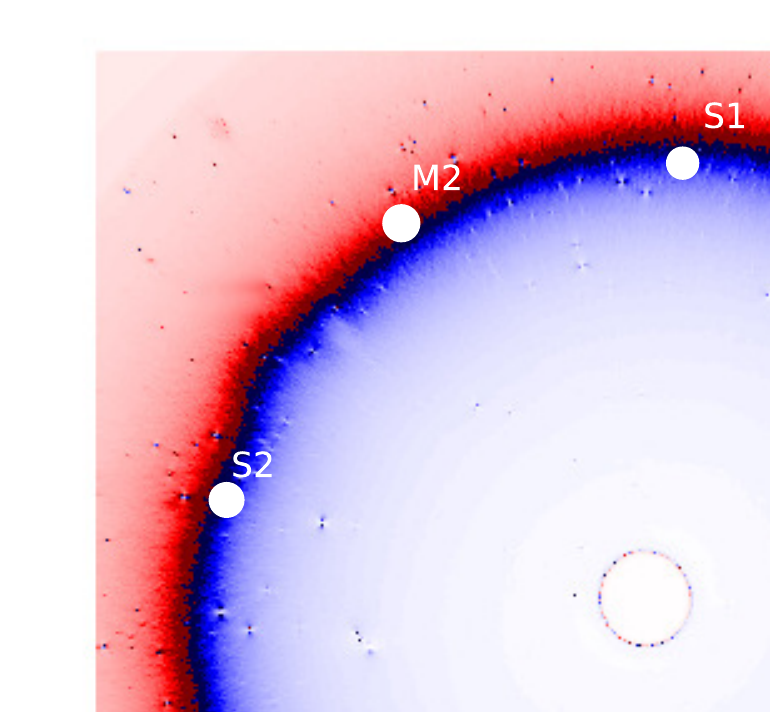}
\caption{\label{fig:magmaps_beforeafter} Left: Original magnification map, with all small scale structure present. Right: Magnification map with 3 globular clusters and one background or satellite galaxy removed. Before their removal, each of these features in the convergence map contributed the equivalent of $10^7 - 10^{7.5} M_{\odot}$.}
\end{figure}
\begin{figure} 
\includegraphics[clip,trim=0.25cm 0cm 0.5cm
5.55cm,width=0.5\textwidth,keepaspectratio]{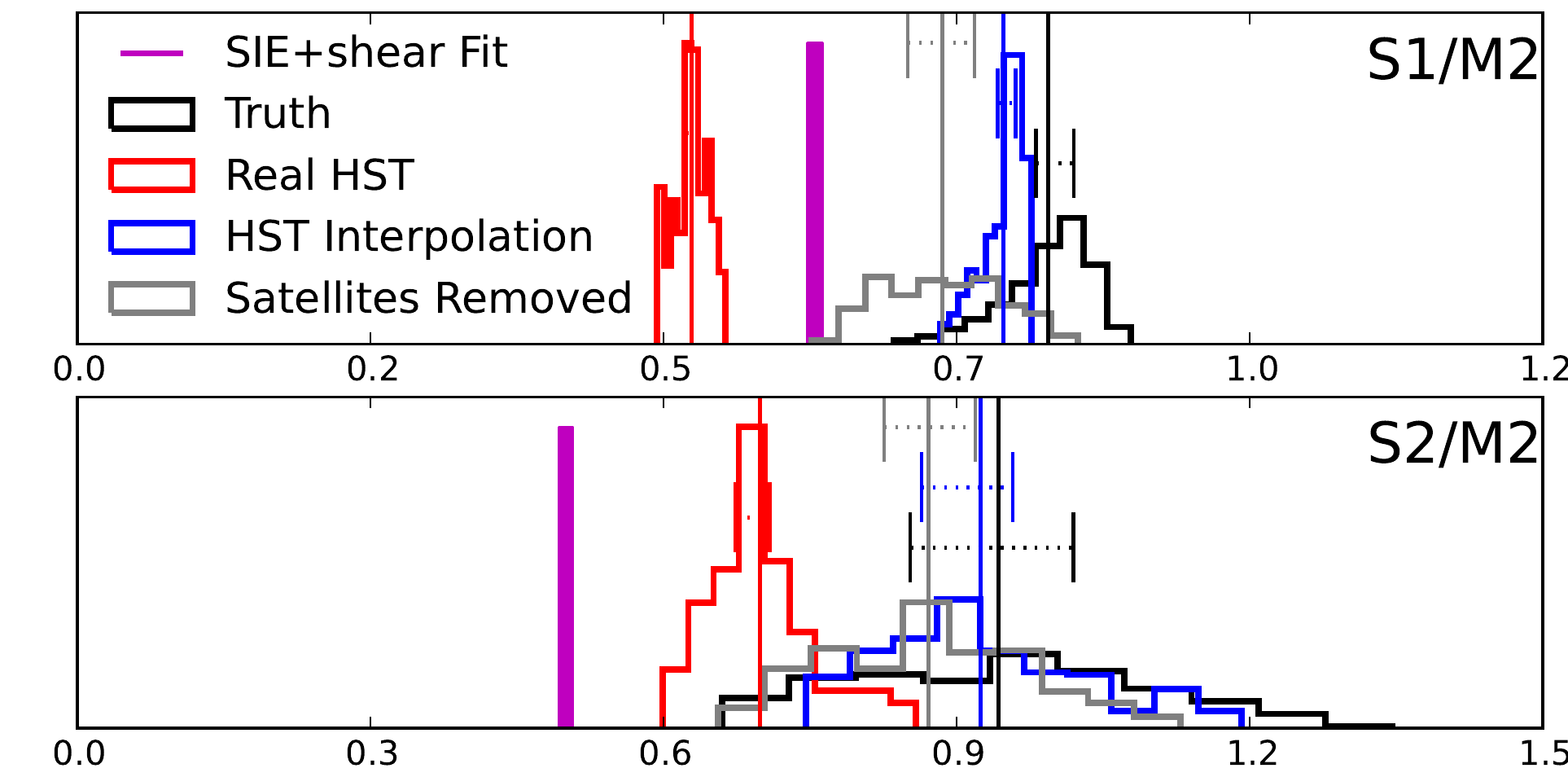}
\caption{\label{fig:newfluxratios} Flux ratios after the removal of small scale structure between the three cusp images in NGC7626.}
\end{figure}
\item NGC7626: The most massive cusp mock lens 
($\sigma_* = 274$ km sec$^{-1}$) with a significant $R_{\rm{cusp}}$ anomaly. NGC7626 is surrounded by globular clusters and luminous satellites. One background galaxy is visible to the lower left, and it induces a flux ratio anomaly in the fold configuration (see the S2/M1 ratio in Figure \ref{fig:fluxratios}), although it is too far from the cusp images to be responsible for the $R_{\rm{cusp}}$ anomaly and does not affect the merging pair in the fold configuration. There are two structures between the M2 and S2 images, seen clearly in the convergence map (Figure \ref{fig:fluxratios}) and in the map of the magnification surface (Figures \ref{fig:magmaps_NGC7626} and \ref{fig:magmaps_beforeafter}) which visibly perturb the critical curve. The structure outside the curve resembles a background spiral galaxy, while the object just inside resembles a large globular cluster. Since NGC7626 does not posses a stellar disk or boxy isophotes that could explain the anomaly, we experimented with removing these features individually, replacing them with smooth interpolations of the convergence map. Specifically, we removed the globular cluster and background galaxy between M2 and S2, a small globular cluster near S1, and a very small cluster near S2. The before/after magnification maps are shown in Figure \ref{fig:magmaps_beforeafter}, and the new flux ratios in Figure \ref{fig:newfluxratios}. After removing these small scale structures, which our normalization procedure assigned convergence equivalent to that produced by a $10^7 M_{\odot}$ perturber, we find that the $R_{\rm{cusp}}$ anomaly shrinks in magnitude to 0.19 from 0.26. In the context of Figure \ref{fig:poserr_vs_Rcuspfold}, this suggests it could be accounted for by an SIE model. We therefore conclude that the main source of anomaly in this system is due to structure in the deflector on scales smaller than the image separation. This hypothesis is supported by examining the residual map in Figure \ref{fig:magmaps_NGC7626}, where the alternating blue and red colors coincide with the location of the perturbing globular clusters and galaxy. NGC7626 highlights that even massive deflectors can suffer flux ratio anomalies if there is sufficient small scale structure near the critical curve, whether it is in the form of dark substructure or luminous matter. However, it is important to remember that that this is seemingly a rare occurrence, and it is possible that this signal will be overwhelmed by the lensing signatures of a full population of dark subhalos, a question we will address in a future paper. 

In the $\theta$ vs. $R_{\rm{cusp}}$ parameter space, RXJ0911+0551 \citep{Kneib00} is the nearest neighbor of NGC7626. NGC7626 is a round deflector with an ellipticity of 0.17, while a best fit SIE model of J0911 \citep{COSMOX} favors a deflector with ellipticity 0.11. The velocity dispersion of J0911, if it is modeled as an SIE with Einstein radius 0.9 arcsec works out to $\sigma_{\rm{SIE}}=239$km s$^{-1}$ after adopting correct lens and source redshifts \citep{Kneib00,Bade97}, while NGC7626 has a velocity dispersion of 274 km s$^{-1}$. Neither J0911 nor NGC7626 display astrometric anomalies with respect to a smooth model when a second deflector galaxy is included in the model for J0911 \citep{COSMOX}. NGC7626 stands out in our set of mock lenses, as it is round with high velocity dispersion, but still displays significant flux ratio anomalies, indicting that high velocity dispersion does not always guarantee benign flux ratios.  
\begin{figure*} 
\includegraphics[clip,trim=2.5cm .5cm 1cm
.25cm,width=0.48\linewidth,keepaspectratio]{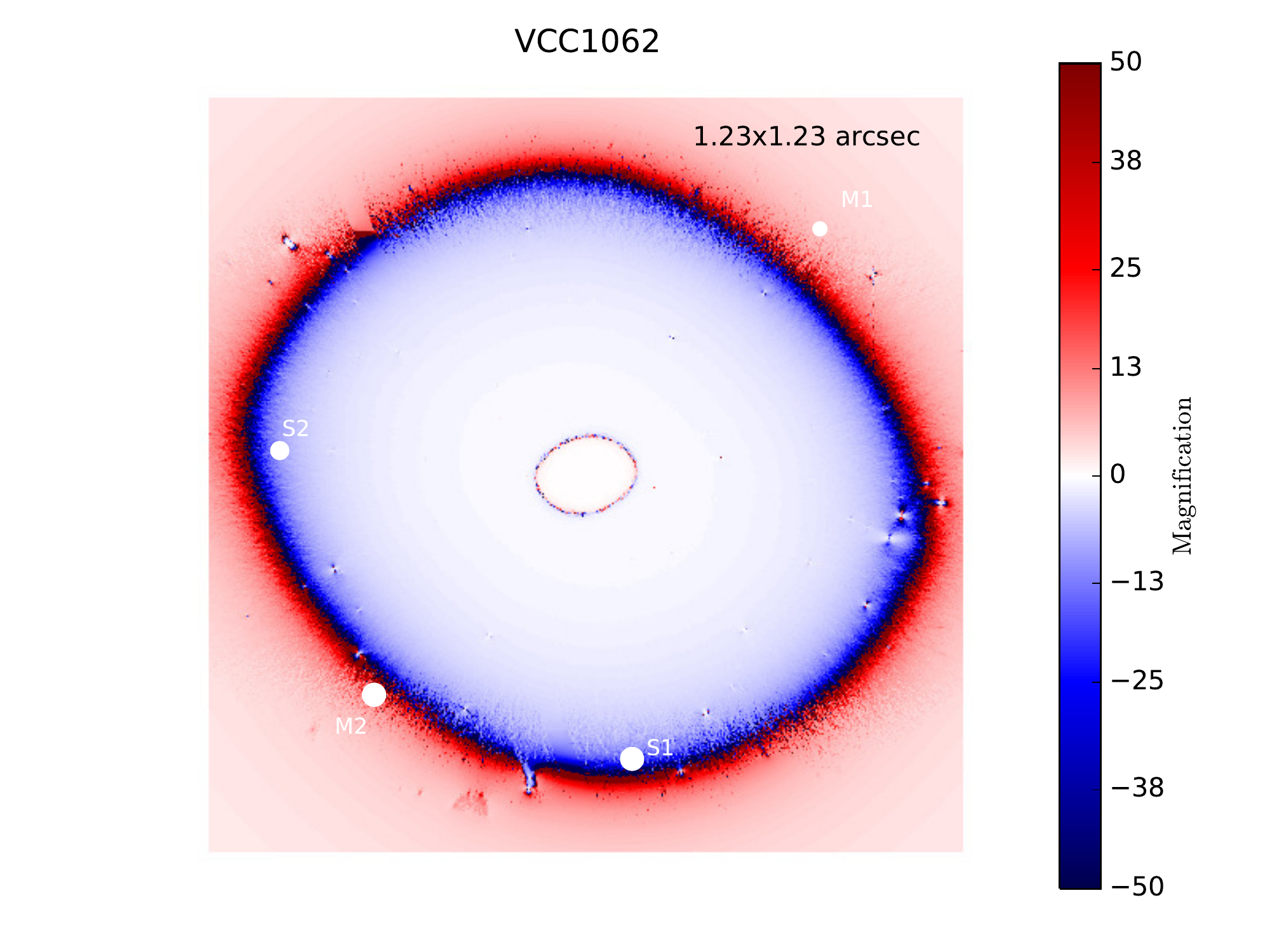}
\includegraphics[clip,trim=2.5cm .5cm 1cm
.25cm,width=0.48\linewidth,keepaspectratio]{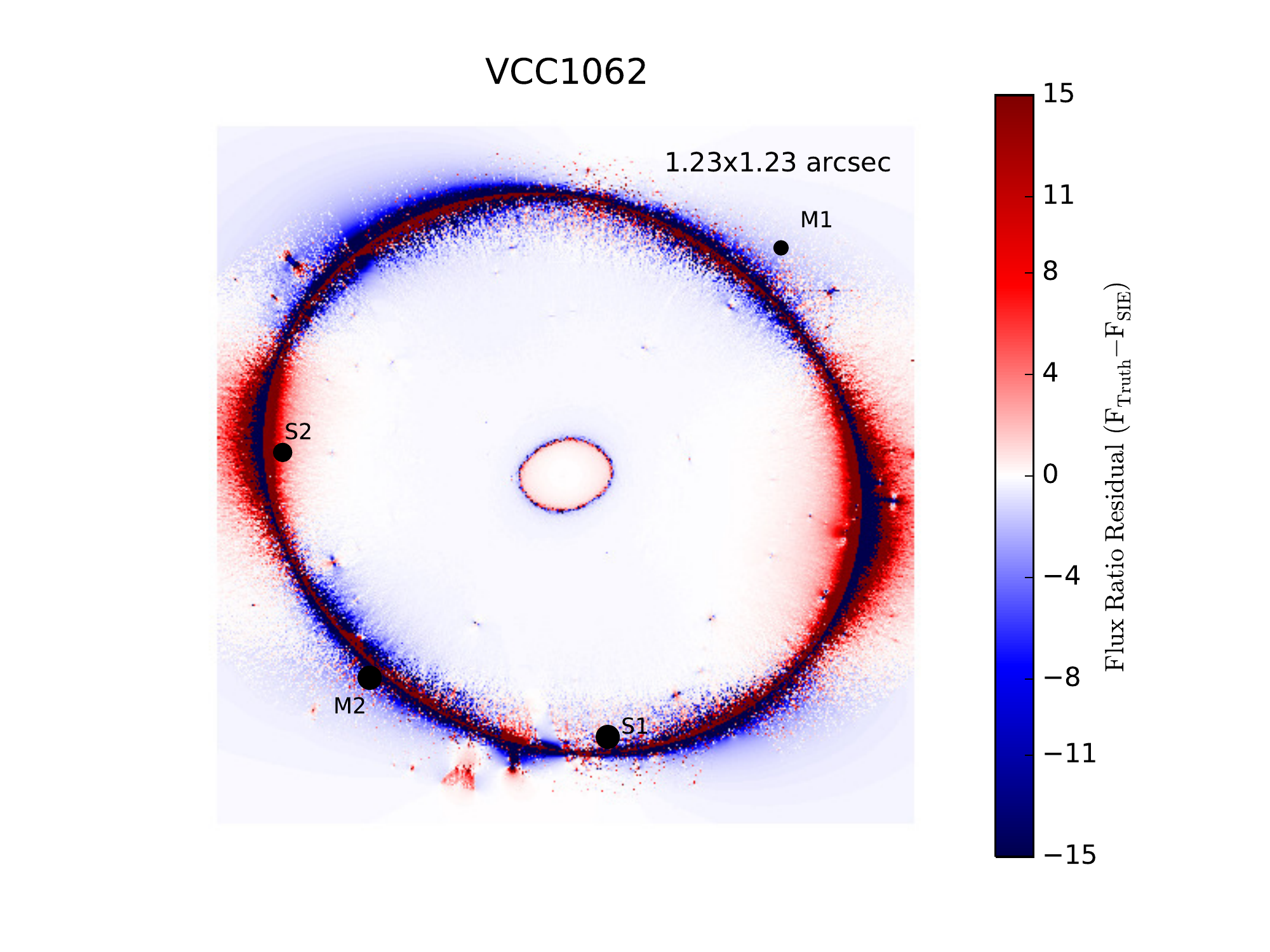}
\caption{\label{fig:magmaps_VCC1062} $R_{\rm{cusp}}$ anomaly. Left: Magnification surface derived from the convergence map. Right: Residuals after subtracting the magnification mag of the best fit SIE model. The effect the the stellar manifests itself as a multipole pattern in the residual map.}
\end{figure*}
\item VCC1062: The external shear applied in this lens forms an angle of 61 $\deg$ with the stellar quadrupole moment position angle, similar to VCC1692, that results in a cusp configuration that is not coaxial with the stellar ellipticity position angle. Coupled with the boxy isophotes (see Figure \ref{fig:C0_vs_Rcuspfold}), this results in a complicated potential that the best fit SIE fails to capture, as seen in the map of flux ratio residuals in Figure \ref{fig:magmaps_VCC1062}. 
\begin{figure*}
\includegraphics[clip,trim=2.5cm .5cm 1cm
.25cm,width=.48\textwidth]{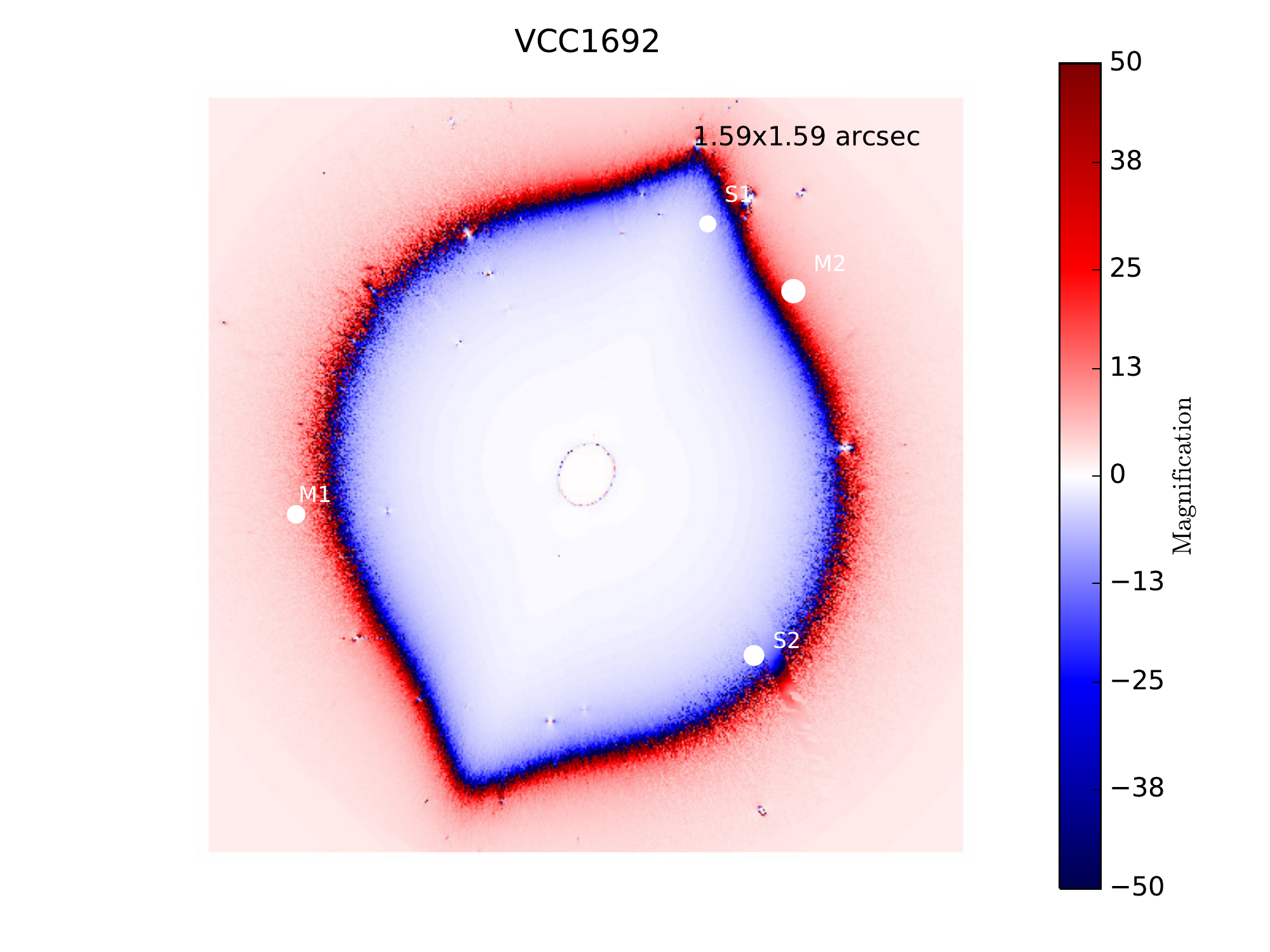}
\includegraphics[clip,trim=2.5cm .5cm 1cm
.25cm,width=.48\textwidth]{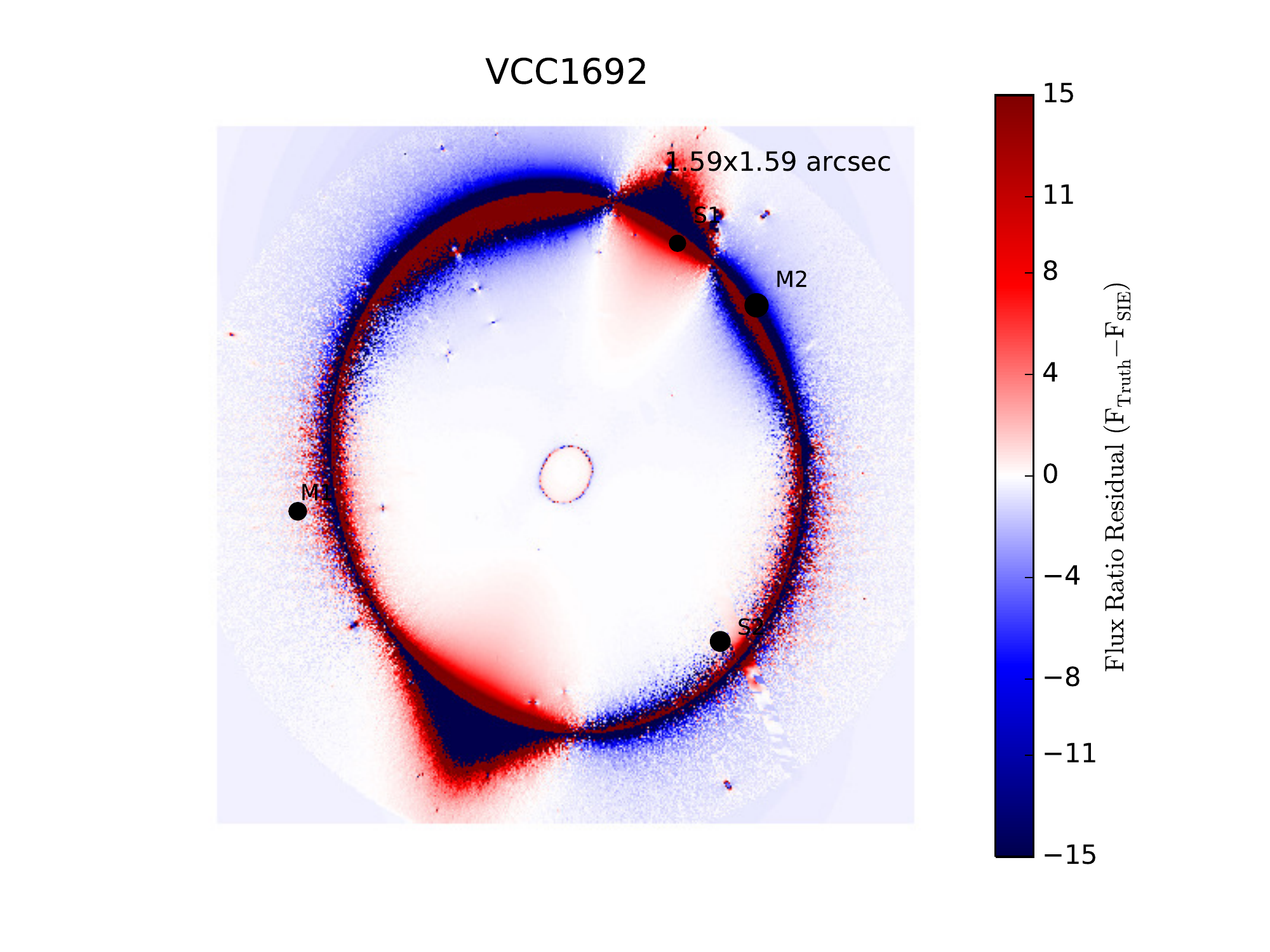}
\caption{\label{fig:magmaps_VCC1692f} $R_{\rm{fold}}$ anomaly. Left: Magnification surface derived from the convergence map. Right: Residuals after subtracting the magnification mag of the best fit SIE model. The merging image pair in this fold configuration happens to land near the portion of the critical curve extended by the stellar disk, a feature the best SIE fails to capture. The resulting residual in the magnification surface gives rise to the large $R_{\rm{fold}}$ anomaly.} 
\end{figure*}
The small velocity dispersion ($\sigma_* = 179$ km/sec) results in a small Einstein radius, which in turn results in images close to the center of the lens. As a result, the images S2 and M2 are located closer to the ends of the elongated baryonic mass distribution where there is more curvature in the potential, making this lens more susceptible to influence from its luminous mass component. While both the SIE and SNFW fail to recover the correct flux ratios, the anomalies are $<40\%$. However, collectively the anomalies lead to a significant $R_{\rm{cusp}}$ anomaly. VCC1062 has an anomaly quite similar to that observed in B0712+472 \citep{Jackson98}. Both deflectors have relatively small stellar velocity dispersion (B0712 has $\sigma_{\rm{SIE}} = 189$ km/sec), which we estimate for B0712 by adopting the correct redshifts as cited in \citet{COSMOX} and utilizing the relationship between image separation and velocity dispersion presented in \citep{KochanekFPlane}. The two lens systems both appear to have highly elliptical baryonic mass distributions, consistent with the presence of edge-on massive disk \citep{Jackson98}. While the small mass and high ellipticity of B0712 suggests the anomaly may be influenced by baryonic matter, the astrometric anomaly noted by \citet{Kawano++04} is not a common feature among our mock lenses, and as such alternate explanations are favored, such as line of sight perturbations, as mentioned by \citet{F+L02}, or dark substructure. Discrepancies between the flux ratios in the optical/near IR and radio data suggest microlensing and/or dust extinction could also be present. Regardless, deeper imaging of this system could help disentangle the possible role of baryonic structure from other sources of flux ratio perturbation. 
Completely different is the case of the real lens B1422+231 \citep{Patnaik++92}. Even though the cusp flux ratio anomaly is similar to that of VCC1062, the cusp image separation $\theta$ is large enough that the measured $R_{\rm{cusp}}$ value \citep{Koo++03a} alone is not inconsistent with lensing by a smooth potential \citep{KGP03,Xuetal2015}. Like many systems in our sample, the astrometric anomalies in B1422 are relatively tame compared to other real lens systems, however the introduction of a dark subhalo near an image corrects the flux ratio anomaly \citep{Nie++14}.
\end{itemize}
\subsubsection{\rm{FOLDS}}

\begin{itemize}

\item VCC1692: The only significant $R_{\rm{fold}}$ anomaly appears in VCC1692, the lens system with the largest $R_{\rm{cusp}}$ anomaly. The S1/M2 flux ratio anomaly is 50\%, likely because of the influence of the stellar disk. This effect is clear in Figure~\ref{fig:magmaps_VCC1692f}, where large residuals between the best fit SIE and the \textit{Truth} model are evident. Unlike the cusp configuration, the \textit{HST Interpolated} distribution agrees with the \textit{Truth} flux ratios, indicating that the curvature of the gravitational potential just off the major axis of the disk is gradual enough that the convolution procedure still captures the disk's effect on the magnification surface. From a modeling standpoint, this implies that the information needed to accurately reproduce the lensing signal of a very disky deflector is lower for fold configurations than for cusp configurations, because cusps images live near the ends of the disk (for major axis cusps), where there is greater curvature. Conversely, folds will tend to straddle the sides of a disk, so one need only resolve a small portion of a relatively straight critical curve dividing the two images.

B1555+375 \citep{Mar+99}, the closest real analogue to VCC1692 in the $R_{\rm{fold}}$ / $\theta_1$ parameter space, has a nearly identical $\theta_1$ and a large $R_{\rm{fold}}$. B1555 is also a very small angular separation lens, with a deflector velocity dispersion estimated from the image separation of 134 km s$^{-1}$, and is highly elliptical. The high ellipticity $\epsilon = 0.54$ and disk feature detected in deep AO imaging \citep{HsuehEtal16} further strengthen the analogy between the mock and the real system. Attempts to model the lens system \citep{Mar+99,MirandaJetzer07} also find that the astrometry of B1555 is consistent with an SIE model, so no extreme astrometric anomalies are present, as is the case with the majority of our mock lenses. Another clue to the nature of the anomaly in B1555 arises from the modeling of VCC1692: if flux ratios are included as constraints in the SIE fit to VCC1692, the resulting astrometric errors and flux ratios appear strikingly similar to that observed in B1555. This suggests that, while formally not a good fit to the data, a single SIE can capture the astrometry and flux ratios of a disky galaxy to within 10 mas and $\approx 70 \%$, respectively. It is likely that this is not a coincidence, as \citet{HsuehEtal16} show that the system can be fit to high precision by explicitly modelling the stellar disk, without the need to invoke dark subhalos. Naturally, this does not mean that dark substructure is not present, just that it is not required.
\newline \indent The lens system MG0414+0534 is not fit for juxtaposition with VCC1692, as the central velocity dispersion, derived from the Einstein of an SIE \citep{Xuetal2015}, is approximately 286 km s$^{-1}$, while lens models favor moderate SIE ellipticity of $\approx 0.2$, indicating that the lensing galaxy is likely very massive and round \citep{Hewitt++92}. While the anomaly is small in magnitude, the proximity of the merging image pair makes it unlikely that a smooth potential provides an adequate description of the lens system \citep{Minezaki++09,Keeton:2005p591,Xuetal2015}. However, \citet{Xuetal2015} point out that other sources of anomaly besides dark substructure may be needed to explain the observed anomaly, citing mass structure along the line of sight as a plausible culprit. Deep imaging of this system would help rule out the possibility that baryons play a significant role, although the non-detection of baryon-induced $R_{\rm{fold}}$ anomalies in our sample suggests the dark matter is responsible. 

B0128+437 \citep{Phillips++00}, a small deflector with $R_{\rm{Ein}}=0.24"$, and low sersic index, consistent with a late-type morphology \citep{LAF10}. \cite{BiggsEtal04} show that lens models favor very elliptical SIE profiles, but fail to fit the observed image positions. While the small size of the lens and elongated nature of the deflector suggest baryons may contribute a non-negligible effect to the flux ratios, the presence of astrometric anomalies suggests non-baryonic substructure may also contribute.

The system B1608+656 \citep{Fas++96} is peculiar in that it consists of two merging galaxies \citep{Fas++02,Koo++03}, with the most massive one having a velocity dispersion of $260\pm15$ s$^{-1}$ \citep{Suy++10}. The B1608 system is complex enough that a  description in terms of simple anomalies is not appropriate and searches for dark matter substructure must take into account this complexity with a detailed model \citep{Suy++09,Suy++10}. 
Even more anomalous than anyone of our mock lenses is the system B1933+503 \citep{Syk++98}, a well known peculiar system with a late-type deflector that contains a prominent stellar disk \citep{Suy++12b}. \citet{K+D04} investigate whether higher order multipole terms in the lens potential can account for the observed anomaly, and conclude that such an explanation is unlikely, which seems to favor a dark substructure as a source of flux ratio perturbation. However, it is possible that the lensing properties of galaxies with very irregular morphology, such as a prominent edge on disk, may require a more different description than can be encapsulated by adding a few higher order multipole terms.
\begin{figure*}
\includegraphics[trim=0cm .5cm 0cm .4cm,clip,width=.7\textwidth]{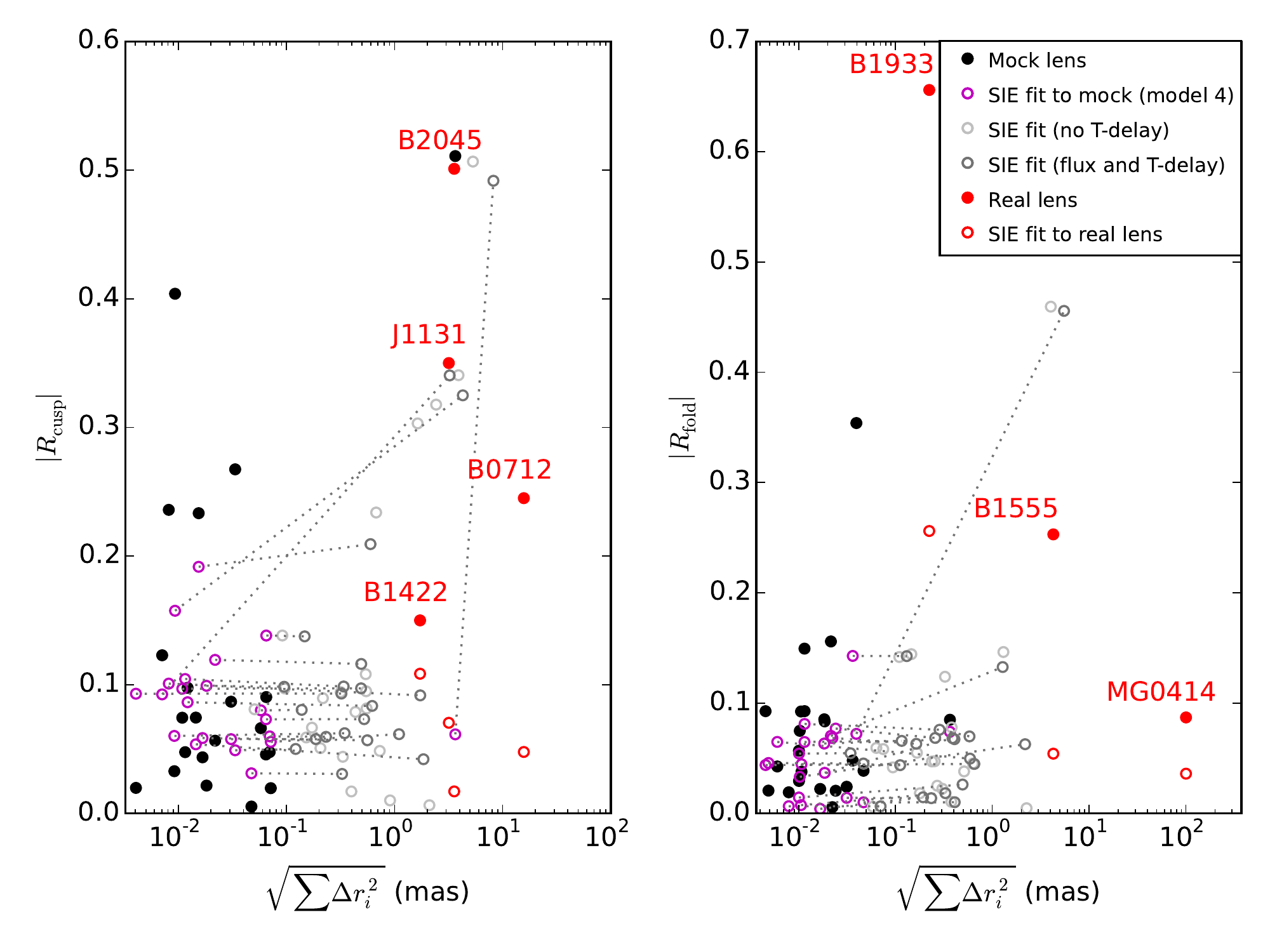}
\caption{\label{fig:poserr_vs_Rcuspfold} $R_{\rm{cusp}}$ and $R_{\rm{fold}}$ parameters for the \textit{Truth} model and best fit SIE \textit{Truth} model (black points and purple circles, respectively), together with the values observed in real lenses (red points), and SIE fits to real lenses (red circles). The real lenses and \textit{Truth} data points are distributed along the x-axis according to the astrometric error of their best fit SIE model summed in quadrature. Counterintuitively, as most of our mock lenses have R values larger than the best fit SIE, a positively sloped dashed line is actually an improved model for the data, although the R value increases. This flux ratio precision comes at a cost of larger astrometric and time-delay errors, irrespective of the exact uncertainty we place on the flux ratios and time delays. Further, without time delay information (light grey points), the code cannot distinguish between major and minor axis cusps, which further complicates the modeling process. In this plot, we impose flux ratio uncertainties of 10 \% (grey points). When fitting the real lens systems, we adopt lens data and observational uncertainties from \citealt{COSMOX} (MG0414, B2045, MG0414), \citealt{Sluse++06} (J1131), \citealt{Jackson98} (B0712), \citealt{Nie++14} (B1422), \citealt{Coh++01} (B1933), and \citealt{HsuehEtal16} (B1555). There is a clear separation between the real lenses and the
mock lenses, with real systems possessing both astrometric
and  flux ratio anomalies, and our set of mock lenses mostly
confined to  flux ratio anomalies 6 30\% and nearly perfect
astrometric precision.
}
\end{figure*}
\end{itemize}
The possibility of a large $R_{\rm{cusp}}$ or $R_{\rm{fold}}$ arising from the baryonic structure of the lens, especially in low mass ellipticals with features such as disks or boxy isophotes, behooves observers to investigate whether the lensing galaxy possesses baryonic mass distributions that require detailed modeling. Indeed, deep imaging of B1555 and B0712 shows that a disk is present, and can account for the apparent anomaly \citep{HsuehEtal16}. Of the lenses in our sample with the largest $R_{\rm{cusp}}$ values, some have stellar disks visible even in the rebinned images. Others do not have visible disks but are significantly elongated, even in the rebinned images. These findings suggest that in most cases an elliptical galaxy at $z=0.5$ can be imaged well enough by the HST for potential sources of baryonic anomaly to be identified and modeled, but care should be taken to account for the interplay between the external shear and stellar ellipticity, which could result in an off-axis cusp, as in VCC1692 and VCC1062 imaged in Figures \ref{fig:fluxratios} and \ref{fig:fluxratios2}.

\subsubsection{Characterizing a baryonic-lensing signal through modeling: astrometric and flux ratio anomalies}  

Among the properties of the baryonic mass of a deflector likely to give rise to flux ratio anomalies, stellar disks or other elogated structures, most often seen in low mass, low stellar velocity dispersion galaxies, constitute the majority of the anomalous systems in our sample. On the other hand, in round, high velocity dispersion systems such as NGC7626 where there is no obvious stellar disk or other large scale feature, anomalies could be induced by compact structures near the critical curve. Regardless of the origin, flux ratio anomalies from luminous matter may be difficult to identify solely by examining flux ratios.
\newline \indent In order to help distinguish a baryonic lensing signal in systems similar to NGC7626 from other sources of anomaly, we highlight a feature of our mock systems, seen even systems with significant flux ratio anomalies, that is not frequently observed in real lens systems. Our mock lenses are characterized by a conspicuous absence of astrometric anomalies, which can be present in real systems at the level of tens of mas for subhalos located near an image, in projection \citep{Chiba02,Che++07}. This suggests that a feature of perturbation by dark matter subhalos, that could be used to distinguish between baryonic and dark matter perturbations, is a flux ratio anomaly coupled to an astrometric anomaly, especially if the introduction of a dark substructure to the lens model simultaneously resolves both discrepancies.   
\newline \indent To compare the astrometric precision of the SIE model fit to our mock lenses with that of an SIE fit to real  lenses, we fit several of the real systems shown in Figure \ref{fig:Rcuspfold_vs_real} with an SIE plus external shear, varying the Einstein radius, ellipticty, shear, position angles, and deflector centroid. For the resulting best fit model we compute the flux ratio and astrometric anomalies. We omit systems such as J0911 and B1608, which require complicated modeling involving two galaxies within the Einstein radius. We repeat the fit for each of our mock systems omitting time delays and enforcing flux ratio constraints, to see if the inclusion or exclusion of either these data significantly impacts the results.
\begin{figure*}
{\includegraphics[trim=0cm 0cm 0cm
0cm,clip,width=.48\textwidth]{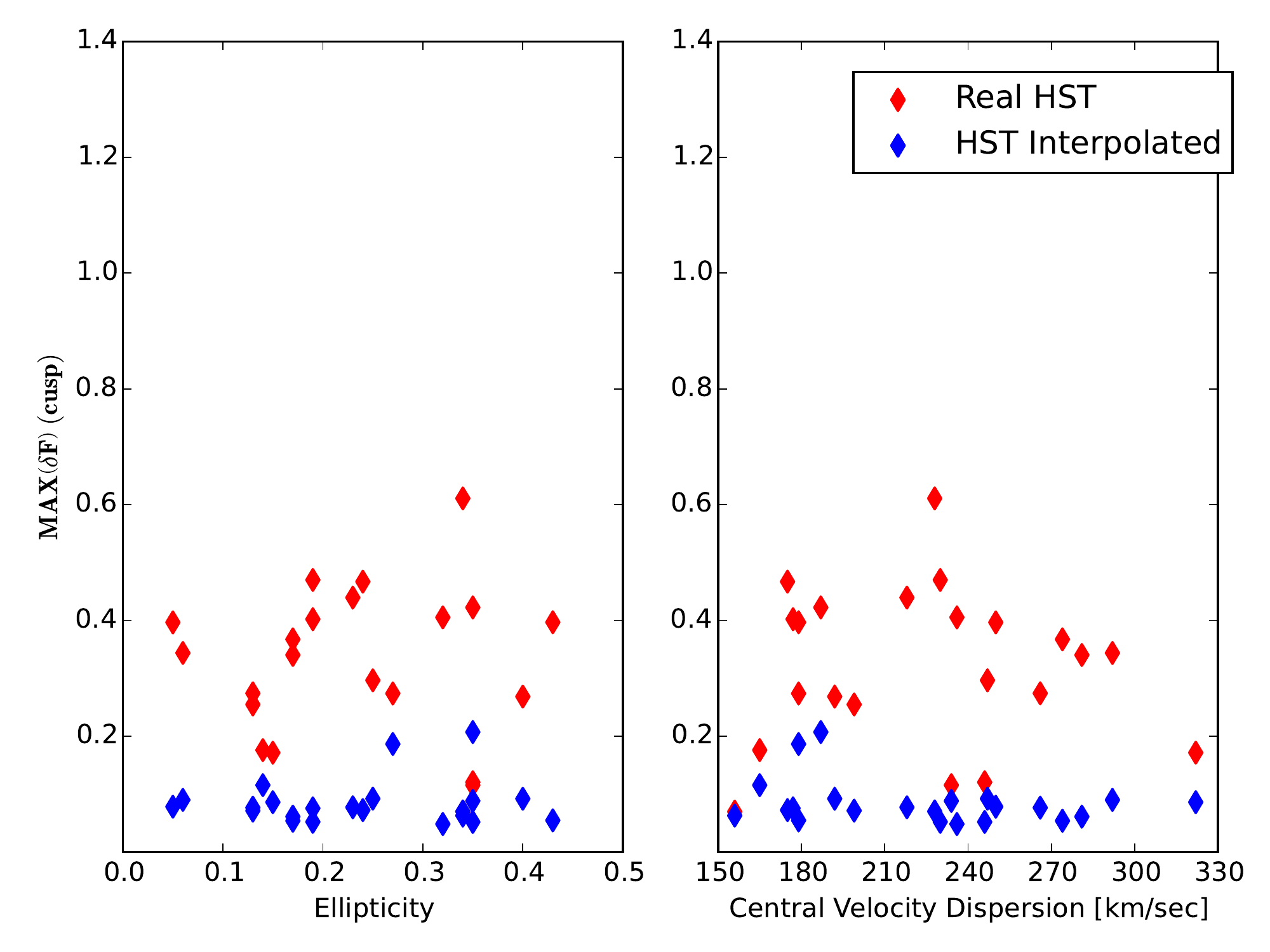}}
{\includegraphics[trim=0cm 0cm 0cm
0cm,clip,width=.48\textwidth]{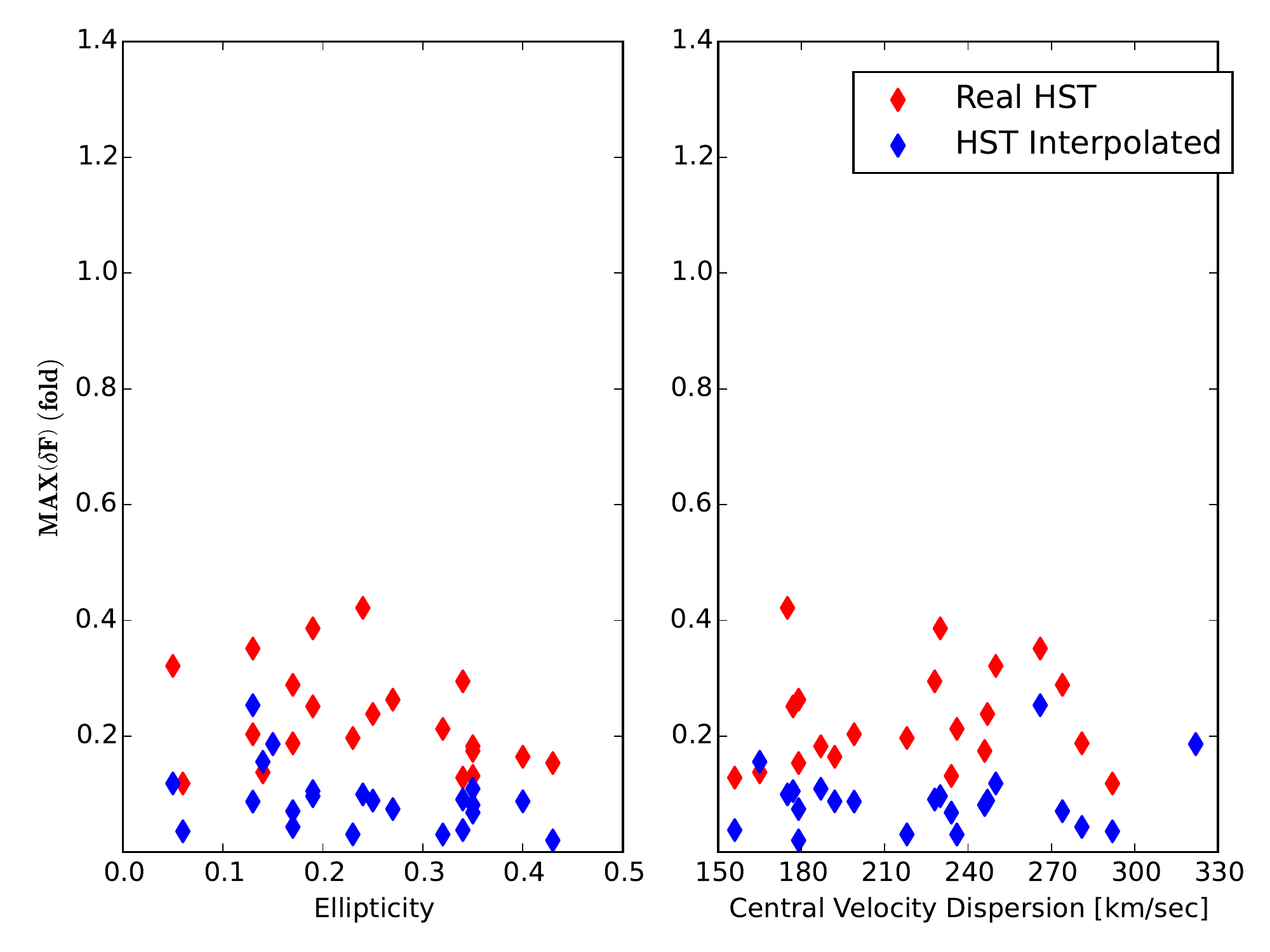}}
\caption{\label{fig:fluxratios_23}Largest flux ratio anomaly for Model 2 (\textit{Real HST}) and Model 3 (\textit{HST Interpolated}) in each lens as a function of ellipticity and central velocity dispersion for cusp configurations (left) and fold configurations (right).}
\end{figure*}
\begin{figure*}
{\includegraphics[trim=0cm 0cm 0cm .3cm,clip,width=8.5cm, height=6.85cm]{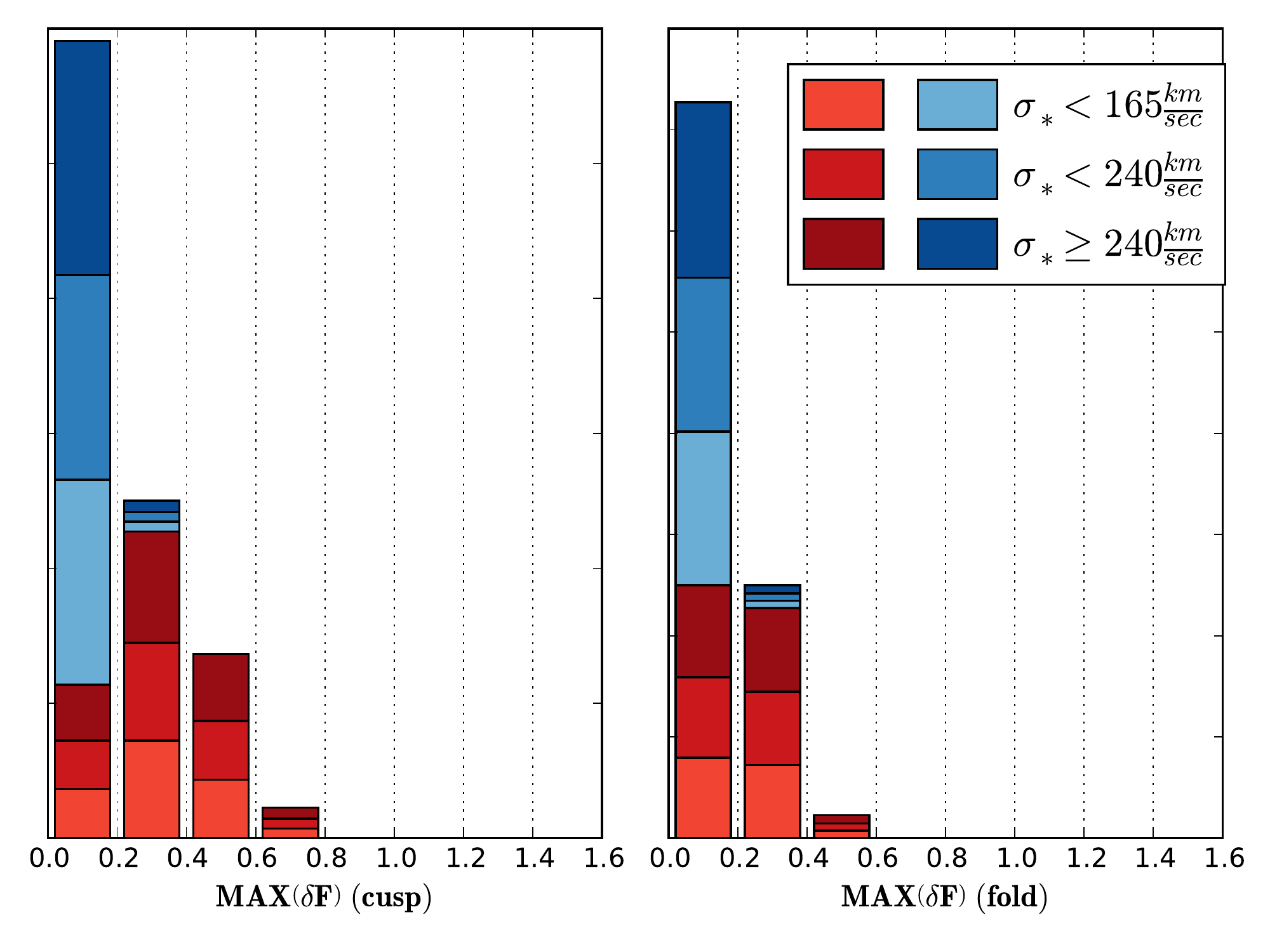}}
{\includegraphics[trim=0cm 0.2cm 0cm 0cm,clip,width=8.5cm, height=7cm]{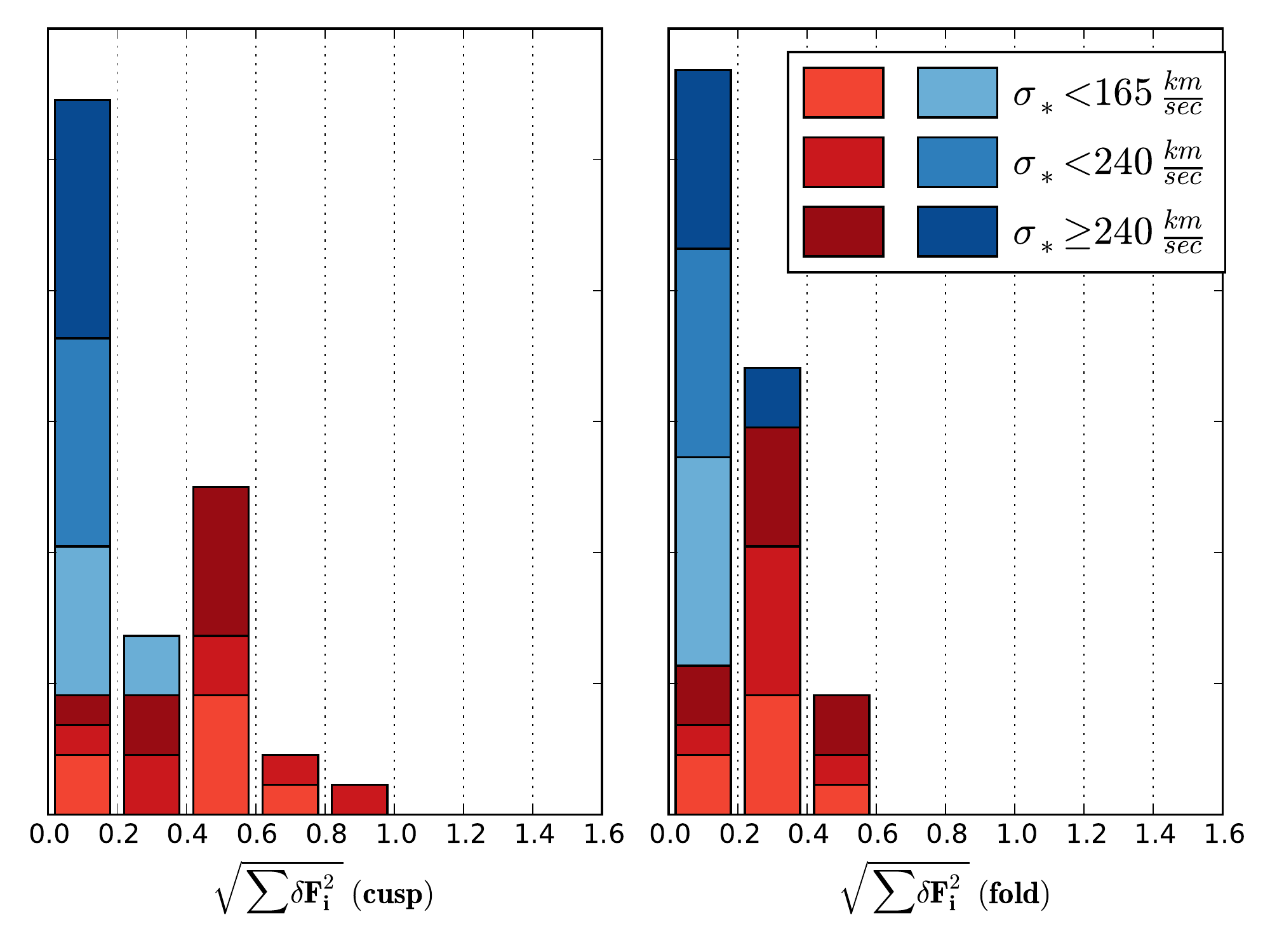}}
\caption{\label{fig:distribution_23}Distributions of the largest anomalies (left) and the three anomalies summed in quadrature (right) for each lens, color coded by the deflector's central velocity dispersion. The color scheme is the same as in Figure~\ref{fig:fluxratios_23}}
\end{figure*} 
\begin{figure*}
{\includegraphics[trim=0cm 0cm 0cm 0cm,clip,width=.48\textwidth]{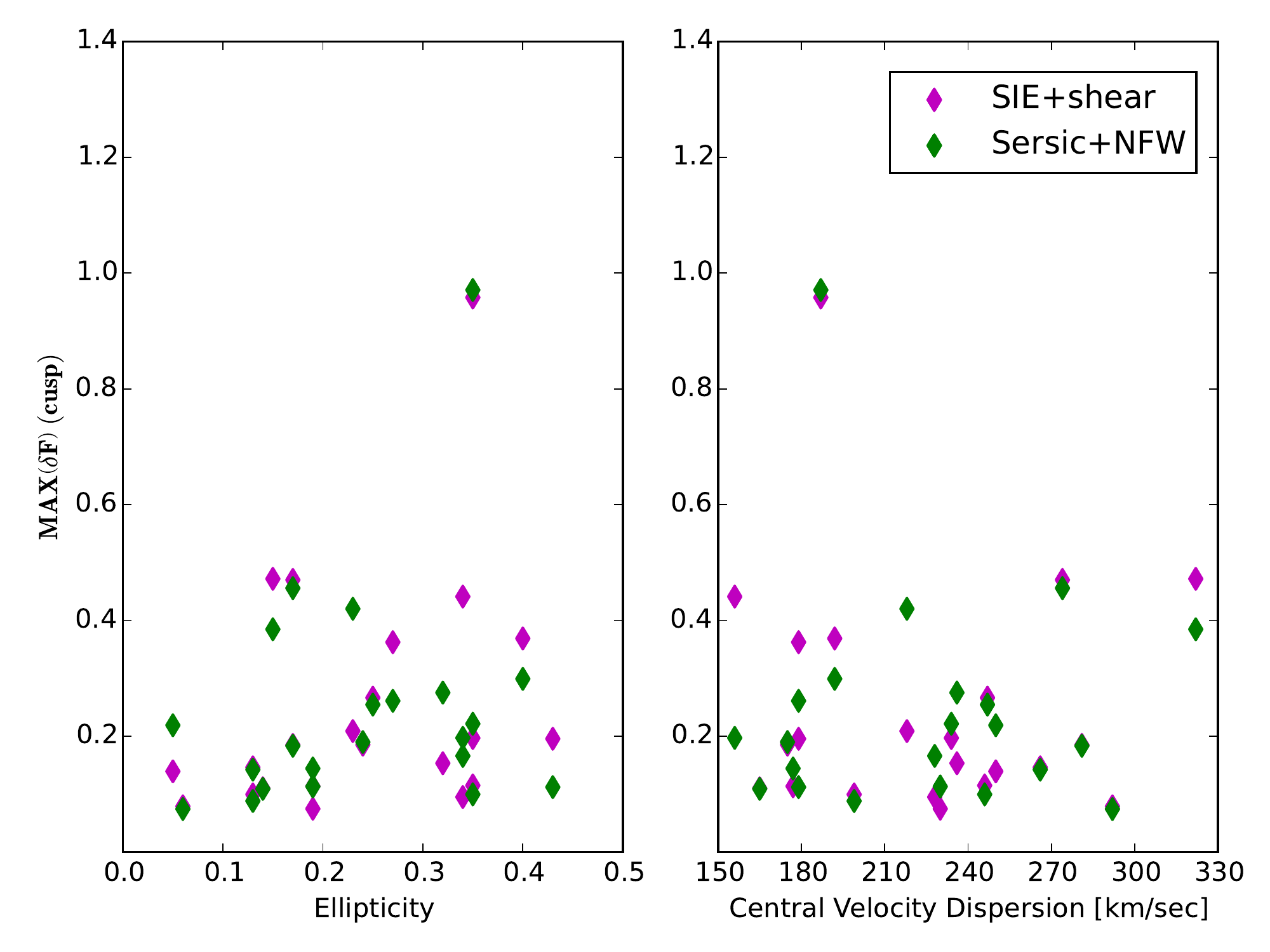}}
{\includegraphics[trim=0cm 0cm 0cm 0cm,clip,width=.48\textwidth]{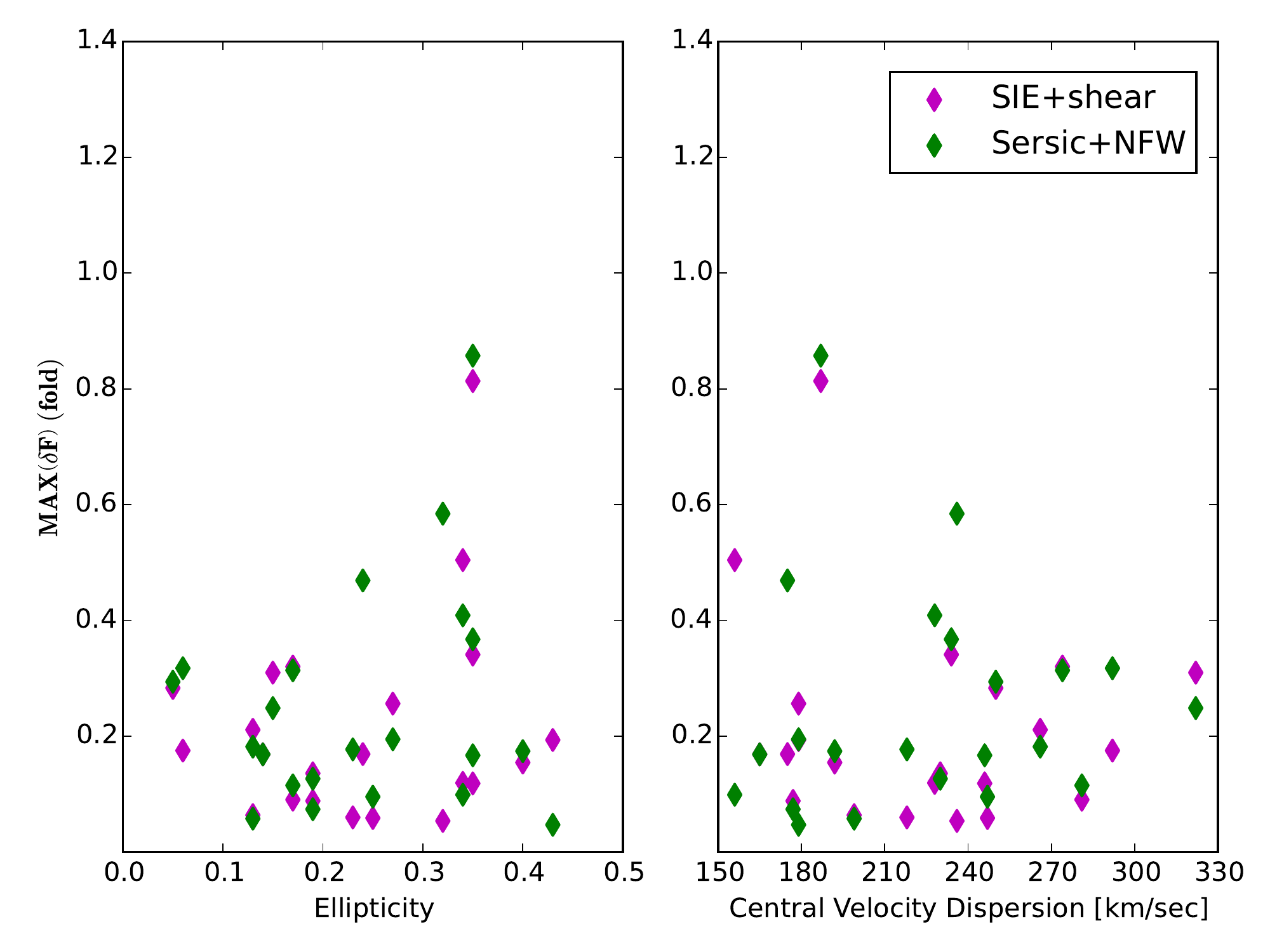}}
\caption{\label{fig:fluxratios_45}Largest flux ratio anomaly for Model 4 (purple) and Model 5 (green) in each lens as a function of ellipticity and central velocity dispersion for cusp configurations (left) and fold configurations (right).}
\end{figure*}
\begin{figure*}
{\includegraphics[trim=0cm 0cm 0cm .3cm,clip,width=8.5cm, height=6.85cm]{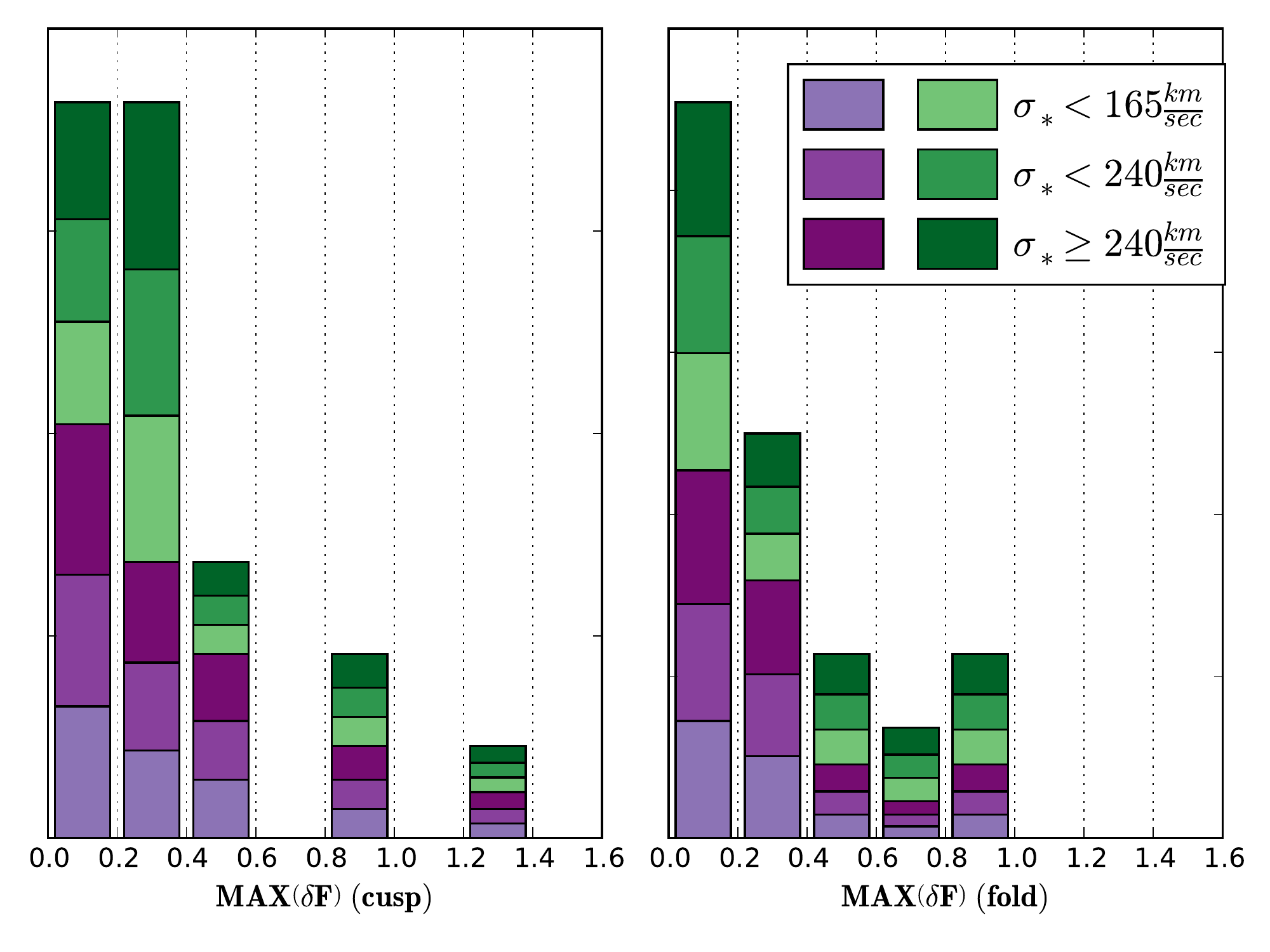}}
{\includegraphics[trim=0cm 0.2cm 0cm 0cm,clip,width=8.5cm, height=7cm]{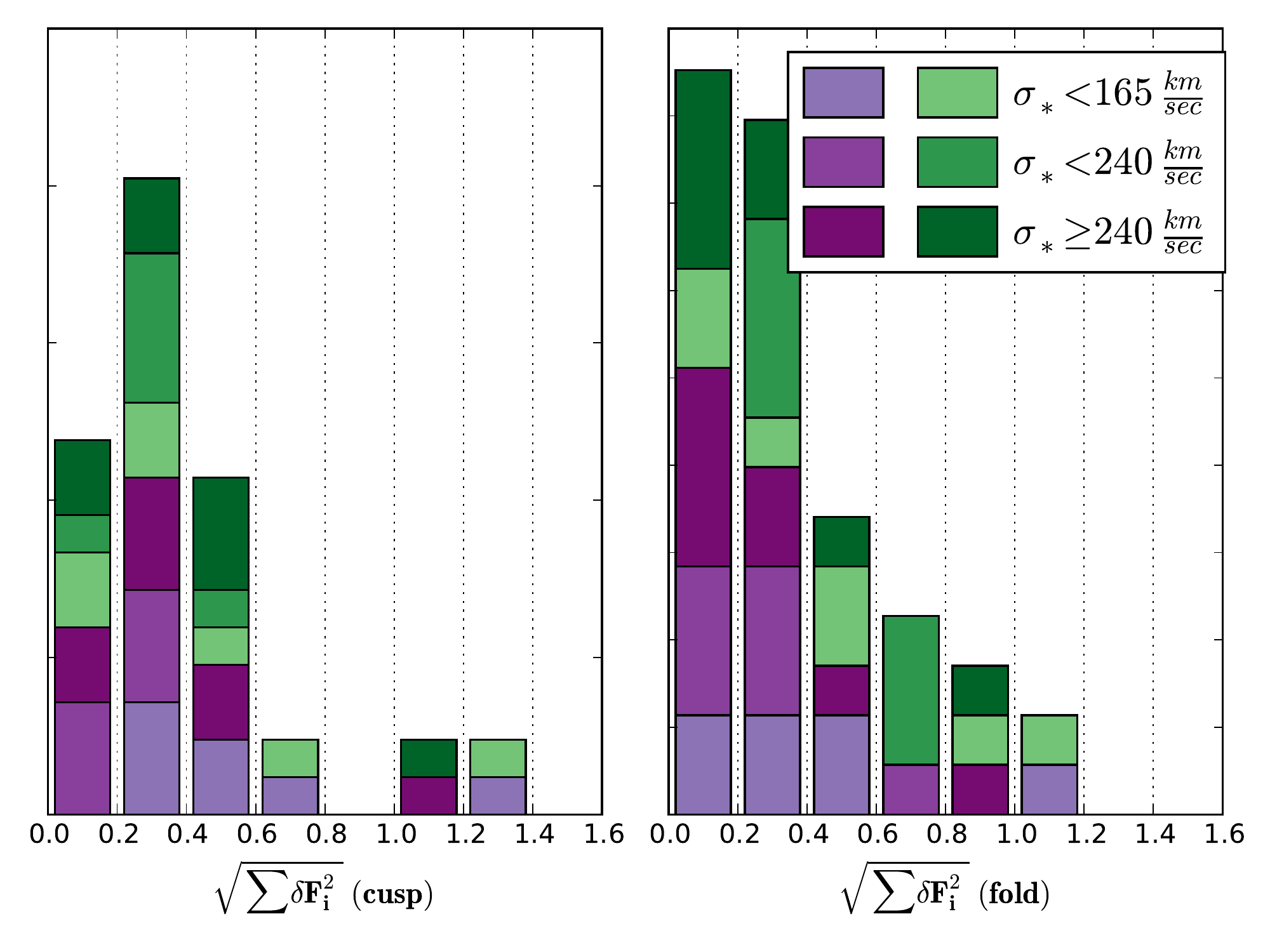}}
\caption{\label{fig:distributions_45}Distributions of the largest anomalies (left) and the three anomalies summed in quadrature (right) for each lens, color coded by the deflector's central velocity dispersion. The color scheme is the same as in Figure~\ref{fig:fluxratios_45}}
\end{figure*}
\begin{figure*}
{\includegraphics[trim=0cm 0cm 0cm 0cm,clip,width=.48\textwidth]{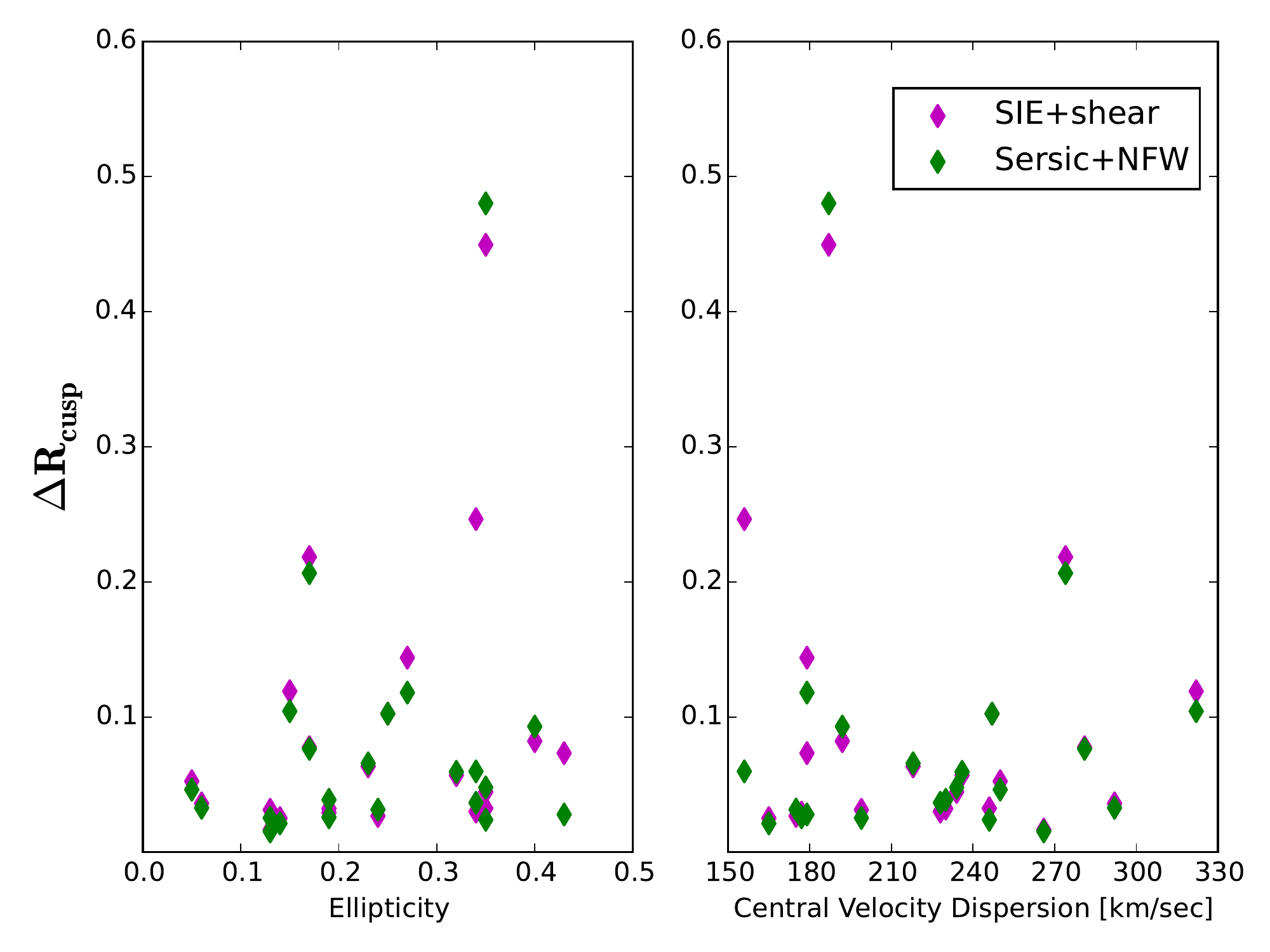}}
{\includegraphics[trim=0cm 0cm 0cm 0cm,clip,width=.48\textwidth]{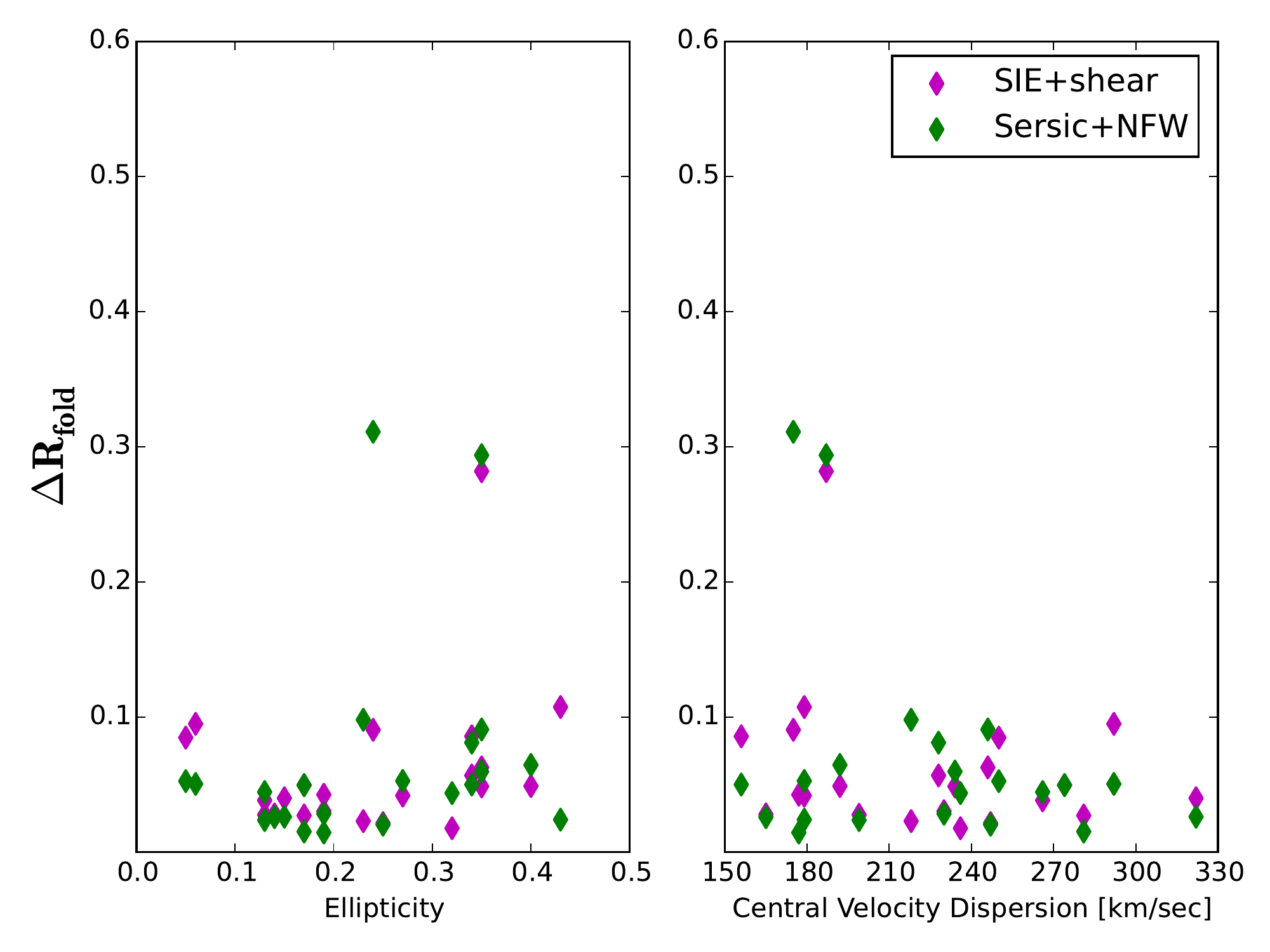}} 
\caption{\label{fig:Rcuspfold_plot}Differences in the R-cusp (left) and R-fold (right) statistics between the SIE and SNFW models, and the \textit{Truth} data.}
\end{figure*}
\newline \indent In Figure \ref{fig:poserr_vs_Rcuspfold}, we plot the $R_{\rm{cusp}}$ or $R_{\rm{fold}}$ values of our mock lenses and some real lens systems, along with a best fit SIE to each of the real systems and mocks as a function of the total astrometric error summed in quadruature. There is a clear separation between the real lenses and the mock lenses, with real systems possessing both astrometric and flux ratio anomalies, and our set of mock lenses mostly confined to flux ratio anomalies $\leq 30 \%$ and nearly perfect astrometric precision. The effect is more pronounced in cusp lenses, although both image configurations follow this general trend. Enforcing flux ratio contraints of 10 \% in the SIE fit to the mock lenses sometimes corrects the flux ratio anomaly at the expensive of astrometric precision, but the resulting points still populate a different region of parameter space than the real lens systems. Additionally, while SIE fits to the mock lenses systems come close to reproducing the observed $R_{\rm{cusp}}$ or $R_{\rm{fold}}$ value, the best fit model of a real lens system differs significantly.
\newline \indent Our analysis aims to characterize the properties of a deflector (low velocity dispersion, high ellipticity, etc.) that may increase the likelihood of observing a baryon-induced flux ratio anomaly. While the trend in Figure \ref{fig:poserr_vs_Rcuspfold} can be used to characterize a purely baryonic lensing signal by the absence of an astrometric anomaly, it should not be adopted as a criterion used to rule out lens systems as candidates for analysis of dark matter substructure, as our analysis does not address the question of whether a dark subhalo will necessarily result in simultaneous astrometric and flux ratio anomalies, and the relative magnitudes of these perturbations. Further, the interpretation of what constitutes an astrometric anomaly is model dependent, and depends on the precision the available data. Regardless of these nuances, given a large sample of lenses, observations of the morphological features of the lensing galaxy, together with an absence of astrometric anomalies in the presence of relatively small flux ratio anomalies, could be used to flag certain systems as more likely than others to exhibit lensing effects induced by luminous matter. This would necessitate additional observations of the lensing galaxy, and detailed modeling of its morphology.  
\subsection{$\bf{R_{\rm{\bf{cusp}}}}$, $\bf{R}_{\rm{\bf{fold}}}$, and flux ratios; Models 2-5.}
\subsubsection{Flux Ratios: Models 2 and 3 (\textit{Real HST} and \textit{HST Interpolated})}
The process of rebinning pixels introduces a significant source of flux ratio anomaly - relative to the \textit{Truth} model - compared to Model 3, as seen in Figures \ref{fig:fluxratios_23} and \ref{fig:distribution_23}. This suggests that in a real lens observed at redshift 0.5, directly using pixel values to infer properties of the lens baryonic mass distribution introduces flux ratio anomalies of about 40\% for cusp configurations, and about 20\% for fold configurations. There is no clear trend between ellipticity, stellar velocity dispersion, and flux ratio anomaly. Based on these findings, we conclude that the source of anomalies for the pixellated models are numerical and associated with computing lensing derivatives from coarsely sampled data. Thus, even in the presence of exquisite data, it is best to interpolate the pixellated data with smooth functions \citep[e.g. the multi-gaussian-expansion method of][]{Cap02} to describe the baryonic component correctly and attribute excess anomalies to dark subhalos. As the smooth model can be interpreted as the best empirical basis one could use to model a lens, given that there exists an average variation of $9.3\%$ and $10.6\%$ in flux ratios between the smooth model and the \textit{Truth} data, for fold and cusp lenses, respectively, we conclude that this is a typical perturbation induced by baryonic structure alone.
\subsubsection{Flux Ratios: Models 4 and 5 (SIE and SNFW)}The relative flux ratio anomalies for both the SIE and SNFW models display a clearer dependence on baryon ellipticity and the galaxy’s central velocity dispersion, with the largest flux ratio anomalies present in highly elongated and low velocity dispersion galaxies. However, there is considerable scatter in the trends, as many highly elliptical and low velocity dispersion targets do not posses significant baryonic-induced anomalies. The largest anomalies are present in cusp configurations. Many of the errors are on the order of about 10\%, comparable to the noise floor derived from the \textit{HST Interpolated} model, and as such should not be formally considered `anomalies'. There is no significant difference between the accuracy of flux ratios recovered by the SIE and SNFW models. 
\subsubsection{$R_{\rm{cusp}}$ and $R_{\rm{fold}}$: Models 4 and 5 (SIE and SNFW)}
The offsets of $R_{\rm{cusp}}$ and $R_{\rm{fold}}$ statistics, shown in Figure \ref{fig:Rcuspfold_plot}, between the \textit{Truth} data and the SIE and SNFW fits reveal a clear trend in anomalies for the $R_{\rm{cusp}}$ statistic, with highly elliptical and low velocity dispersion targets possessing the largest anomalies. However, this relationship is not deterministic, as some low velocity dispersion or significantly elongated deflectors do not display anomalies. The largest offsets in the $R_{\rm{fold}}$ statistic, however, do not appear to be correlated with ellipticity or velocity dispersion, which suggests that this statistic is less sensitive to the baryonic structure of the lensing galaxy. On the other hand, our results demonstrate that the $R_{\rm{cusp}}$ statistic is recovered almost exactly by smooth lens models in galaxies with low ellipticity. This suggests that image magnifications in the cusp configuration are more strong affected by baryon ellipticity due to the proximity of stellar mass to the lensed images in elongated deflectors. However, very massive galaxies with high velocity dispersions will result in images far enough away from the majority of the stellar mass, presumably leaving their magnifications unaffected.
\section{Summary and conclusions}\label{sect:conclusions}
Motivated by the growing sizes of known lensed quasars samples and the
interest in the lens systems as a probe of dark matter substructure,
we have carried out a systematic study of ``baryonic anomalies''. We
have used a sample of high resolution images of nearby early-type
galaxies as a starting point to create mock gravitational lens
systems, and then we have studied how well the arrival time,
positions, and fluxes of the lensed images are reproduced by lens
models based on the observed surface brightness distribution and on
commonly used functional forms. Our findings can be summarized as
follows:

\begin{itemize}

\item Arrival times and image positions are virtually unaffected  by baryonic substructure and can be recovered within the uncertainties by both empirical lens models and simply parametrized models. We conclude that astrometric anomalies are unlikely to arise from baryonic lensing effects, and can therefore be used to distinguish between the lensing signal of luminous matter and a dark subhalo, which \citet{Che++07} show can induce astrometric anomalies of order 10 mas. While the absence of astrometric anomalies is a common feature among our mock lenses, the non-detection of astrometric anomalies does not mean that dark substructure is not present. Rather, we claim that in a large sample of lenses, highly elongated deflectors with low stellar velocity dispersion and no astrometric anomalies are the most likely lens systems to posses lensing signals from baryonic structure, and warrant further study and detailed modeling.
\item The baryonic structure of a lensing galaxy can introduce a source of flux ratio anomaly in strong lensing that is more pronounced in highly elongated galaxies, and galaxies with low central velocity dispersion. We interpret this as evidence that the baryonic anomalies are dominated by large-scale features such as embedded disks, or isophotal twisting. Our analysis suggests that small-scale features like globular clusters or compact dwarf satellite galaxies contribute substantially to the anomaly in NGC7626, although the non-detection of anomalies from small scale structure in the majority of our mock systems suggests this would be a sub-dominant effect in a large sample of lens systems.
\item As our sample of mock lenses contains a disproportionately large number of small deflectors with low velocity dispersions, our analysis likely over-estimates the frequency and magnitude of flux ratio anomalies induced by the stellar mass of a deflector. In light of this, the fact that only 4 out of the 22 mock lenses we study display anomalies indicates that baryon induced flux ratio anomalies are a rare occurrence. Further, the magnitude of these anomalies are significantly smaller than those observed in real systems like B2045 and J1131. Together, these facts suggest that a baryonic lensing signal alone would not dominate the signal from dark substructure, although in some systems the perturbation can be non-negligible.
\item By comparing the \textit{Truth} data with the \textit{Real HST} and \textit{HST Interpolated} models, we show that the process of rebinning pixels introduces a large source of error in lens modeling, while convolution with a smoothing kernel produces, on average, flux ratio anomalies of 9.3\% and 10.6\% for cusp and fold configurations, respectively. Given that the smooth model represents the best smooth profile that may be devised to model a lens from observational data, we conclude that a relative flux ratio anomaly of roughly 10\% constitutes an effective minimum flux ratio precision in strong lensing for deflectors at typical redshift $0.5$, given the resolution of current optical and near infrared data. This anomaly arises from galactic-scale structures as well as small scale luminous satellites which are likely associated with dark subhalos. 
\item The most morphologically complex realistic lenses may have anomalous $R_{\rm{cusp}}$  and 
$R_{\rm{fold}}$ statistics between 0 and 0.5 purely due to
luminous structure and substructure. The broad range of values makes it
imperative to study in detail the distribution of luminous matter
while interpreting these systems. The case of B2045 \citep{Fas++99b},
which posses an
 anomalous $R_{\rm{cusp}}$ value nearly identical to
the mock lens
 VCC1692, is a good illustration. Whereas the mock lens
VCC1692 is
 highly elongated ($\epsilon=0.35$) and has a low stellar velocity
dispersion, the
 deflector in B2045 is nearly round and has a much
higher velocity
 dispersion. Thus, even though the two systems have
approximately the same R$_{\rm{cusp}}$ anomaly, the physical origin is
likely different: In the case of VCC1692 it is explained by the
presence of a disk, whereas this is not a viable explanation for
B2045, supporting the hypothesis that a dark subhalo lurks near a
lensed image. In addition, B2045 has image positions that cannot be
recovered by a smooth potential, while the positions of VCC1692 can be
recovered with a $\chi^2$ value of 1.2 (see Table
\ref{table:fitting_list}), illustrating how astrometric anomalies can
be used to identify a lens likely susceptible to baryon-induced flux ratio anomalies, as opposed to dark substructure or line-of-sight induced anomalies.
\item Real anomalous lenses and our mocks generally separate well in the space of flux-ratio and astrometric anomalies, indicating that a joint analysis of fluxes and positions is essential to disentangle galaxy-scale luminous structure from true dark matter substructure signatures.
\end{itemize}
Our results are consistent with and generalize earlier work by
\citet{M+S98}, \citet{Chi02},  and \citet{Kawano++04}, who showed with
simple models that small scale baryonic substructure such as globular
clusters and m=3 multipole terms constitute a subdominant source of
flux ratio anomalies. Similarly, our conclusion that large scale
features such as disks and isophotal twisting are a non-neglibile
source of uncertainties is consistent with earlier theoretical work by
\citet{MHB03}, and recent work \citet{HsuehEtal16}, in which flux
ratio anomalies of $\approx 40\%$ (with respect to a SIE model) were
observed in a lens with a pronounced stellar disk. The good quality of
the \citet{HsuehEtal16} data made it possible to observe and model the
disk, correcting the anomaly, and illustrates the importance of deep
imaging of the lens galaxy. However, when deep imaging is not
available, or when the lensed images are so bright to completely
dominate the lens galaxy, this potential source of error can
controlled and mitigated by restricting samples to deflector galaxies
with high central velocity dispersions, which tend to have low
ellipticity and be true elliptical galaxies with no disk
\citep{Mor++07,Cap++11}.

Baryon induced anomalies are enhanced in low mass systems due to both kinematic and morphological features, and also due to the lensing geometry. Low mass systems will tend to have smaller Einstein radii, which will result in images located closer to the center of the lens, where baryonic mass dominates. As such, the flux ratios between images in these lenses will be more strongly influenced by stellar mass.
 
In conclusion, in order to minimize the impact of stellar mass on flux ratios, we recommend that one restrict lensing studies to the most massive galaxies with large Einstein radii and low ellipticity, and allow for a residual noise floor to absorb both perturbations by undetected structure in the lensing galaxy, and the intrinsic uncertainties introduced by modeling lenses with smooth potentials. If possible, one should also obtain deep and high resolution images of the deflector to look for irregular morphological features, while simultaneously modeling both flux ratio and astrometric data. Furthermore, one must keep in mind the fundamental distinction between purely baryonic anomaly-inducing components, like a stellar disk or a globular cluster, and compact satellite galaxies. Whereas the former class of objects is purely noise from the point of view of dark matter studies, the latter is typically associated with the elusive subhalos.
We leave to future work the analysis of how the detailed morphology of real deflectors affects the inference of the properties a population of dark subhalos. 

\section*{Acknowledgments}

TT, DG, and AA acknowledge support from NSF grant AST-1450141
``Collaborative Research: Accurate cosmology with strong gravitational
lens time delays''. TT gratefully acknowledges support by the Packard
Foundation through a Packard Research Fellowship. CRK acknowledges support from grant HST-AR-14305.002-A, which was provided by NASA through a grant from the Space Telescope Science Institute, which is operated by the Association of Universities for Research in Astronomy, Incorporated, under NASA contract NAS5-26555. We thank Chris
Fassnacht, Phil Marshall and Paul Schechter, for useful comments and
suggestions on this work. We acknowledge the use of the  HyperLeda database (http://leda.univ-lyon1.fr).

\bibliographystyle{mnras}
\bibliography{references}
%\begin{thebibliography}{}
%\bibitem[Bovy et al.(2011)]{bov11b} Bovy, J., Hennawi, J.~F., Hogg, D.~W., et al.\ 2011, \apj, 729, 141
%\end{thebibliography}
%\clearpage

\begin{table*}
\centering
\caption[width=\textwidth]{Columns 1 and 2 list galaxy name and redshift. The following columns list the stellar velocity dispersion, half light radius, stellar ellipticity, position angle, S{\'e}rsic index, external shear, and external shear position angle. Asterisks denote quantities obtained by fitting with {\tt {galfit} }, while the rest are obtained from the literature. The references are as follows: 1) HyperLeda online catalog \cite{Makarov++2014} 2) \cite{Fer++06b} 3) \citep{Ma++2014}}
\resizebox{\textwidth}{!}{%
\begin{tabular}{||c c c c c c c c c c c c||}
\hline
Name & Redshift & $\sigma_*$ & $R_{\rm{1/2}}$ & $\epsilon$ & $\theta_{\epsilon}$ & $n$ & $\gamma$ & $\theta_{\gamma}$ & References\\
 &  & km/sec & arcsec &  & degrees & & & degrees & \\
\hline
VCC1664 & 0.0038 & 156 & 15.8 & 0.34 & 46.8 & 3.982 & 0.08 & 40 & 1, 2\\

VCC1297 & 0.0038 & 165 & 2.33 & 0.20 & -32.1 & 2.732 & 0.08 & 25 & 1, 2\\

VCC798 & 0.0038 & 175 & 170.8 & 0.24 & 30.2 & 6.813 & 0.05 & 55 & 1, 2\\

VCC2092 & 0.0038 & 177 & 24.13 & 0.19 & 71.2 & 4.279 & 0.05 & 15 & 1, 2\\
 
VCC1231 & 0.0038 & 179 & 16.89 & 0.43 & -86.3 & 2.955 & 0.08 & 35 & 1, 2\\

VCC1062 & 0.0038 & 179 & 16.92 & 0.27 & 86.8 & 3.291 & 0.05 & 25 & 1, 2\\

VCC1692 & 0.0038 & 187 & 9.5 & 0.35 & -21.7 & 2.458 & 0.05 & 39 & 1, 2\\

VCC2000 & 0.0038 & 192 & 10.45 & 0.40 & -84.9 & 3.977 & 0.05 & 75 & 1, 2\\

VCC355 & 0.0038 & 199 & 9.78 & 0.13 & -9.4 & 3.725 & 0.05 & 38 & 1, 2\\

NGC4872 & 0.0231 & 218 & 28.25* & 0.23* & 104.0* & 6.190* & 0.05 & 70 & 1\\

VCC1903 & 0.0038 & 228 & 106.8 & 0.34 & -16.8 & 6.852 & 0.05 & 17 & 1, 2\\

VCC881 & 0.0038 & 230 & 411.84 & 0.19 & -57.2 & 7.016 & 0.05 & 85 & 1, 2\\

IC4051 & 0.0231 & 234 & 21.01 & 0.35 & 97.2 & 3.210* & 0.05 & 71 & 1, 3\\

NGC5322 & 0.008 & 236 & 26.6 & 0.32* & 97* & 3.690* & 0.05 & 57 & 3\\

NGC1132 & 0.0231 & 246 & 16.1 & 0.35* & 146.0* & 2.15* & 0.05 & 135 & 3\\

VCC731 & 0.0050 & 247 & 115.4 & 0.25 & 43.9 & 5.871 & 0.05 & 70 & 1, 2\\

VCC1632 & 0.0038 & 250 & 85.12 & 0.05 & -60.7 & 7.088 & 0.05 & 55 & 1, 2\\\

NGC4874 & 0.0231 & 266 & 23.8 & 0.13 & 49.3 & 1.860* & 0.05 & 45 & 1, 3\\

NGC7626 & 0.0130 & 274 & 20.1 & 0.17 & 20.5 & 5.240* & 0.05 & 121 & 1, 3\\

NGC5557 & 0.0130 & 281 & 16.2 & 0.17* & 97.0* & 3.910* & 0.05 & 23 & 3\\

NGC1272 & 0.0180 & 292 & 20.7 & 0.06* & 76.0* & 2.440* & 0.05 & 23 & 3\\

NGC6482 & 0.0130 & 322 & 10.1 & 0.15 & 68.3 & 2.870* & 0.05 & 68 & 1, 3\\
\hline

\end{tabular}}
\label{table:gal_list}

\end{table*} 

\clearpage
\begin{table*}
\centering
\caption{Result of the SIE and SNFW model fits to the \textit{Truth} data from each lens. From left to right, we display the galaxy name, lens configuration, reduced $\chi^{2}$ for the fit with a SIE, Einstein radius of the SIE, ellipticity, position angle, shear, and shear angle, $\chi^2$ of the SNFW model fit, the normalization of the S{\'e}rsic profile, ellipticty and position angle of the S{\'e}rsic profile, S{\'e}rsic index, NFW halo normalization $\kappa_s$, external shear, external shear angle, and finally the NFW scale radius. A hyphen indicates that a parameter was held fixed to a value either obtain from the literature or fit with {\tt {galfit} } (see Table 1). The reduced $\chi^2$ are very small because we do not add measurement noise to our data.}
\resizebox{\textwidth}{!}{%
\begin{tabular}{||c c c c c c c c c c c c c c c c c c c c c||}
\hline
Name & type & $\chi^2_{SIE}$ & $\chi^2_{SIE}$ & $R_{\rm{Ein}}$ & $\epsilon$ & $\theta_{\epsilon}$ & $\gamma$ & $\theta_{\gamma}$ & $\chi^2_{SNFW}$ & $\chi^2_{SNFW}$ & $N$ & $\epsilon$ & $\theta_{\epsilon}$ & $R_{\rm{1/2}}$ & $n$ & $\kappa_{s}$ & $\gamma$ & $\theta_{\gamma}$ & $R_{s}$\\
 &  & (position) & (time delay) & arcsec &  & degrees &  & degrees & (position) & (time delay) & $\Sigma/\Sigma_{\rm{crit}}$ &  & degrees & arcsec & & $\Sigma/\Sigma_{\rm{crit}}$ & & degrees & arcsec\\
\hline
VCC1664 & CUSP & 0.000 & 0.003 & 0.39 & 0.32 & -40.4 & 0.02 & 6.3 & 0.000 & 0.000 & 6.66 & 0.70 & -42.3 & - & 1.220 & 0.303 & 0.09 & 41.1 & 1.020\\
  & FOLD & 0.000 & 0.019 & 0.39 & 0.30 & -40.7 & 0.03 & 19.0 & 0.000 & 0.001 & 126.48 & 0.63 & -42.5 & - & 2.808 & 0.277 & 0.08 & 40.9 & 1.010\\
VCC1297 & CUSP & 0.000 & 0.000 & 0.44 & 0.09 & 27.0 & 0.05 & 21.8 & 0.141 & 0.000 & 193.42 & - & - & - & - & 2.634 & 0.08 & 23.8 & 0.150\\
  & FOLD & 0.000 & 0.000 & 0.44 & 0.09 & 20.6 & 0.05 & 26.6 & 0.291 & 0.000 & 24.93 & - & - & - & - & 2.804 & 0.08 & 24.2 & 0.140\\
VCC798 & CUSP & 0.000 & 0.000 & 0.49 & 0.15 & 34.4 & 0.03 & -26.9 & 0.000 & 0.000 & 145.92 & 0.18 & 36.6 & - & 4.382 & 0.041 & 0.03 & -33.3 & 10.940\\
  & FOLD & 0.000 & 0.003 & 0.49 & 0.18 & 26.6 & 0.04 & -16.5 & 0.002 & 0.053 & 1.82 & 0.44 & 37.5 & - & 1.788 & 0.075 & 0.05 & 30.9 & 11.250\\
VCC2092 & CUSP & 0.000 & 0.000 & 0.50 & 0.09 & -2.7 & 0.04 & 29.6 & 0.000 & 0.000 & 4.07 & 0.16 & -1.8 & - & 1.361 & 0.213 & 0.05 & 22.0 & 1.550\\
  & FOLD & 0.000 & 0.002 & 0.50 & 0.09 & 1.3 & 0.04 & 26.9 & 0.000 & 0.000 & 134.53 & 0.28 & 7.7 & - & 3.374 & 0.252 & 0.05 & 19.6 & 1.550\\
VCC1231 & CUSP & 0.000 & 0.001 & 0.53 & 0.18 & 8.1 & 0.11 & 28.1 & 0.004 & 0.053 & 175.03 & - & - & - & - & 0.304 & 0.09 & 36.7 & 1.110\\
  & FOLD & 0.000 & 0.003 & 0.52 & 0.22 & 6.7 & 0.12 & 24.6 & 0.002 & 0.123 & 216.16 & - & - & - & - & 0.267 & 0.09 & 34.8 & 1.120\\
VCC1062 & CUSP & 0.000 & 0.002 & 0.51 & 0.17 & -7.8 & 0.06 & 15.2 & 0.000 & 0.001 & 79.09 & 0.41 & -11.3 & 0.50 & - & 0.181 & 0.04 & 23.6 & 2.520\\
  & FOLD & 0.000 & 0.016 & 0.50 & 0.27 & -5.7 & 0.11 & 11.4 & 0.018 & 0.021 & 61.79 & 0.25 & 21.6 & 1.18 & - & 0.009 & 0.11 & 4.8 & 5.930\\
VCC1692 & CUSP & 0.883 & 0.761 & 0.54 & 0.45 & 8.7 & 0.18 & 19.8 & 1.239 & 0.050 & 37.73 & 0.37 & 17.3 & 0.68 & - & 0.001 & 0.13 & 31.6 & 3.410\\
  & FOLD & 0.001 & 0.017 & 0.57 & 0.13 & 2.9 & 0.02 & -42.1 & 0.302 & 0.505 & 174.09 & - & - & - & - & 0.636 & 0.04 & 23.9 & 0.530\\
VCC2000 & CUSP & 0.000 & 0.000 & 0.60 & 0.08 & -19.2 & 0.03 & -6.2 & 0.725 & 0.819 & 107.05 & - & - & - & - & 0.925 & 0.05 & 78.3 & 0.520\\
  & FOLD & 0.000 & 0.002 & 0.60 & 0.10 & 6.1 & 0.04 & -25.3 & 0.000 & 0.004 & 54.16 & 0.18 & -34.9 & - & 1.422 & 0.170 & 0.07 & -15.1 & 0.670\\
VCC355 & CUSP & 0.000 & 0.000 & 0.65 & 0.01 & -12.1 & 0.05 & 37.7 & 0.443 & 0.008 & 106.28 & - & - & - & - & 0.903 & 0.05 & 35.4 & 0.570\\
  & FOLD & 0.000 & 0.000 & 0.65 & 0.02 & -12.4 & 0.05 & 38.4 & 1.011 & 0.008 & 42.98 & - & - & - & - & 0.887 & 0.05 & 35.0 & 0.590\\
NGC4872 & CUSP & 0.000 & 0.003 & 0.77 & 0.24 & 8.4 & 0.04 & 36.8 & 0.000 & 0.010 & 91.82 & 0.58 & 15.8 & 0.20 & - & 0.355 & 0.06 & -18.9 & 1.220\\
  & FOLD & 0.000 & 0.004 & 0.76 & 0.25 & 9.3 & 0.05 & 35.5 & 0.000 & 0.044 & 24.29 & 0.42 & 4.0 & - & 3.229 & 0.101 & 0.03 & 40.4 & 10.560\\
VCC1903 & CUSP & 0.000 & 0.000 & 0.83 & 0.26 & -17.3 & 0.05 & -44.9 & 0.001 & 0.007 & 7.26 & 0.27 & 40.1 & - & 1.976 & 0.060 & 0.04 & -16.7 & 6.870\\
  & FOLD & 0.000 & 0.017 & 0.83 & 0.26 & -18.8 & 0.06 & 43.6 & 0.001 & 0.061 & 3.41 & 0.22 & -15.8 & - & 1.346 & 0.006 & 0.05 & -44.2 & 6.890\\
VCC881 & CUSP & 0.000 & 0.001 & 0.84 & 0.26 & 35.1 & 0.07 & -32.1 & 0.001 & 0.001 & 25.80 & 0.24 & 30.3 & - & 3.404 & 0.010 & 0.05 & -34.0 & 26.820\\
  & FOLD & 0.000 & 0.001 & 0.84 & 0.24 & 33.0 & 0.06 & -31.3 & 0.000 & 0.004 & 14.26 & 0.22 & 28.9 & - & 2.996 & 0.001 & 0.05 & -38.1 & 26.450\\
IC4051 & CUSP & 0.000 & 0.024 & 0.88 & 0.26 & 15.8 & 0.06 & 37.3 & 0.001 & 0.036 & 26.95 & 0.23 & 14.8 & - & 3.359 & 0.000 & 0.06 & 37.4 & 26.260\\
  & FOLD & 0.000 & 0.003 & 0.88 & 0.27 & 18.1 & 0.07 & 38.4 & 0.000 & 0.000 & 79.90 & 0.26 & 9.8 & 1.30 & - & 0.048 & 0.06 & -6.6 & 6.570\\
NGC5322 & CUSP & 0.000 & 0.001 & 0.89 & 0.20 & 6.0 & 0.06 & 35.7 & 0.054 & 0.221 & 329.38 & - & - & - & - & 0.139 & 0.06 & 50.3 & 3.790\\
  & FOLD & 0.000 & 0.000 & 0.89 & 0.18 & 3.6 & 0.05 & 38.1 & 0.000 & 0.001 & 38.14 & 0.30 & -0.4 & - & 2.692 & 0.181 & 0.05 & -34.0 & 3.590\\
NGC1132 & CUSP & 0.000 & 0.010 & 0.97 & 0.23 & -37.3 & 0.02 & 22.3 & 0.494 & 0.311 & 8.72 & - & - & - & - & 0.150 & 0.03 & -52.4 & 5.090\\
  & FOLD & 0.000 & 0.019 & 0.97 & 0.23 & -37.0 & 0.02 & 20.2 & 0.021 & 0.413 & 6.85 & - & - & - & - & 4.038 & 0.06 & -46.8 & 0.180\\
VCC731 & CUSP & 0.000 & 0.002 & 1.00 & 0.20 & 36.2 & 0.05 & 2.4 & 0.000 & 0.000 & 27.85 & 0.19 & 40.2 & - & 3.384 & 0.014 & 0.04 & 1.6 & 26.400\\
  & FOLD & 0.000 & 0.011 & 1.00 & 0.17 & 42.0 & 0.04 & -1.2 & 0.000 & 0.016 & 4.72 & 0.16 & -44.2 & - & 1.662 & 0.001 & 0.03 & 1.1 & 9.880\\
VCC1632 & CUSP & 0.000 & 0.000 & 1.01 & 0.07 & -32.5 & 0.07 & -32.6 & 0.056 & 1.625 & 943.00 & - & - & - & - & 0.314 & 0.03 & 57.3 & 3.730\\
  & FOLD & 0.000 & 0.002 & 1.01 & 0.07 & -21.3 & 0.07 & -29.3 & 0.000 & 0.006 & 18.04 & 0.26 & -42.7 & - & 3.116 & 0.227 & 0.03 & -31.0 & 5.260\\
NGC4874 & CUSP & 0.001 & 0.020 & 1.18 & 0.02 & 34.6 & 0.06 & 43.9 & 0.103 & 0.012 & 3.32 & - & - & - & - & 0.130 & 0.03 & 42.9 & 9.010\\
  & FOLD & 0.000 & 0.006 & 1.18 & 0.03 & 26.4 & 0.06 & 42.2 & 0.337 & 0.034 & 3.40 & - & - & - & - & 0.127 & 0.03 & 42.2 & 9.070\\
NGC7626 & CUSP & 0.004 & 0.006 & 1.23 & 0.09 & 35.6 & 0.07 & 35.4 & 0.852 & 0.936 & 39.21 & - & - & - & - & 0.369 & 0.04 & -54.7 & 3.490\\
  & FOLD & 0.000 & 0.085 & 1.23 & 0.09 & -40.1 & 0.07 & 41.8 & 0.000 & 0.059 & 16.12 & 0.13 & -43.0 & - & 1.933 & 0.102 & 0.07 & 40.1 & 4.290\\
NGC5557 & CUSP & 0.000 & 0.005 & 1.28 & 0.09 & 9.1 & 0.07 & 19.8 & 0.181 & 0.152 & 564.82 & - & - & - & - & 0.231 & 0.06 & 23.3 & 3.500\\
  & FOLD & 0.000 & 0.000 & 1.28 & 0.10 & 9.5 & 0.07 & 19.7 & 0.000 & 0.020 & 14.67 & 0.28 & -21.0 & - & 2.600 & 0.347 & 0.04 & 23.1 & 3.360\\
NGC1272 & CUSP & 0.000 & 0.029 & 1.41 & 0.07 & 15.9 & 0.07 & 19.4 & 0.428 & 0.337 & 6.60 & - & - & - & - & 0.291 & 0.04 & 19.4 & 4.470\\
  & FOLD & 0.000 & 0.019 & 1.40 & 0.07 & 40.4 & 0.07 & 27.9 & 0.000 & 0.010 & 3.29 & 0.64 & 4.9 & - & 2.160 & 0.299 & 0.04 & 26.8 & 5.140\\
NGC6482 & CUSP & 0.002 & 0.164 & 1.69 & 0.17 & -0.8 & 0.05 & 34.4 & 0.001 & 0.188 & 29.40 & 0.25 & -7.0 & - & 3.343 & 0.073 & 0.03 & 36.6 & 26.420\\
  & FOLD & 0.017 & 2.410 & 1.70 & 0.16 & -13.7 & 0.03 & 40.9 & 0.015 & 1.672 & 39.70 & 0.27 & -14.8 & 1.75 & - & 0.146 & 0.04 & 8.4 & 8.800\\

\end{tabular}}
\label{table:fitting_list}

\end{table*} 

\appendix

\onecolumn

\section{\bf From surface brightness to surface mass density}
\label{app:A}

The dimensionless surface mass densities of an NFW halo and an image of a galaxy are given by:
\begin{align*}
\kappa_D \left(r\right) &= 2\kappa_s \frac{1-F\left(r / r_{s}\right)}{\left(r / r_{s}\right)^2-1} = \kappa_s g\left(r\right) & \kappa_B &= \lambda \ c\left(r\right)
\end{align*}
Where

\[
F\left(x\right) = \begin{cases}
\frac{\tanh^{-1}\left[\sqrt{1-x^2} \ \right]}{\sqrt{1-x^2}}; \ \ &x \leq 1 \\
1 \ \ &x=1 \\
\frac{\tan^{-1}\left[\sqrt{x^2-1} \ \right]}{\sqrt{x^2-1}}; \ \ &x \geq 1

\end{cases}
\] 

\noindent Where we adopt a polar coordinate system, and where $x\equiv r/r_s$, with $r_s$ the scale radius of the NFW halo. $\lambda$ is a normalization factor we apply to the images obtained from the HST with units of $[\rm{convergence]}/[\rm{pixel \ count}]$, and c(r) is a 2 dimensional image with pixel values corresponding to photon counts. There are two free parameters $\left(\kappa_s, \lambda\right)$, and two constraints on the surface mass densities of dark matter and baryons $\left(\kappa_D,\kappa_B\right)$: 
\begin{itemize}
\item the average convergence (baryons plus NFW) within the Einstein radius $R_{\rm{Ein}}$ is 1, a standard result for lenses with circular symmetry:
\begin{equation}
\nonumber {\bar\kappa_s}  = \frac{1}{\pi R_{\rm{Ein}}^2} \int dA_{R_{\rm{Ein}}} \kappa \left(x\right) = \frac{1}{\pi R_{\rm{Ein}}^2} \int \left(\kappa_B + \kappa_{D}\right)dA_{R_{\rm{Ein}}} = 1
\end{equation}

\item The contribution to the total convergence within $R_0 = \frac{R_{1/2}}{2}$ from the NFW halo is some fraction $f$:
\begin{equation}
\nonumber \frac{\int dA_{R_0} \ \kappa_{D}}{\int dA_{R_0} \left(\kappa_B + \kappa_{D}\right)} = f  
\end{equation}
\end{itemize}
Inserting the expressions for $\kappa_D$ and $\kappa_B$ into (1) and (2) yields two equations in two unknowns:

\begin{align*}
\frac{1}{\pi R_{\rm{Ein}}^2} \int dA_{R_{\rm{Ein}}} \Big[\kappa_s g\left(r\right) + \lambda \ c\left(r\right)\Big] &= 1 \\
\left[\int dA_{R_0} \kappa_s g\left(r\right)\right]\left[\int dA_{R_0} \left[\kappa_s g\left(r\right) + \lambda c\left(r\right)\right]\right]^{-1}&= f
\end{align*}
\noindent It is useful to introduce the notation:

\begin{align*}
\nonumber \int dA_R \ g\left(r\right) &= 2\pi R_s^2 \ G\left(n\right); \ G\left(n\right) \equiv \log\left(\frac{n^2}{4}\right)+\frac{2\tanh^{-1}\left(\sqrt{1-n^2}\right)}{\sqrt{1-n^2}}; \ n\equiv \frac{R}{R_S} \\
\nonumber 
\int dA_R \ c\left(r\right) &= \pi R^2 \left(\frac{1}{N}\sum_{i,j =0}^{N}  D_{ij}\right) =  \pi R^2 \langle C\left(r\right) \rangle \ \ ; \ \left[r< \ R\right]
\end{align*}
Where the integral over the count map is expressed as a discrete sum over pixels within the radial limit of integration. \\ \\
Solving for $\kappa_s$ and $\lambda$:
\begin{align*}
\kappa_s &= \frac{R_{\rm{Ein}}^2}{2R_s^2} \left[G\left(n_1\right)+G\left(n_2\right)\left(\frac{1-f}{f}\right)\left(\frac{R_{\rm{Ein}}}{R_0}\right)^2  \dfrac{\langle C \rangle _{R_{\rm{Ein}}}}{\langle C \rangle _{R_{0}}}\right]^{-1} \\
\lambda &= \left(\frac{1-f}{f}\right) \frac{R_{\rm{Ein}}^2}{R_0^2} \left[\frac{G\left(n_1\right)}{G\left(n_2\right)}\langle C \rangle _{R_{0}} + \left(\frac{1-f}{f}\right) \left(\frac{R_{\rm{Ein}}}{R_0}\right)^2 \langle C \rangle _{R_{\rm{Ein}}} \right]^{-1}
\end{align*}

\noindent where $\left[n_1 \equiv \frac{R_{\rm{Ein}}}{R_s}, n_2 \equiv \frac{R_0}{R_s}\right]$. The scale radius $R_s$ is taken to be $5 R_{1/2}$, while the number f is taken to be $\frac{1}{3}$, consistent with the results of \citet{Aug++10}. With these choices, the only free parameters are the Einstein radius $R_{\rm{Ein}}$, which we obtain via the measured central velocity dispersion of the galaxy, and the half-light radius for each galaxy found in the literature.\\

\begin{table*}
\centering
\label{table:stellar_mass}
\caption{Stellar mass estimates derived via our normalization procedure and by \citet{Gal++08b}.}
\begin{tabular}{||c c c||}
\hline
Galaxy & $\rm{Log}_{10} M_{\odot}$ (from convergence map) & $\rm{Log}_{10} M_{\odot}$ (from Gallo, Treu et al 2008)\\
\hline
VCC1664 & 10.8 & 10.6 \\
VCC1692 & 10.8 & 10.6 \\
VCC2000 & 10.6 & 10.4 \\
VCC881 & 11.5 & 11.9 \\
VCC731 & 11.3 & 11.7 \\
VCC1903 & 11.3 & 11.3 \\ 
VCC1231 & 10.9 & 10.8 \\ 
VCC355 & 10.6 & 10.3 \\
VCC1062 & 10.6 & 10.7 \\
\hline 
\end{tabular}
\end{table*}

\section{\bf SIMULATING AN EXTENDED SOURCE}
\label{app:B}

\indent To get around the issue of background noise in the optical images that introduces an artificial micro-lensing signal, which results in a large scatter in the distribution of image magnifications, we model the source as an extended object of diameter 5 parsecs in the source plane.\\
\indent To model an extended source, we take the original source position as the center of a 2-d Gaussian, characterized by a covariance matrix 
\[
\Sigma_0 = \begin{bmatrix}
\sigma_0^2 & 0 \\ 
0 & \sigma_0^2 \\
\end{bmatrix}
\]
where we take $\sigma_0 = 2.5$ pc. We take the area of the source to be 
\begin{equation}
\nonumber A_{src} = \pi \sigma_0^2
\end{equation}
which corresponds to a circle in source plane of diameter 5 parsecs. We draw 100 random source positions from this distribution and solve the lens equation with gravlens for each one, yielding 4 images positions for each source position. This process results in 4 clusters of 100 points each, with each cluster described by its own covariance matrix describing an ellipse in the image plane. The area of this ellipse is given by
\begin{equation}
\nonumber A_{img} = \pi \sqrt{\lambda_1 \lambda_2}
\end{equation}
where $\left(\lambda_1,\lambda_2\right)$ are the eigenvalues of the covariance matrix describing the 100 $\left(x,y\right)$ coordinates for each of the four images. The magnification for each image is then given by the ratio of $A_{img}$ to the area of the source:
\begin{equation}
\nonumber M_{i} = \frac{A_{img}}{A_{src}}
\end{equation}

\begin{figure*}
\includegraphics[trim=0cm 0cm 0cm 0cm,clip,width=.48\textwidth]{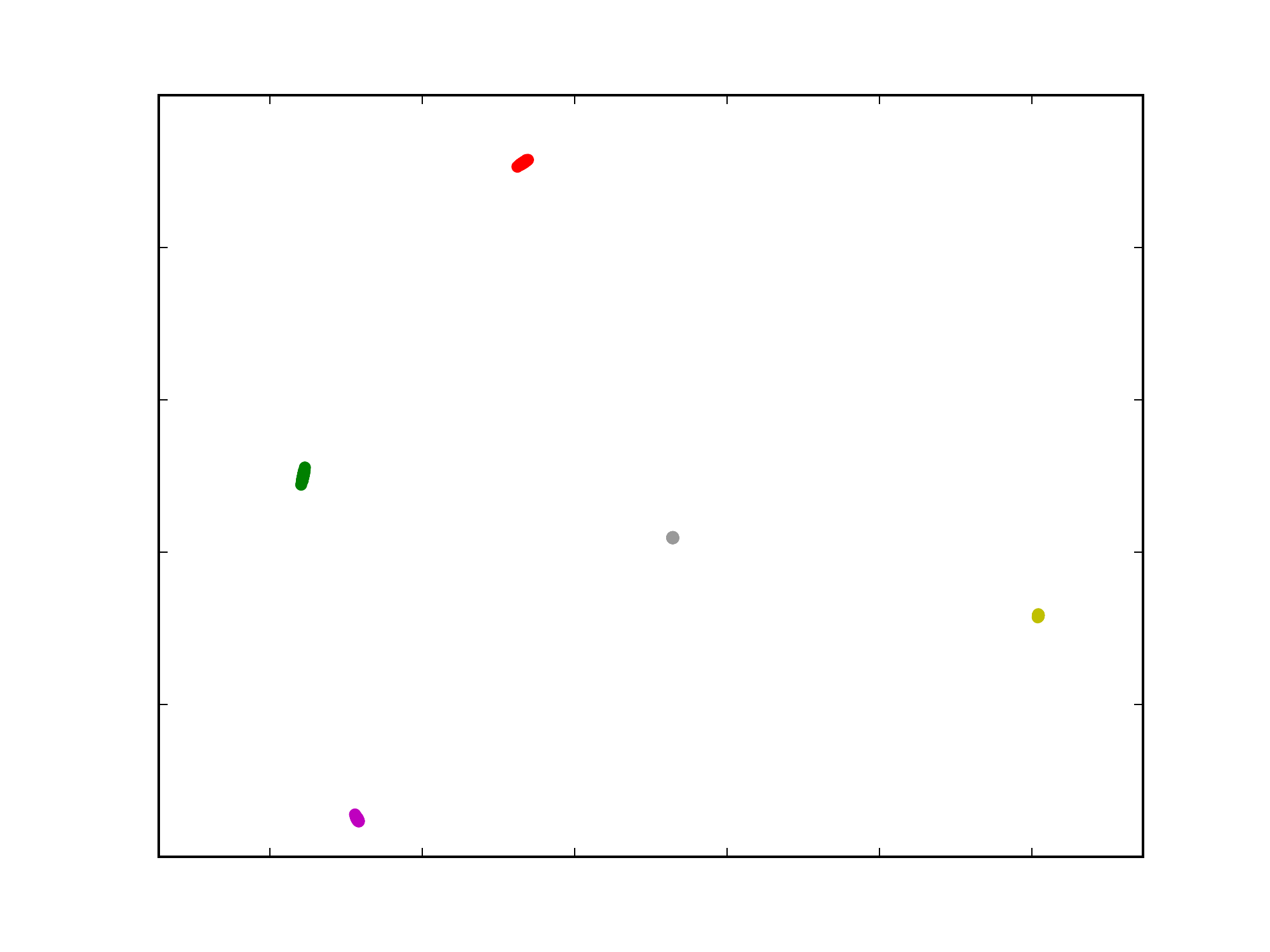}
\includegraphics[trim=0cm 0cm 0cm 0cm,clip,width=.48\textwidth]{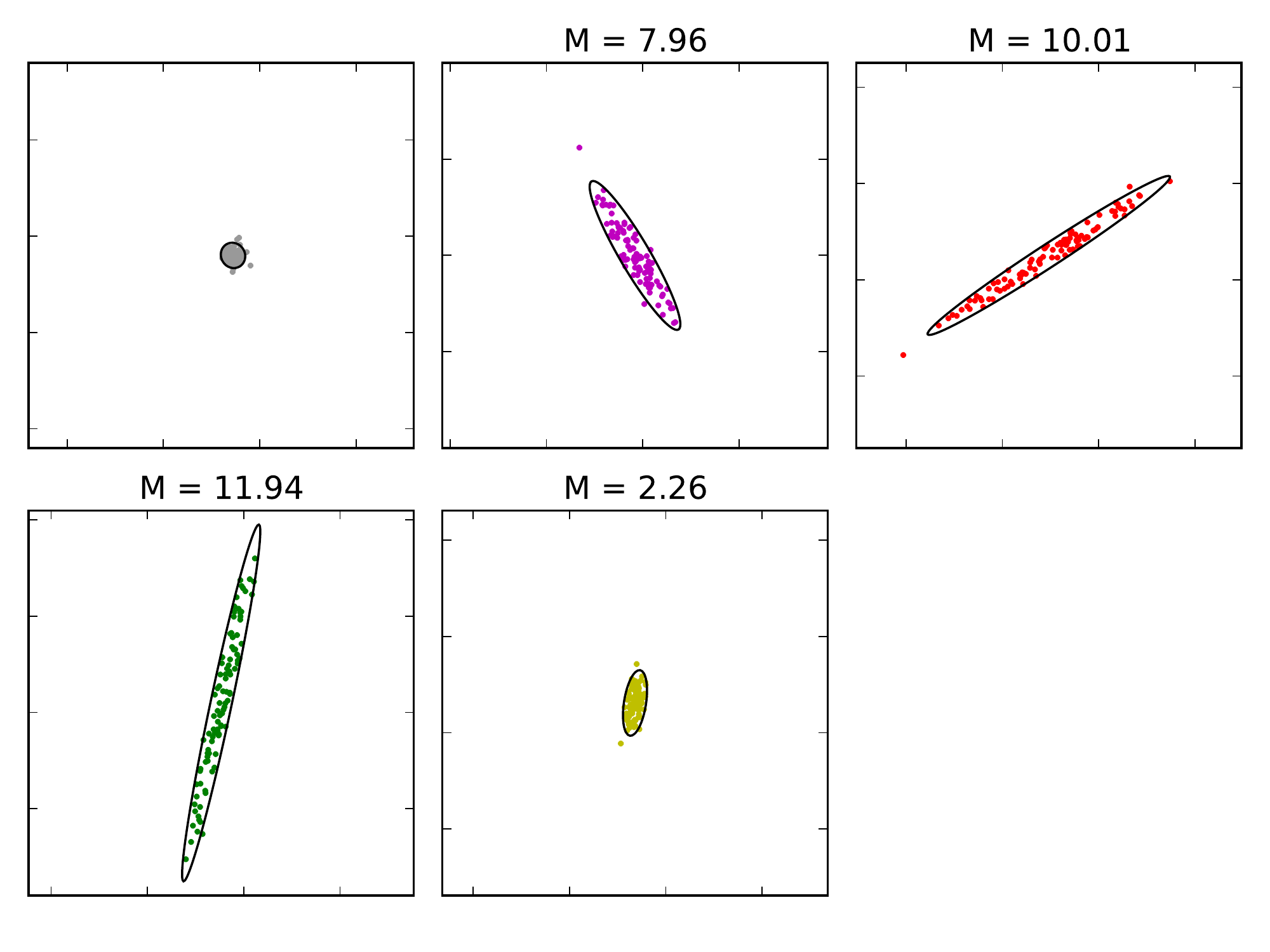}
\caption{The individual panels on the right hand side show zoomed-in images of the clusters of points in the left panel. The grey points are drawn from a circular Gaussian distribution, centered at a reference source position, simulating an extended background source of diameter 5 pc. For each of these source points, we use \textit{gravlens} to solve the lens equation, resulting in 4 additional points, each representing an image produced by the lens system. After repeating this procedure 100 times, the area of the ellipse describing the covariance matrix for each set of points is used to compute the compute the magnification. This procedure is repeated for each of the 250 randomly sampled reference source positions. The ellipses in the right-hand panels correspond to 90\% confidence intervals.}
\end{figure*}

\end{document}